\documentclass[final]{siamltex}
\usepackage[sans]{dsfont}
\usepackage[T1]{fontenc}
\usepackage[english]{babel}
\usepackage{amssymb}
\usepackage{latexsym}
\usepackage{mathrsfs}
\usepackage{amscd}
\usepackage{graphicx}
\usepackage{float}
\usepackage{showkeys}
\usepackage{amsmath}
\usepackage{amsfonts}
\usepackage{enumerate}

\newtheorem{thm}{Theorem}[section]
\newtheorem{rmk}{Remark}[section]
\newtheorem{alg}{Algorithm}[section]

\title{Binary interaction algorithms for the simulation of
flocking and swarming dynamics\thanks{This work was partially supported by National Institute for High Mathematics, Scientific Computing Group (INDAM-GNCS), Rome, Italy.}}

\author{G. Albi\footnote{University of Ferrara, Department of
Mathematics, Via Machiavelli 35, I-44121 Ferrara, Italy. {\tt
\{giacomo.albi,lorenzo.pareschi\}@unife.it}}\and L. Pareschi\footnotemark[2]}

\begin{document}
\maketitle
\begin{abstract}
Microscopic models of flocking and swarming takes in account large
numbers of interacting individuals. Numerical resolution of large
flocks implies huge computational costs. Typically for $N$
interacting individuals we have a cost of $O(N^2)$. We tackle the
problem numerically by considering approximated binary interaction
dynamics described by kinetic equations and simulating such
equations by suitable stochastic methods. This approach permits to
compute approximate solutions as functions of a small scaling parameter $\varepsilon$ at a reduced complexity of $O(N)$ operations. Several numerical results show
the efficiency of the algorithms proposed.
\end{abstract}

\begin{keywords}Kinetic models, mean field models, Monte Carlo methods, flocking, swarming, collective behavior
\end{keywords}

\begin{AMS}
65C05, 65Y20, 82C80, 92B05
\end{AMS}

\pagestyle{myheadings}
\thispagestyle{plain}
\markboth{G. ALBI AND L. PARESCHI}{Binary collision algorithms for flocking and swarming}

\section{Introduction}
The study of mathematical models describing collective behavior and synchronized motion of animals, like bird flocks, fish schools and insect swarms, has attracted a lot of attention in recent years~\cite{MR2765734, MR2507454, MR2744705, MR2744704, MR2761862, MR2324245, d2006self, motsch2011new, vic1995}. In biological systems such behaviors are observed in every level of the food
chain, from the swarm intelligence of the zoo
plankton, to bird flocking and fish schools,
to mammals moving in formation~\cite{fernandez2004flock, hildenbrandt2010self, okubo86}. Beside biology, emerging collective behaviors play a relevant role in several applications involving the dynamics of a large number of individuals/particles which range from computer science~\cite{reynolds1987flocks}, physics~\cite{MR918448} and engineering~\cite{lee2009fast} to social sciences and economy~\cite{MR2300700}. We refer to~\cite{MR2761862} for a recent review of some of the mathematical topics and the applications involved.
    
    Naturally occurring synchronized motion
has inspired several directions of research
within the control community. A well-known application is related to formation flying missions and missions involving the coordinated control of several autonomous
vehicles~\cite{Phd2009}. There are several current projects which are dealing with the formation flying and coordinated
control of satellites, like the DARWIN project of the European Space Agency (ESA) with the goal of launching a space-based
telescope aiding in the search for possible life-supporting planets, or the PRISMA project led by the Swedish Space Corporation (SSC) which will be the first real formation flying space mission launched~\cite{Damico2008}. 
    
In this manuscript we will focus on general models which are capable to reproduce flocking, swarming and other collective  behaviors. Most of the classical models describing these phenomena are based on the simple definition of three interacting zones, the so called \emph{three-zones model}~\cite{Aoki_1982,huth1994simulation}.
    
\begin{figure}[ht]
    \centering
        \includegraphics[scale=0.4]{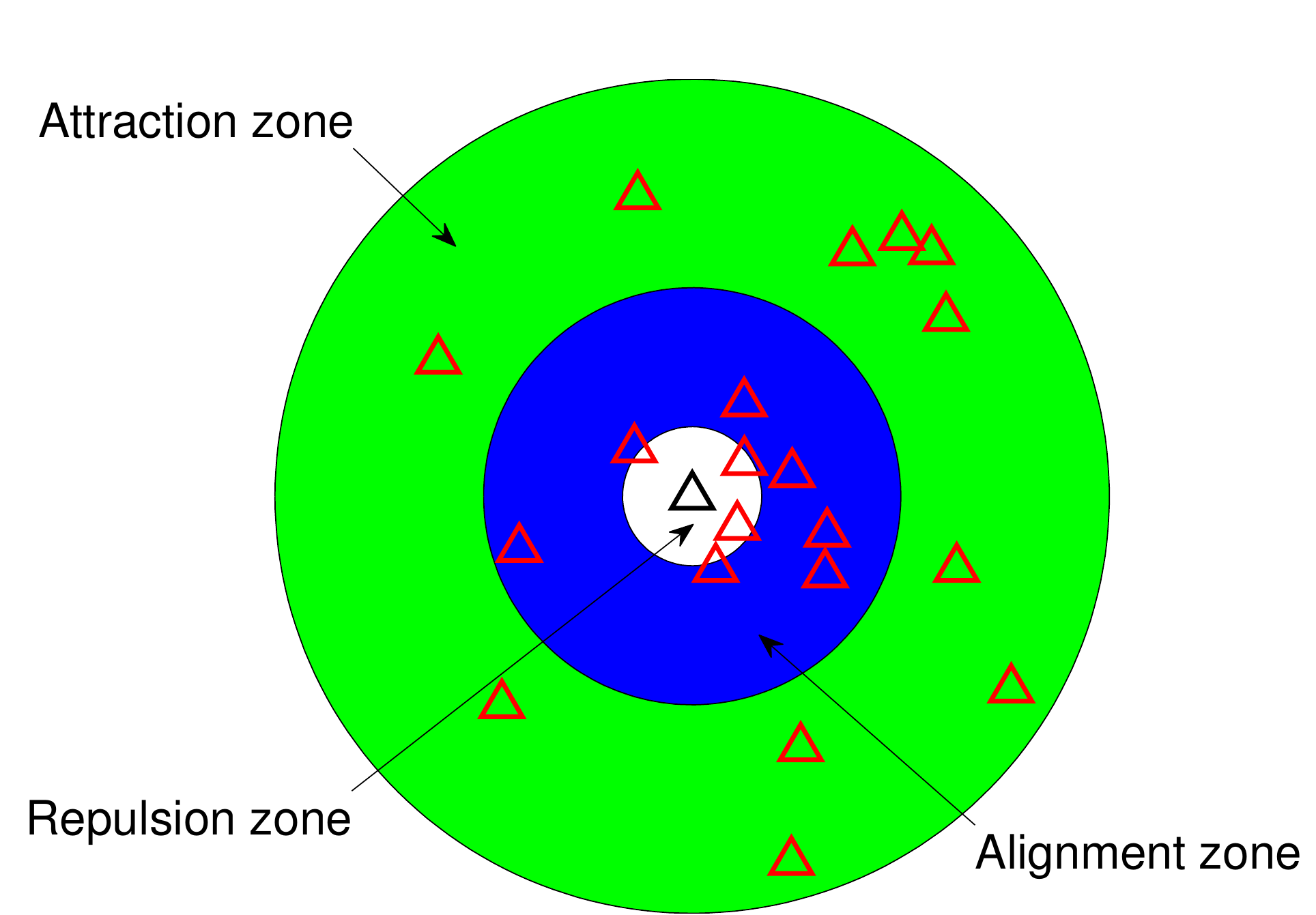}
\caption{Sketch of the three-zone model}
    \label{fig:ZoneModel}
    \end{figure}

    Let us briefly summarize the three-zones assumptions. We define three regions around each individual: a short-range repulsion zone, an intermediate velocity alignment zone and a long-range attraction zone (see Figure~\ref{fig:ZoneModel}). Each interaction between individuals is evaluated accordingly to the relative position in the model.
    \begin{itemize}
        \item \emph{Repulsion zone:} when individuals are too close each other they tend to move away from that area.
        \item \emph{Alignment zone:} individuals try to identify the possible direction of the group and to align with it.
        \item\emph{Attraction zone:} when individuals are too far from the group they want to get closer.
    \end{itemize}
    
Typically different interaction models are taken in the different zones~\cite{MR2765734, d2006self} or the modelling is focused on a specific zone, like the alignment/consensus dynamic~\cite{MR2324245, motsch2011new}. Of course the particular shape and size of the zones depends on the specific application considered. For example recent studied on birds flocks suggest that each bird modifies its position, relative to few individuals directly surrounding it, no matter how close or how far away those individuals are~\cite{ballerini2008interaction}. It is not clear however if this applies also to other kind of animals.

    Studying this kind of dynamics for large system of individuals implies a considerable effort in numerical simulations, 
    microscopic models based on real data my take into account very large numbers of interacting individuals (from several hundred thousands up to millions). Computationally the problems have the same structure of many classical problems in computational physics which require the evaluation of all pairwise long range interactions in a large
    ensembles of particles.
    The $N$-body problem of gravitation (or electrostatics) is a classical example.  Such problems involve the evaluation of summations of the type
    \begin{equation}
        S^N_i=\sum_{j=1}^{N}w_j K(x_i,x_j),\qquad \forall\,\, i.
    \end{equation}
    A direct evaluation of such sums at $N$ target points clearly requires $O(N^2)$ operations and algorithms which reduce
    the cost to $O(N^{\alpha})$ with $1\leq\alpha< 2$, $O(N\log N)$ etc. are
    referred to as \emph{fast summation methods}.
     For uniform grid data the most famous of these is certainly the Fast Fourier Transform (FFT). In the case of general data
most fast summation methods are approximate methods based on analytical considerations, like
    the \emph{Fast Multipole} method~\cite{MR918448}, \emph{Wavelets Transform} methods~\cite{rokh91} and, more recently,  dimension
    reduction using \emph{Compressive Sampling} techniques~\cite{MR2300700}, or based on some \emph{Monte Carlo} strategies at different levels~\cite{caflisch1998monte}. Extensions of the above mentioned approaches to kinetic equations are discussed for example in~\cite{mp06,lemou98,lemou03,fornasier2011particle, MR1865188}.

    From a mathematical modeling point of view, these problem have been developed extensively in the kinetic research community (see~ \cite{MR2374464,MR2425606,MR2744704}) where the derivation of kinetic and an hydrodynamic equations represent a first step towards the reduction of the computational complexity. Of course, passing from a microscopic description based on phase-space particles $(x_i(t),v_i(t))$ to a mesoscopic level where the object of study is a particle distribution function $f(x,v,t)$ redefines the model in a new one where new methods of solution are required.

    In this paper we are going to follow this research path in two main directions: first we review the derivation of the different kinetic approximations from the original microscopic
    models and then we introduce and analyze several stochastic Monte Carlo methods to approximate the kinetic equations.
    Monte Carlo methods are the most well-known approach for the numerical solution of the Boltzmann equation of rarefied gases in the short-range interactions, and many efficient algorithms have been presented~\cite{bird1963approach, MR845926,caflisch1998monte,MR1865188}. On the other hand the literature on efficient Monte Carlo strategies for long-range interactions, and thus Landau-Fokker-Plank equations, is much less developed but of great interest in the field of plasma physics~\cite{bobylev2000theory,MR2812253}.

Here, inspired by the techniques introduced in~\cite{bobylev2000theory, MR2812253} for plasma, we develop direct simulation Monte Carlo methods based on a binary collision dynamic described by the corresponding kinetic equation. The methods permit to approximate the microscopic dynamic at a cost directly proportional to the number of sample particles involved in the computation, thus avoiding the quadratic computational cost. The limiting behavior characterizing the mean-field interaction process of the particles system is recovered under a suitable asymptotic scaling of the binary collision process. In such a limit we show that the Monte Carlo methods here developed are in very good agreement with the direct evaluation of the original microscopic model but with a considerable gain of computational efficiency.

The rest of the manuscript is organized as follows.
In the first section we present some of the classical microscopic models for flocking and swarming. Generalization of the notion of visual cone~\cite{MR2744704} are also discussed.
Since the interaction is non local, the derivation of the limiting mean-field kinetic equation is made through a \emph{Povzner-Boltzmann} kinetic equation~\cite{MR0142362} in the anologous situation of the so-called \emph{grazing collision
limit}~\cite{MR2596552}. To solve the resulting Boltzmann-like mesoscopic partial differential
equation we introduce different stochastic binary interaction algorithms and 
compare their computational efficiency and accuracy with a direct evaluation of the microscopic models and a stochastic approximation of the mean-field kinetic model. We show that the new approach permits to
reduce the overall cost from $O(N^2)$ to $O(N)$ operations. In particular we show that the choice $\varepsilon=\Delta t$, where $\varepsilon$ is the small scaling parameter leading to the mean field kinetic model, originates binary interaction algorithms consistent with the limiting behavior of the particle system.
Furthermore, in contrast with classical methods~\cite{bobylev2000theory, MR2812253}, the nature of the approximating equations is such that the resulting Monte Carlo algorithms are fully mesh less.  
In the last section of the paper we report several simulations in two and three space dimensions of different microscopic models solved by the binary Monte Carlo method in the above scaling.


\section{Microscopic models}

In this section we review some well-known microscopic models of flocking and swarming (see~\cite{MR2324245,d2006self,motsch2011new} and the references therein).
We are interested in the study of a dynamical system composed of $N$ individuals with the following general structure
    \begin{equation}
        \left\{
        \begin{array}{l}
        \dot{x}_i=v_i \\\\
        \displaystyle \dot{v}_{i}=S(v_i)+\frac1{N}\sum_{j=1}^N \left[H(x_i,x_j)(v_j-v_i)+A(x_i,x_j)+R(x_i,x_j)\right]\psi_{\alpha}(x_i,x_j,v_i)
        \end{array}\right.
         i= 1,\ldots,N,
    \label{model}
    \end{equation}
where $(x_i,v_i)$ lives in $\mathbb{R}^{2d}$, $d\geq1$, $S(v_i)$ describes a self-propelling term, $H(x_i,x_j)$ the alignment process, $A(x_i,x_j)$ the attraction dynamic and the term $R(x_i,x_j)$ the short-range repulsion. In (\ref{model}) the multiplicative factor $\psi_{\alpha}(x_i,x_j,v_i)\in [0,1]$ takes into account the effects of space perception as a function of some vector of parameters $\alpha$.  


\subsection{Cucker and Smale model}
    Cucker and Smale model is a pure alignment model, no repulsion or attraction or other effects are
    taken in account, see \cite{MR2324245, MR2295620} and \cite{MR2596552}. The classical model reads as follow
    \begin{equation}
        \left\{
        \begin{array}{l}
        \dot{x}_i=v_i \\\\
        \displaystyle\dot{v}_{i}=\frac{1}{N}\sum_{j=1}^N H(\left|x_j-x_i\right|)(v_j-v_i)
        \end{array}\right.
        \qquad\qquad i= 1,\ldots,N,
    \label{CuckerSmale}
    \end{equation}
    where $H(\left|x_j-x_i\right|)$ is a function that measures the
    strength of the interaction between individuals $i$ and $j$ , and depends on the mutual distance, under the assumption that
   closer individuals have more influence than the far distance ones. 

    A typical choice of function $H$ is the following
    \begin{equation}
        H(r)=\frac{K}{(\varsigma^2+r^2)^\gamma},
    \label{Hxy}
    \end{equation}
    where $K,\varsigma>0$ are positive parameters and $\gamma\geq0$.
\begin{figure}[t]
    \centering
        \includegraphics[scale=0.7]{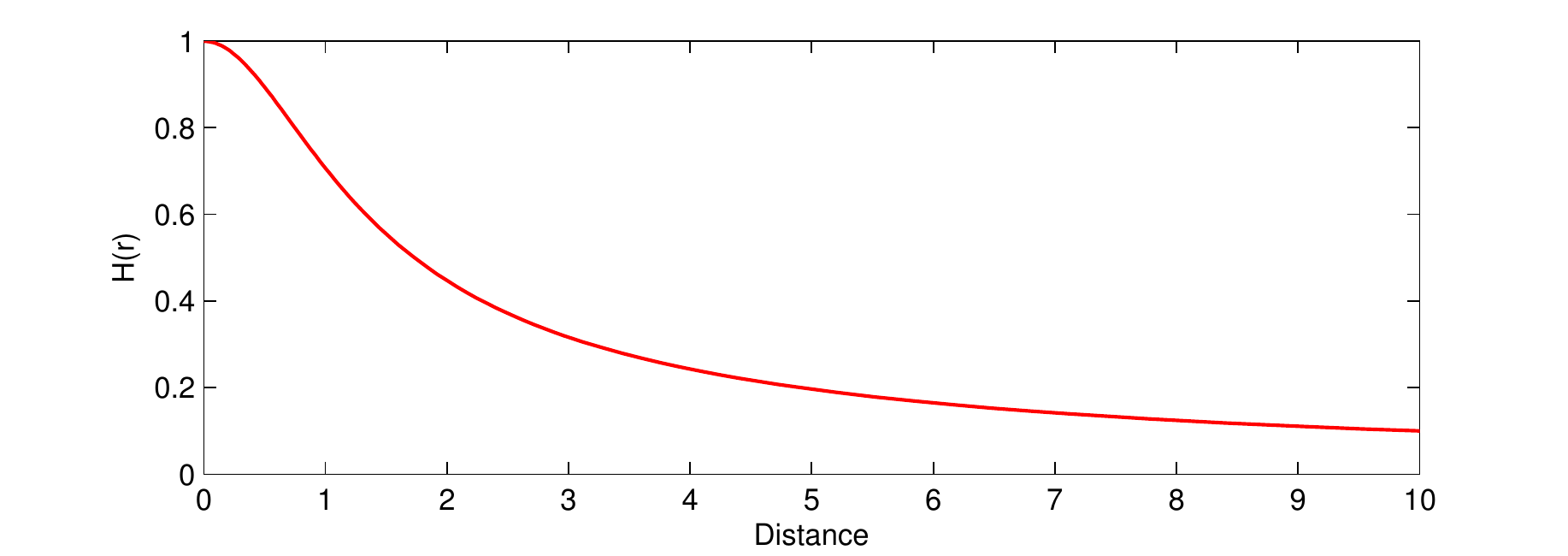}    
    \caption{Profile of the function $H$ in Cucker-Smale model}
    \end{figure}
    Under this assumptions it can be shown that well-posedness holds for the initial value problem of (\ref{CuckerSmale}) and the
    solution is mass and momentum preserving, with compact support for position and velocity, see \cite{MR2536440}.

    Moreover in \cite{MR2324245,MR2596552} it was established that the parameter $\gamma$ discriminate the
    behavior of the solution, in the following way

    \begin{thm}
    Let $\Gamma(t)=\frac{1}{2}\sum_{i\neq j}|x_i(t)-x_j(t)|^2$ and $\Lambda(t)=\frac{1}{2}|v_i(t)-v_j(t)|^2$.
    If $\gamma\leq\frac{1}{2}$ then
    \begin{enumerate}[(i)]
        \item exist a positive costant $B_0$ such that: $\Gamma(t)\leq B_0$ for all $t\in\mathbb{R}$.
        \item $\Lambda(t)$ converge towards zero as  $t\rightarrow\infty$.
        \item The vector $x_i-x_j$ tends to a limit vector $\hat{x}_{ij}$, for all $i,j=1,\ldots,N$.
    \end{enumerate}
    \end{thm}

    In other words, the velocity support collapses exponentially to a single point and the flock holds the same disposition. From
    this theorem we recover the notion of \emph{unconditional flocking} in the regime $\gamma\leq \frac{1}{2}$.
    If $\gamma>\frac{1}{2}$ in general unconditional flocking doesn't follow, but under some conditions on initial data
     flocking condition is reached, see \cite{MR2744704}.
     
    Note that standard Cucker-Smale model prescribes perfectly symmetric interactions and takes in account only the {alignment dynamic}. As a result total momentum is preserved by the dynamics. The introduction of a limited space perception (like a visual cone) breaks symmetry and momentum conservation. This choice corresponds to take a function for the strength of the interaction of the type 
    \begin{equation}
      H_\alpha(x_i,x_j,v_i)=H(|x_i-x_j|)\psi_\alpha(x_i,x_j,v_i),
    \label{csH2}
\end{equation}    
     where the parameter vector $\alpha$ is related, for example, to the width of the visual cone. 


\subsection{D'Orsogna-Bertozzi et al. model}
    The microscopic model introduced by D'Orsogna, Bertozzi et al.~\cite{d2006self} considers a self-propelling, attraction and repulsion dynamic and reads 
    \begin{equation}
        \left\{
        \begin{array}{l}
        \dot{x}_i=v_i \\\\
        \displaystyle\dot{v}_{i}=(a-b|v_i|^2)v_i-\frac{1}{N}\sum_{j\neq1}\nabla_{x_i} U(\left|x_j-x_i\right|)
        \end{array}\right.
        \qquad\qquad i= 1,\ldots,N,
    \label{model_Bert}
    \end{equation}
    where $a$, $b$ are nonnegative parameters, $U:\mathbb{R}^d\longrightarrow\mathbb{R}$ is a given
    potential modeling the short-range repulsion and long-range attraction, and $N$ is the
    number of individuals.
    Function $U$ gives us the {attraction-repulsion dynamic} typically described by a \emph{Morse potential}
    \begin{equation}
      U(r)=-C_A e^{-r/l_A}+C_R e^{-r/l_R},
    \label{pot}
    \end{equation}
    where $C_A, C_R, l_A, l_R$ are positive constants measuring the strengths and the characteristic lengths of the attraction and repulsion.
     In (\ref{model_Bert}) the term $(a-b|v_i|^2)v_i$ characterizes \emph{self-propulsion} and \emph{friction}. Asymptotically this term give us a desired velocity, in fact for large times the velocity of every single particle tends to $\sqrt{a/b}$.
     
    The most interesting case in biological applications occurs when the constants in the Morse potential satisfy the following inequalities $C:=C_R/C_A>1$ and $l:=l_R/l_A<1$, which correspond to the long range attraction and short range repulsion.
    Moreover the choice of the parameters fixes the evolution of the $N$ particles system towards a particular equilibrium. The following distinction holds: if $Cl^d>1$ then crystalline patterns are observed and for $Cl^d<1$ the motion of particles converges to a circular motion of constant speed, where $d\geq 2$ is the space dimension. In \cite{d2006self} a further study of the parameters can be found.
    
\begin{figure}[t]
    \centering
        \includegraphics[scale=0.5]{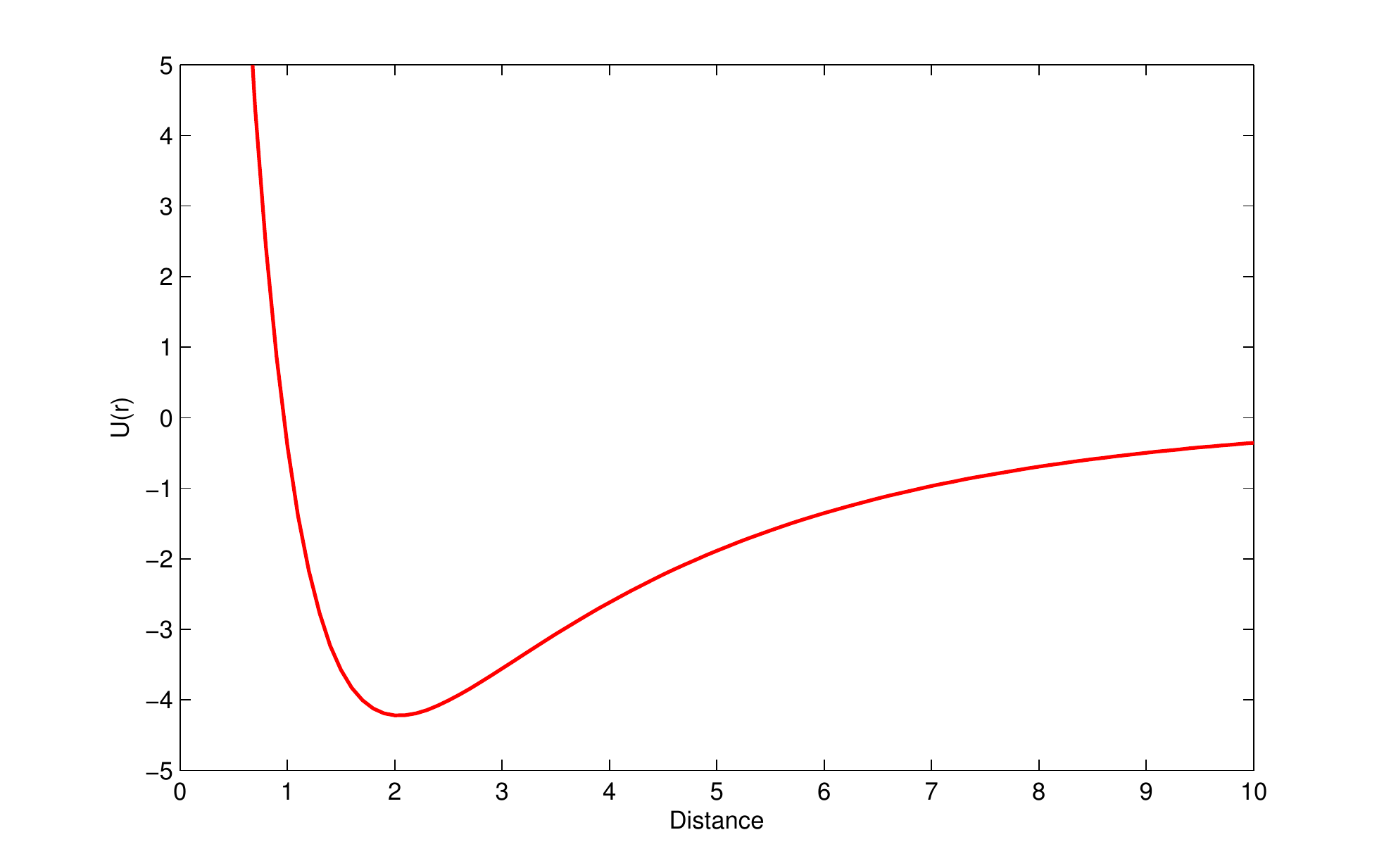}   
    \caption{The Morse potential in D'Orsogna-Bertozzi et al. model}
    \end{figure}

    \subsection{Motsch-Tadmor model}
    In a recent work \cite{motsch2011new} the authors propose a modification of the classical Cucker-Smale model as follows
    \begin{equation}
        \left\{
        \begin{array}{l}
        \dot{x}_i=v_i \\\\
        \displaystyle\dot{v}_{i}=\frac1{N}\sum_{j=1}^{N} h(x_i,x_j)(v_j-v_i),
        \end{array}\right.
        \qquad\qquad i= 1,\ldots,N,
    \label{Motsch-Tadmor}
    \end{equation}
    where $h$ is defined by $$h(x_i,x_j)=\frac{H(|x_i-x_j|)}{\bar{H}(x_i)},\qquad \bar{H}(x_i)=\frac1{N}\sum_{k=1}^{N}H(|x_i-x_k|).$$
     The model differs from the classical one, since the influence between two particles $H(|x_j-x_i|)$ is weighted by the average  influence on the single individual $i$.
     In this way the function $h(x_i,x_j)$ lose in general any kind of symmetry property of the original Cucker-Smale dynamic.
     
     We emphasize, however, that in our general setting this model is included in the Cucker-Smale alignment dynamic of type (\ref{csH2}) with a particular choice for the function defining the space perception of the form
\begin{equation}
\psi_{\alpha}(x_i,x_j,v_i)=\frac{1}{\bar{H}(x_i)}.
\end{equation}
This can be interpreted as a higher perception level of zones where the individuals have a higher concentration and a lower interest in zones where individuals are more scattered.     
     
    \subsection{Perception cone, topological interactions and roosting force}
\begin{figure}[t]
    \centering
        \includegraphics[scale=0.50]{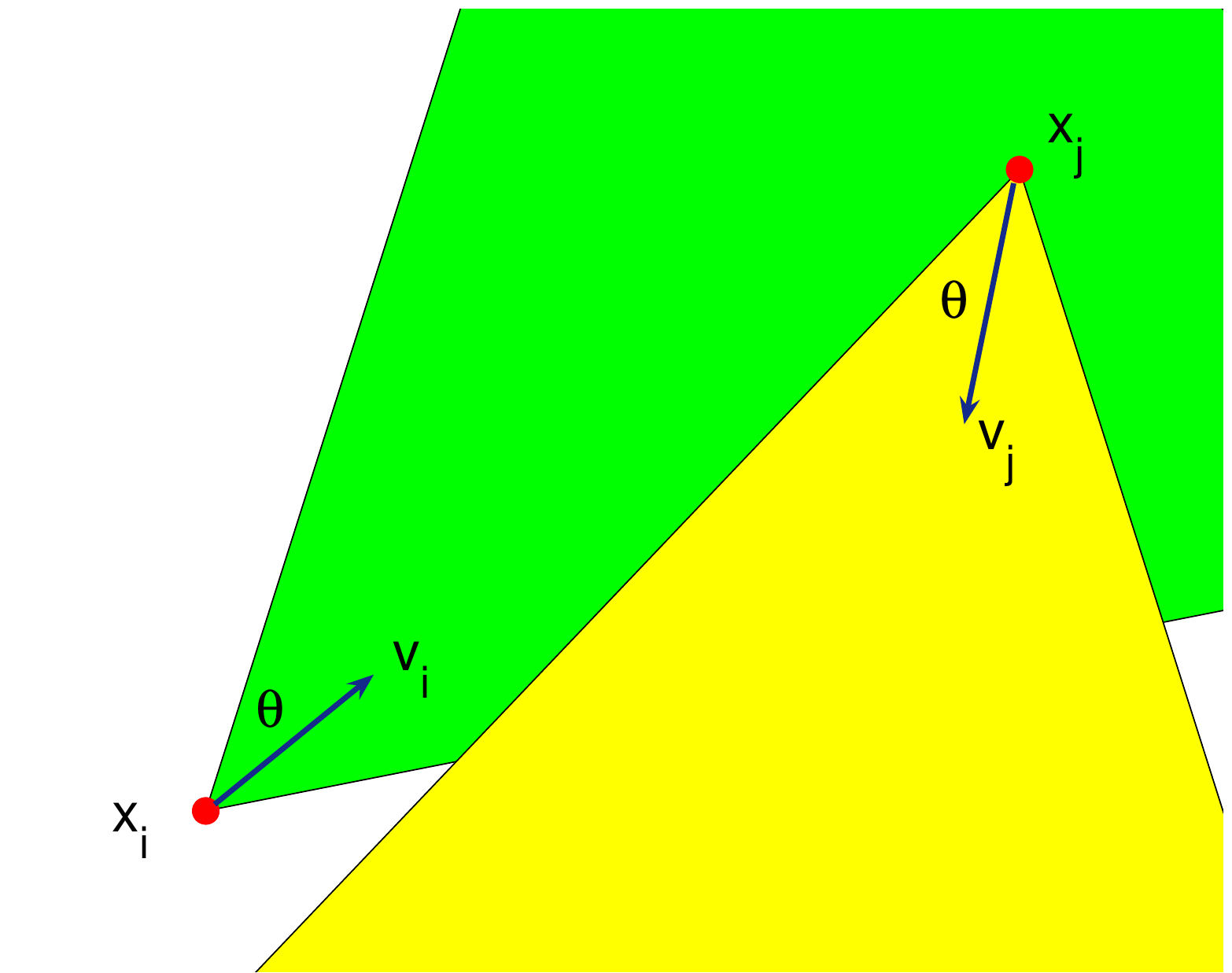}
    \caption{One of the possible configurations in the interaction with a perception cone. Individual $j$ is perceived by individual $i$ but not vice versa.}
    \label{fig:ZoneModelCone}
    \end{figure}

    For interacting animals like birds, fishes, insects the visual perception of the single individual plays a fundamental role~\cite{MR2027139, fernandez2004visual, fernandez2004flock}.
    In \cite{MR2744704} the authors introduce in the dynamic a further rule: the \emph{visual cone}.
    A {visual cone} identifies the area in which interaction is possible and blind area where can not be interaction.
    Mathematically speaking the {visual cone} depends on an angle, $\theta$, that give us the visual width. Together with
    position and velocity the visual area can be described as follows
    \begin{equation}
        \Sigma(x_i,v,\theta)=\left\{y\in\mathbb{R}^d:\frac{(x_i-y)\cdot v_i}{\left|(x_i-y)\right|\left|v_i\right|}\geq\cos(\theta/2)\right\}.
    \label{eq:cono}
    \end{equation}
    As already discussed the introduction of a visual cone breaks the typical symmetry of the interaction (see Figure \ref{fig:ZoneModelCone}).

    The drawback of this choice is that a single individual that
    has no one in his visual cone, never changes his direction.
    For real situations this assumption is clearly too strong, since many other stimuli are received by the surrounding. We cannot ignore other perceptions like hearing, smell and visual memory.
    For example fishes use their visual perception mostly on large/medium distance whereas on medium/short distance they rely on their lateral line.
    These observations lead naturally to improve the idea of a visual cone by introducing a \emph{perception cone} as follows: we assign two different weights
    measuring the strength/probability of the interaction. A weight $p_1$ in the case of strong perception and $p_2$ in case of weak perception, with $0\leq p_2\leq p_1\leq 1$.
    Note that taking $p_1=1$ and $p_2=0$ we have the standard visual cone. 
    For example, in the simulation section we consider a perception cone $\psi_\alpha$, $\alpha=(\theta,p_1,p_2)$, with the following form
	\begin{equation}
\psi_\alpha(x_i,x_j,v_i)=p_2+(p_1-p_2)\mathds{1}_{\Sigma(x_i,v_i,\theta)}(x_j),
	\label{HHH}
	\end{equation}
	where $\mathds{1}_{\Sigma(x_i,v_i,\theta)}(\cdot)$ is the indicator function of the set $\Sigma(x_i,v_i,\theta)$ defined according to 
	(\ref{eq:cono}).
	
    Related efforts to improve the dynamic consider also different ingredients like \emph{topological interactions} where
    individuals interact only with the closest individuals and with a limited number of them, see \cite{ballerini2008interaction, MR2744705}.
    Another variant concerns the introduction of a term describing a \emph{roosting force}~\cite{carrillo2010self, MR2765734}. In fact, flocking
    phenomena tends to stay localized in a particular area, this force acts orthogonal to the single velocity, giving each
    particle a tendency towards the origin.	

\section{Kinetic equations}
For a realistic numerical simulation of a flock the number of interacting individuals can be rather large, thus we need to solve a very large system of ODEs, which can constitute a serious difficulty.
    An alternative way to tackle this problem is to consider a nonnegative distribution function $f(x,v,t)$ describing the number density of individuals at time $t\geq 0$ in position $x\in\mathbb{R}^d$ with velocity $v\in\mathbb{R}^d$. The evolution of $f(x,v,t)$ is characterized by a kinetic equation which takes into account the motion of individuals due to their own velocity and the velocity changes due to the interactions with other individuals.
    Following~\cite{MR2744704} we consider here binary interaction Boltzmann-type and mean-field kinetic approximation of the microscopic dynamics. 

\subsection{Boltzman-Povzner kinetic approximations}

In agreement with (\ref{CuckerSmale}) and (\ref{csH2}), we consider a microscopic binary interaction between two individuals with positions and velocities $(x,v)$ and $(y,w)$ according to
\begin{equation}
\left\{
\begin{array}{l}
v^*=(1-\eta(H_\alpha(\left|x-y\right|,v))v+\eta H_\alpha(\left|x-y\right|,v)w,\\\\
w^*=\eta H_\alpha(\left|x-y\right|,w)v+(1-\eta(H_\alpha(\left|x-y\right|,w))w,
\end{array}\right.
\label{Binary}
\end{equation}
where $v^*,w^*$ are the \emph{post-interaction} velocities and
$\eta$ a parameter that measures the strength of the interaction.
Analogous binary interactions can be introduced for other
swarming and flocking dynamics like D'Orsogna-Bertozzi.

    We describe the interaction of the sistem with following integro-differential equation of Boltzmann type
    \begin{eqnarray}
    \nonumber
    (\partial_t f+v\cdot\nabla_x f )(x,v,t)=\frac{1}{\varepsilon}Q(f,f)(x,v,t),\\[-.25cm]
 \label{Q_Boltz}
    \\[-.25cm]
    \nonumber
        Q(f,f)=\int_{\mathbb{R}^{2d}}(\frac{1}{J}f(x,v_*,t)f(y,w_*,t)-f(x,v,t)f(y,w,t))dw dy,
    \end{eqnarray}
    where $(v_*,w_*)$ are the \emph{pre-interacting} velocity that generate the couple $(v,w)$ according to (\ref{Binary}), $J$ is the Jacobian of the transformation of $(v,w)$ to $(v_*,w_*)$. Without visual limitation the Jacobian reads $J=(1-2\eta H(|x-y|))^d$.
    Note that, at variance with classical Boltzmann equation the interaction is non local as in Povzner kinetic model \cite{MR0142362}.

    Let us introduce the time scaling \begin{equation}t\to t/\varepsilon,\qquad\eta=\lambda\varepsilon,\end{equation} where $\lambda$ is a constant and $\varepsilon$ a small parameter. The scaling corresponds to assume that the parameter $\eta$ characterizing the strength of the microscopic interactions is small, thus the frequency of interactions has to increase otherwise the collisional integral will vanish.
    This corresponds to large
    scale interaction frequencies and small interaction strengths, in agreement with a classical mean-field limit and similarly to the so-called \emph{grazing collision limit} of the Boltzmann equation for granular gases~\cite{McNamara1995}.

\subsection{Derivation of the mean-field kinetic model}
    First of all let us remark that the dynamic (\ref{Binary}) doesn't preserve the momentum, as consequence of the velocity dependent function $H_\alpha$ we have
\begin{equation}
v^*+w^*=v+w-\eta(H_\alpha(\left|x-y\right|,w)-H_\alpha(\left|x-y\right|,v)).
\end{equation}
Moreover under the assumptions $|H_\alpha(r,v)|\leq 1$ and $\eta\leq 1/2$ , it is easy to prove that the support of velocity is limited by initial condition
\begin{equation}
v^{*}=(1-\eta H_\alpha(\left|x-y\right|,v))v+\eta H_\alpha(\left|x-y\right|,v)w\leq \max\{|v|,|w|\}.
\end{equation}

    Considering now the weak formulation of (\ref{Q_Boltz}) the Jacobian term disappears and we get the rescaled equation
    \begin{eqnarray}
        \nonumber
        \frac{\partial}{\partial t}\int_{\mathbb{R}^{2d}}\phi(x,v)f(x,v,t)dv
        dx+\int_{\mathbb{R}^{2d}}(v\cdot\nabla\phi(x,v))f(x,v,t)dv dx=\\[-.25cm]
        \label{weak}
        \\[-.25cm]\nonumber
        \frac{1}{\varepsilon}\int_{\mathbb{R}^{4d}}(\phi(x,v^*)-\phi(x,v))f(x,v,t)f(y,w,t)dv dx dw dy,
    \end{eqnarray}
    for $t>0$ and for all $\phi\in C^\infty_0(\mathbb{R}^{2d})$, such that
    \begin{equation}
        \lim_{t\rightarrow 0}\int_{\mathbb{R}^{2d}}\phi(x,v)f(x,v,t)dv dx=\int_{\mathbb{R}^{2d}}\phi(x,v)f_0(x,v,t)dv dx,
    \end{equation}
    where $f_0(x,v)$ is the starting density.

For small values of $\varepsilon$ we have $v^*\approx v$ thus we can consider the Taylor expansion of $\phi(x,v^*)$
    around $v$ up to the second order we obtain the following formulation to the collisional integral
    \begin{multline}
        \frac{1}{\varepsilon}\int_{\mathbb{R}^{4d}}{(\phi(x,v^*)-\phi(x,v))f(x,v,t)f(y,w,t)dv dx dw dy}=\\
        =\lambda\underbrace{\int_{\mathbb{R}^{4d}}(\nabla_v\phi(x,v)\cdot(w-v))H_\alpha(x,y,v)f(x,v,t)f(y,w,t)dv dx dw dy}_{:=I_1(f,f)}\\
        +\lambda^2\varepsilon\underbrace{\int_{\mathbb{R}^{4d}}\left[\sum_{i,j=1}^d
        \frac{\partial^2\phi(x,\tilde{v})}{\partial v_i^2}(w_j-v_j)^2\right] (H_\alpha(x,y,v))^2f(x,v,t)f(y,w,t)dv dx dw dy}_{:=        I_2(f,f)}
    \label{Taylor}
    \end{multline}
    for some $\tilde{v}=\tau v+(1-\tau )v^*$, $0\leq\tau\leq 1$. In the limit $\varepsilon\to 0$ the term $I_2(f,f)$ vanishes since the second momentum of the solution is not increasing and $H_\alpha(x,y,v)\leq 1$ hence~\cite{MR2596552}  
    \begin{equation}
        \left|I_2(f,f)\right|\leq2\|\phi(x,v)\|_{C^2_0}\int_{\mathbb{R}^{2d}}|v|^2f_0(x,v)dx dv.
    \end{equation}
    Thus in the limit the second-order term can be neglected and $I_1(f,f)$ constitutes an approximation of the collisional integral $Q(f,f)$, in the strong divergence form
    \begin{equation}
        I_{1}(f,f)=-\nabla_v\cdot\int_{\mathbb{R}^{2d}}{(w-v)H_\alpha(x,y,v)f(y,w,t)f(x,v,t)dw dy},
    \label{kin2}
    \end{equation}
    or equivalently in convolution form~\cite{MR2596552}     \begin{equation}
        I_{1}(f,f)=\nabla_v\cdot\left\{f(x,v,t)[(H_\alpha(x,y,v)\nabla_v e(v))*f](x,v,t)\right\},
    \end{equation}
    where $e(v)=|v|^2/2$ and $*$ is the $(x,v)$-convolution. As observed in~\cite{MR2596552}, the operator $I_1(f,f)$ preserves the dissipation proprieties of original Boltzmann operator.

Finally we get the mean-field kinetic equation
    \begin{equation}
        \partial_t f+v\cdot\nabla_x f=-\lambda\nabla_v[\xi(f)f]
    \label{kinetic}
    \end{equation}
    \begin{equation*}
        \xi(f)=\int_{\mathbb{R}^{2d}}H_\alpha(x,y,v)(w-v)f(y,w,t)dw dy.
    \end{equation*}
    As noted in \cite{MR2765734}, the continuos kinetic model (\ref{kinetic}) and the microscopic one (\ref{CuckerSmale})-(\ref{csH2}) are really the same when we take the discrete $N$-particle distribution \[f(x,v,t)=\frac{1}{N}\sum_{i=1}^N\delta(x-x_i(t))\delta(v-v_i(t)),\]
where $\delta(\cdot)$ denotes the Dirac-delta function.    
    
\begin{rmk} 
\begin{itemize}
\item Kinetic formulation for the D'Orsogna Bertozzi et al. with perception cone can be derived in the same way and yields the mean-field model
    \begin{equation}
        \partial_t f+v\cdot\nabla_x f+\nabla_v(S(v)f)=-\lambda\nabla_v\cdot[\int_{\mathbb{R}^{2d}}Z_\alpha(x,y,v)f(y,w,t) dy dw]f(x,v,t),
    \label{kinetic_Bertozzi}
    \end{equation}
    where $Z_\alpha(x,y,v)=[A(x,y)+R(x,y)]\psi_{\alpha}(x,y,v)$ represents the attraction repulsion term.
\item In~\cite{Couzin2010} the authors observed that a certain degree of randomness helps the coherence in the collective swarm behavior. Following~\cite{MR2744704}, if we add in (\ref{CuckerSmale}) a nonlinear noise term depending on function $H_\alpha$ and perform essentially the same derivation of the above paragraph we obtain the kinetic equation
    \begin{equation}
        \partial_t f+v\cdot\nabla_x f=-\lambda\nabla_v[\xi(f)f]+\sigma\Delta_v (H_\alpha*\rho)f,
    \label{kinetic_diff}
    \end{equation}
where $\rho=\rho(x,t)$ represent the mass of the system and $\sigma\geq 0$ the strength of the noise. If $H_\alpha(x,y,v)\equiv H(x,y)$, the right hand side can be written as a Fokker-Plank operator $$\nabla_v\cdot (\sigma(H*\rho)\nabla_v f-\lambda\xi(f)f),$$
and thus a \emph{global Maxwellian function} is a steady state solution for the equation (\ref{kinetic_diff}).

\end{itemize}
\end{rmk}
    
\subsection{Alternative formulations}
In this section we present some alternative formulations of the Boltzmann equation describing the binary interaction dynamics for alignment. All the formulations share the property that in the mean-field limit originate the same kinetic model (\ref{kinetic}). 

The Boltzmann equation (\ref{Q_Boltz}) has much in common with a classical Boltzmann equation for Maxwell molecules, in the sense that the collision frequency is independent of the velocity and position of individuals. 
An alternative Boltzmann-like kinetic approximation is obtained
with the interaction operator
\begin{equation}
Q(f,f)=\int_{\mathbb{R}^{2d}}H_{\alpha}(x,y,v)\left(\frac{1}{J}f(x,v_*)f(y,w_*)-f(x,v)f(y,w)\right) dw dy,
\label{Q_Boltz2}
\end{equation}
where now
\begin{equation}
\left\{
\begin{array}{l}
v^*=(1-\eta)v+\eta w,\\\\
w^*=\eta v+(1-\eta)w.
\end{array}\right.
\label{Binary2}
\end{equation}
From the modeling viewpoint here the function $H_\alpha$ is interpreted as the frequency of interactions instead of the strength of the same interactions.

Clearly the two formulations (\ref{Q_Boltz}) and (\ref{Q_Boltz2}) are not equivalent in general. It is easy to verify that formally we obtain the same mean-field limit (\ref{kinetic}). Note however that now the second order term in the expansion (\ref{Taylor}) is slightly different and reads 
\[
I_2(f,f):=\int_{\mathbb{R}^{4d}}\left[\sum_{i,j=1}^d
        \frac{\partial^2\phi(x,\tilde{v})}{\partial v_i^2}(w_j-v_j)^2\right] H_\alpha(x,y,v) f(x,v,t)f(y,w,t)dv dx dw dy. 
\]        
Since $H_\alpha > (H_\alpha)^2$, in practice we may expect a slower  convergence to the mean-field dynamic for small values of $\varepsilon$.  

Let us finally introduce some stochastic effect in the visual cone perception by defining 
$$
H_{\alpha}(x_i,x_j,v_i)=\zeta H(x_i,x_j),
$$
where $\zeta$ is a random variable distributed accordingly to
some $b_{\alpha}(\zeta,x_i,x_j,v_i)\geq 0$ s.t. \begin{equation}\int b_{\alpha}(\zeta,x_i,x_j,v_i)\,d\zeta=1, \quad\forall\,\,x_i,x_j,v_i.
\label{eq:bb}\end{equation}
Then the collision term in the form (\ref{Q_Boltz}) becomes 
\begin{equation}
Q(f,f)=\int_{\mathbb{R}^{2d+1}}b_{\alpha}(\zeta,x,y,v)\left(\frac1{J}f(x,v_*)f(y,w_*)-f(x,v)f(y,w)\right) dw dy d\zeta,
\label{eq:CollStoc1}
\end{equation}    
whereas in the space dependent interaction frequency form (\ref{Q_Boltz2}) reads
\begin{equation}
Q(f,f)=\int_{\mathbb{R}^{2d+1}}B_{\alpha}(\zeta,x,y,v)\left(\frac1{J}f(x,v_*)f(y,w_*)-f(x,v)f(y,w)\right) dw dy d\zeta,
\label{eq:CollStoc2}
\end{equation}
    where $B_{\alpha}(\zeta,x,y,v)= b_{\alpha}(\zeta,x,y,v)H(x,y)$.
    Again it can be shown that thanks to (\ref{eq:bb}) the limit asymptotic behavior
    $\varepsilon\to 0$ is unchanged. We omit the details.
    
We conclude the section reporting an example of distribution for the random variable $\zeta$ which corresponds to the stochastic analogue of (\ref{HHH})
    $$
        \zeta = \left\{
        \begin{array}{ll}
        1\qquad \textrm{with probability}\quad p_1,\quad &\textrm{for} \,\,y\in\Sigma(x,v,\theta),\\
        0\qquad \textrm{with probability}\quad 1-p_1,\quad &\textrm{for}\,\, y\in\Sigma(x,v,\theta),\\
        1\qquad \textrm{with probability}\quad p_2,\quad &\textrm{for} \,\,y\in\mathbb{R}^d\diagdown\Sigma(x,v,\theta),\\
        0\qquad \textrm{with probability}\quad 1-p_2,\quad &\textrm{for}\,\, y\in\mathbb{R}^d\diagdown\Sigma(x,v,\theta).
        \end{array}\right.
    $$

\section{Monte Carlo methods}
Following~\cite{bobylev2000theory,MR2812253} we introduce different numerical approaches for the above kinetic equations based on Monte Carlo methods. The main idea is to approximate the dynamic by solving the Boltzmann-like models for small value of $\varepsilon$. We will also develop some direct Monte Carlo techniques for the limiting kinetic equation (\ref{kinetic}). For the sake of simplicity we describe the algorithms in the case of the collision operator (\ref{Q_Boltz}), extensions to the other possible formulations presented in Section 3 are also discussed along the section. As we will see, thanks to the structure of the equations, the resulting algorithms are fully meshless.
 
\subsection{Asymptotic binary interaction algorithms}

As in most Monte Carlo methods for kinetic equations, see~\cite{MR1865188}, the starting point is a splitting method based on evaluating in two different steps the transport and
    collisional part of the scaled Boltzmann-Povzner equation 
    \begin{equation*}
    \tag{T}
        \frac{\partial f}{\partial t}=-v\cdot\nabla_x f
    \end{equation*}
    \begin{equation*}
        \frac{\partial f}{\partial t}=\frac{1}{\varepsilon}Q_\varepsilon(f,f)
    \tag{C}
    \label{Coll}
    \end{equation*}
where we used the notation $Q_\varepsilon(f,f)$ to denote the scaled Boltzmann operator (\ref{Q_Boltz}). We emphasize that the solution to the collision step for small values of $\varepsilon$ has very little in common with the classical fluid-limit of the Boltzmann equation. Here in fact the whole collision process depends on space and on the small scaling parameter $\varepsilon$. In particular, in the small $\varepsilon$ limit the solution is expected to converge towards the solution of the mean-field model (\ref{kinetic}).    
    
    By decomposing the collisional operator in equation (\ref{Coll}) in its gain and loss parts we can rewrite the collision step as
    \begin{equation}
        \frac{\partial f}{\partial t}=\frac{1}{\varepsilon}\left[Q_\varepsilon^{+}(f,f)-\rho f\right],
    \label{Coll_mod}
    \end{equation}
    where $\rho>0$ represent the total mass and $Q^{+}_\varepsilon$  the \emph{gain} part of the collisional operator. Without loss of generality in the sequel we assume that 
    \begin{equation*}
        \rho=\int_{\mathbb{R}^{2d}}f(x,v,t)dxdv=1.
    \end{equation*}
In order to solve the trasport step we use the exact free flow of the sample particles $(x_i(t),v_i(t))$ in a time interval $\Delta t$
\begin{equation}
x_i(t+\Delta t)=x_i(t)+v_i(t)\Delta t,
\end{equation}
and thus describe the different schemes used for the interaction process in the form (\ref{Coll_mod}).

\subsubsection{A Nanbu-like asymptotic method}
    Let us now consider a time interval $[0,T]$ discretized in $n_{tot}$ intervals of size $\Delta t$. We denote by $f^n$ the  approximation of $f(x,v,n\Delta t)$.
    
    Thus the forward Euler scheme writes
    \begin{equation}
        f^{n+1}=\left(1-\frac{\Delta t}{\varepsilon}\right)f^{n}+\frac{\Delta t}{\varepsilon}{Q_\varepsilon^{+}(f^n,f^n)},
    \label{EE}
    \end{equation}
    where since $f^n$ is a probability density, thanks to mass conservation, also $Q_\varepsilon^{+}(f^n,f^n)$ is a probability density.
    Under the restriction $\Delta t\leq\varepsilon$ then also $f^{n+1}$ is a probability density, since it is a convex combination of
    probability densities.
    
    From a Monte Carlo point of view equation (\ref{EE}) can be interpreted as follows: an individual with velocity $v$ at position $x$ will not interact with other individuals with
    probability $1-\Delta t/\varepsilon$ and it will interact with others with probability $\Delta t/\varepsilon$ according to the   interaction law stated by $Q_\varepsilon^{+}(f^n,f^n)$. Since we aim at small values of $\varepsilon$ the natural choice as in~\cite{bobylev2000theory} is to take $\Delta t=\varepsilon$. The major difference compare to standard Nanbu algorithm here is the way particles are sampled from $Q_{\varepsilon}^+(f^n,f^n)$ which does not require the introduction of a space grid. A simple algorithm for the solution of (\ref{EE}) in a time interval $[0,T]$, $T=n_{tot}\Delta t$, $\Delta t = \varepsilon$ is sketched in the sequel.\medskip
\begin{alg}[\textbf{Asymptotic Nanbu I}]

    \begin{enumerate}

        \item Given $N$ samples $(x^0_k,v_k^0)$, with $k=1,\ldots,N$ from the initial distribution $f_0(x,v)$;
        \item \texttt{for} $n=0$ \texttt{to} $n_{tot}-1$               
        \begin{enumerate}
         \item \texttt{for} $i=1$ \texttt{to} $N$; 
                \item select an index $j$ uniformly among all possible individuals $(x_k^n,v_k^n)$ except $i$;
                \item evaluate $H_{\alpha}(|x_i^n-x_j^n|,v_i^n)$;
                \item compute the velocity change $v_i^*$ using the first relation in (\ref{Binary}) with $\eta=\varepsilon$;
                \item set $(x_i^{n+1},v_i^{n+1})=(x_i^n,v_i^{*})$.
    \item[]\texttt{end for}
    \end{enumerate}
    \item[]\texttt{end for}
    \end{enumerate}
    \label{ANMC}
\end{alg}\medskip
Next we show how the method extends to the case of collision operator of the type (\ref{Q_Boltz2}). In this case an acceptance-rejection strategy is used to select interacting individuals since the forward Euler scheme reads
 \begin{equation}
        f^{n+1}=\left(1-\frac{\Delta t}{\varepsilon}\right)f^{n}+\frac{\Delta t}{\varepsilon}{P_\varepsilon^{+}(f^n,f^n)},
    \label{EE2}
    \end{equation}
where $P_\varepsilon^{+}(f^n,f^n)=Q_\varepsilon(f,f)+f\geq 0$ is again a probability density.

Now using the fact that $H_\alpha\leq 1$ we can adapt the classical acceptance-rejection technique~\cite{MR1865188} to get the following method for (\ref{EE2}) with $\Delta t=\varepsilon$\medskip
\begin{alg}[\textbf{Asymptotic Nanbu II}]
\begin{itemize}
  \item[] In Algorithm \ref{ANMC} make the following change
    \begin{itemize}
                \item[(d)] if $H_{\alpha}(|x_i^n-x_j^n|,v_i^n)>\xi$, $\xi$ uniform in $[0,1]$ then compute the velocity change $v_i^*$ for each individual $i$ of pair $(i,j)$ using the first relation in (\ref{Binary2}) with $\eta=\varepsilon$;
                \item[(e)] set $(x_i^{n+1},v_i^{n+1})=(x_i^n,v_i^{*})$ if the individual has changed its velocity, otherwise set $(x_i^{n+1},v_i^{n+1})=(x_i^n,v_i^{n})$.              
    \end{itemize}
    \end{itemize}
    \label{ANMC2}
\end{alg}\medskip 

Note that in this version two individuals interact always with the same strength in the velocity change but with a different probability related to their distance. As a result the total number of interactions depend on the distribution of individuals and on average is equal to $\bar{H}_{\alpha}N<N$ where
\[
\bar{H}_{\alpha}=\frac1{N^2}\sum_{i,j=1}^NH_{\alpha}(x_i,x_j,v_i).
\]
Thus the method computes less interactions then the one described in Algorithm \ref{ANMC}. In fact, in regions where individuals are scattered very few interactions will be effectively computed by the method. The efficiency of the method can be further improved if one is able to find an easy invertible function $1\geq W_{\alpha}(x_i,x_j,v_i)\geq H_{\alpha}(x_i,x_j,v_i)$ or is capable to compute directly the inverse of $H_{\alpha}(x_i,x_j,v_i)$. We refer to~\cite{MR1865188} for further details on these sampling techniques.   

A symmetric version of the previous algorithms which preserves at a microscopic level other interaction invariants, like momentum in standard Cucker-Smale model, is obtained as follows\medskip
\begin{alg}[\textbf{Asymptotic symmetric Nanbu}]
  \begin{enumerate}
  \item Given $N$ samples $(x^0_k,v_k^0)$, with $k=1,\ldots,N$ from the initial distribution $f_0(x,v)$;
  \item \texttt{for} $n=0$ \texttt{to} $n_{tot}-1$  
  \begin{enumerate}
  \item set $N_c = Iround({N}/{2})$;
  \item select $N_c$ random pairs $(i,j)$ uniformly without repetition among all possible pairs of individuals at time level $n$.
  \item evaluate $H_{\alpha}(|x_i^n-x_j^n|,v_i^n)$ and $H_{\alpha}(|x_i^n-x_j^n|,v_j^n)$;
  \item[(d)] For Algorithm \ref{ANMC}: compute the velocity changes $v_i^*$, $v_j^*$ for each pair $(i,j)$ using relations (\ref{Binary}) with $\eta=\varepsilon$;
  \item[(d)] For Algorithm \ref{ANMC2}: 
  \begin{enumerate}
  \item if $H_{\alpha}(|x_i^n-x_j^n|,v_i^n)>\xi_i$ $\xi_i$ uniform in $[0,1]$ then compute the velocity change $v_i^*$ for each pair $(i,j)$ using the first relation in (\ref{Binary2}) with $\eta=\varepsilon$;
  \item if $H_{\alpha}(|x_i^n-x_j^n|,v_j^n)>\xi_j$ $\xi_j$ uniform in $[0,1]$ then compute the velocity change $v_j^*$ for each pair $(i,j)$ using the second relation in (\ref{Binary2}) with $\eta=\varepsilon$;
    \end{enumerate}
  \item[(e)] set $(x_i^{n+1},v_i^{n+1})=(x_i^n,v_i^{*})$, $(x_j^{n+1},v_j^{n+1})=(x_j^n,v_j^{*})$ for all the individuals that changed their velocity,
  \item[(f)] $(x_h^{n+1},v_h^{n+1})=(x_h^n,v_h^{n})$ for all the remaining individuals.
\end{enumerate}
\item[]\texttt{end for}
  \end{enumerate}
  \label{ANMCS}
\end{alg}\medskip
The function $Iround(\cdot)$ denotes the integer stochastic rounding defined as
\[
Iround(x)=
\begin{cases}
[x]+1,& \xi < x-[x],\\
[x],&\hbox{elsewhere}
\end{cases}
\]
where $\xi$ is a uniform $[0,1]$ random number and $[\cdot]$ is the integer part.

\subsubsection{A Bird-like asymptotic method}
The most popular Monte Carlo approach to solve the collision step in Boltzmann-like equations is due to Bird~\cite{bird1963approach}. The major differences are that the method simulate the time continuous equation and that individuals are allowed to interact more then once in a single time step. As a result the method achieves a higher time accuracy~\cite{MR1865188}. 

Here we describe the algorithm for the collision operator described by (\ref{Q_Boltz}). The method is based on the observation that the interaction time is a random variable exponentially distributed. Thus for $N$ individuals one introduces a local random time counter given by
\begin{equation} 
\Delta t_c(\xi)=-\frac{\ln(\xi)\varepsilon}{N},
\label{ltc1}
\end{equation}
with $\xi$ a random variable uniformly distributed in $[0,1]$.

A simpler version of the method is based on a constant time counter $\Delta t_c$ corresponding to the average time between interactions. In fact, in a time interval $[0,T]$ we have 
\begin{equation}
\Delta t_c=\frac{T}{N_c} =\frac{\varepsilon}{N},
\label{ltc2}
\end{equation}
since $N_c=NT/\varepsilon$ is the number of average interactions in the time interval. Of course taking time averages the two formulations (\ref{ltc1}) and (\ref{ltc2}) are equivalent.

From the above considerations, using the symmetric formulation and the time counter $\Delta t_c=2\varepsilon/N$, we obtain the following method
in a time interval $[0,T]$, $T=N_{tot}\Delta t_c$ 
\medskip

 \begin{alg}[\textbf{Asymptotic Bird I}]
    \begin{enumerate}
        \item Given $N$ samples $(x_k,v_k)$, with $k=1,\ldots,N$ from the initial distribution $f_0(x,v)$
        \item \texttt{for} $n=0$ \texttt{to} $N_{tot}-1$ 
                \begin{enumerate}
                    \item select a random pair $(i,j)$ uniformly among all possible pairs;
                    \item evaluate $H_{\alpha}(|x_i-x_j|,v_i)$ and $H_{\alpha}(|x_i-x_j|,v_j)$;
  \item compute the velocity changes $v_i^*$, $v_j^*$ for each pair $(i,j)$ using relations (\ref{Binary}) with $\eta=\varepsilon$;
                \item set $v_i=v_i^*$ and $v_j=v_j^*$; 
                \end{enumerate}
        \item[]\texttt{end for} 
\end{enumerate}
\label{ABMC}
\end{alg}\medskip
Note that in the above formulation the method has much in common with Algorithm \ref{ANMCS} except for the fact that multiple interactions are allowed during the dynamic (no need to tag particles with respect to time level) and that the local time stepping is related to the number of individuals. As a result in the limit of large numbers of individuals the method converges towards the time continuous Boltzmann equation (\ref{Q_Boltz}) and not to its time discrete counterpart (\ref{EE}), as it happens for Nanbu formulation. Since in Algorithm \ref{ANMCS} we have $n_{tot}=N_{tot}/N_c$, the computational cost of the methods is the same.

Similarly Bird's approach can be extended to collision operator in the form (\ref{Q_Boltz2}) by introducing the following changes\medskip 
\begin{alg}[\textbf{Asymptotic Bird II}]
\begin{itemize}
  \item[] In Algorithm \ref{ABMC} make the following change
    \begin{itemize}
    \item[(c)] 
    \begin{itemize}
    \item if $H_{\alpha}(|x_i^n-x_j^n|,v_i^n)>\xi_i$ $\xi_i$ uniform in $[0,1]$ then compute the velocity change $v_i^*$ for each pair $(i,j)$ using the first relation in (\ref{Binary2}) with $\eta=\varepsilon$;
  \item if $H_{\alpha}(|x_i^n-x_j^n|,v_j^n)>\xi_j$ $\xi_j$ uniform in $[0,1]$ then compute the velocity change $v_j^*$ for each pair $(i,j)$ using the second relation in (\ref{Binary2}) with $\eta=\varepsilon$;
    \end{itemize}
    \end{itemize}
    \end{itemize}
    \label{ABMC2}
\end{alg}\medskip 

%
    
Finally we sketch the algorithm to implement the stochastic perception cone present in (\ref{eq:CollStoc1}) and (\ref{eq:CollStoc2}), that can be easily introduced in all the previous algorithms.\medskip
 \begin{alg}[\textbf{Interaction with stochastic perception cone}]
    \begin{itemize}

    \item \texttt{if} $x_j\in\Sigma(x_i,v_i,\theta)$
        \begin{itemize}
            \item with probability $p_1$ perform the interaction between $i$ and $j$ and compute $v'_i$
        \end{itemize}
        \texttt{else}
            \begin{itemize}
            \item with probability $p_2$ perform the interaction between $i$ and $j$ and compute $v'_i$
        \end{itemize}
\item \texttt{if} $x_i\in\Sigma(x_j,v_j,\theta)$
        \begin{itemize}
            \item with probability $p_1$ perform the interaction between $i$ and $j$ and compute $v'_j$
        \end{itemize}
        \texttt{else}
            \begin{itemize}
            \item with probability $p_2$ perform the interaction between $i$ and $j$ and compute $v'_j$
        \end{itemize}
    \end{itemize}
\end{alg}\medskip
    Note that this reduces further the total number of interactions in the algorithms just described. In contrast, for the deterministic case we simply change the relative interaction strengths using respectively $\eta=p_1\varepsilon$ and $\eta=p_2\varepsilon$ in the binary interaction rules.

\subsection{Mean-field interaction algorithms} 
    Let us finally tackle directly the limiting mean field equation. The interaction step now corresponds to solve
    $$
    \partial_t f=-\nabla_v\left[f\int_{\mathbb{R}^{2d}}H_\alpha(x,y,v)(v-w)f(y,w,t)dw dy\right].
    $$
As already observed, in a particle setting this corresponds to compute the original $O(N^2)$ dynamic. We can reduce the computational cost using a Monte Carlo evaluation of the summation term as described in the following simple algorithm.\medskip    

\begin{alg}[\textbf{Mean Field Monte Carlo}]
    \begin{enumerate}
        \item Given $N$ samples $v_k^0$, with $k=1,\ldots,N$ computed from the initial distribution $f_0(x,v)$ and $M\leq N$;
        \item \texttt{for} $n=0$ \texttt{to} $n_{tot}-1$
                \begin{enumerate}
                \item \texttt{for} $i=1$  \textrm{to} $N$
                \item sample $M$ particles $j_{1},\ldots,j_{M}$ uniformly without repetition among all particles;
                \item compute
                \[
                \bar{H}_{\alpha}^n(x_i)=\frac1{M}\sum_{k=1}^M H_\alpha(x_i,x_{j_k},v_i^n),\quad \bar{v}^n_i=\frac{1}{M}\sum_{k=1}^{M} \frac{H_\alpha(x_i^n,x_{j_k}^n,v^{n}_i)}{\bar{H}_{\alpha}^n(x_i)}v_{j_k},
                \]
                \item compute the velocity change
        $$
            v^{n+1}_i = v^{n}_i(1-\Delta t \bar{H}_{\alpha}^n(x_i))+\Delta t \bar{H}_{\alpha}^n(x_i)\bar{v}_i^n.        $$
        \end{enumerate}
        \quad\texttt{end for}
        \\
        \texttt{end for}
    \end{enumerate}
    \label{MFMC}
\end{alg}\medskip
The overall cost of the above simple algorithm is $O=(MN)$, clearly for $M=N$ we obtain the explicit Euler scheme for the original $N$ particle system. In this formulation the method is closely related to asymptotic Nanbu's Algorithm \ref{ANMC}. It is easy to verify that taking $M=1$ leads exactly to the same numerical method. On the other hand for $M>1$ the above algorithm can be interpreted as an averaged asymptotic Nanbu method over $M$ runs since we can rewrite point $(d)$ as
      $$
      v^{n+1}_i = \frac{1}{M}\sum_{k=1}^{M}\left[\left(1-\Delta t H_\alpha(x_i^n,x_{j_k}^n,v^{n}_i)\right)v^{n}_i+\Delta t H_\alpha(x_i^n,x_{j_k}^n,v^{n}_i)v^{n}_{j_k}\right],\qquad i=1,\ldots,N.
      $$
The only difference is that averaging the result of Algorithm~\ref{ANMC} does not guarantee the absence of repetitions in the choice of the indexes $j_1,\ldots,j_M$.
Thus the choice $\Delta t=\varepsilon$ in Algorithm \ref{ANMC} originates a numerical method consistent with the limiting mean-field kinetic equation. 
Following this description we can construct other Monte Carlo methods for the mean field limit taking suitable averaged versions of the corresponding algorithms for the Boltzmann models. Here we omit for brevity the details. 
\begin{rmk}~
\begin{itemize}
\item In Algorithm~\ref{MFMC} the size of $\Delta t$ can be taken larger then the corresponding $\Delta t=\varepsilon$ in Algorithm~\ref{ANMC}. However, as we just discussed, since large values of $\Delta t$ in the mean-field algorithm are essentially equivalent to large values of $\varepsilon$ in the Boltzmann algorithms we don't expect any computational advantage by choosing larger values of $\Delta t$ in Algorithm~\ref{MFMC}.
\item We remark that changing the time discretization method from Explicit Euler in (\ref{EE}) and (\ref{EE2}) to other methods, like semi-implicit methods or method designed for the fluid-limit, permits to avoid the stability restriction $\Delta t \leq \varepsilon$. Even this approach however does not lead to any computational improvement since a strong deterioration in the accuracy of the solution is observed when $\Delta t > \varepsilon$. Here we don't explore further this direction. 
\end{itemize}
\end{rmk}

\section{Numerical Tests}
\begin{figure}[ht]
\centering
\includegraphics[width=0.45\textwidth]{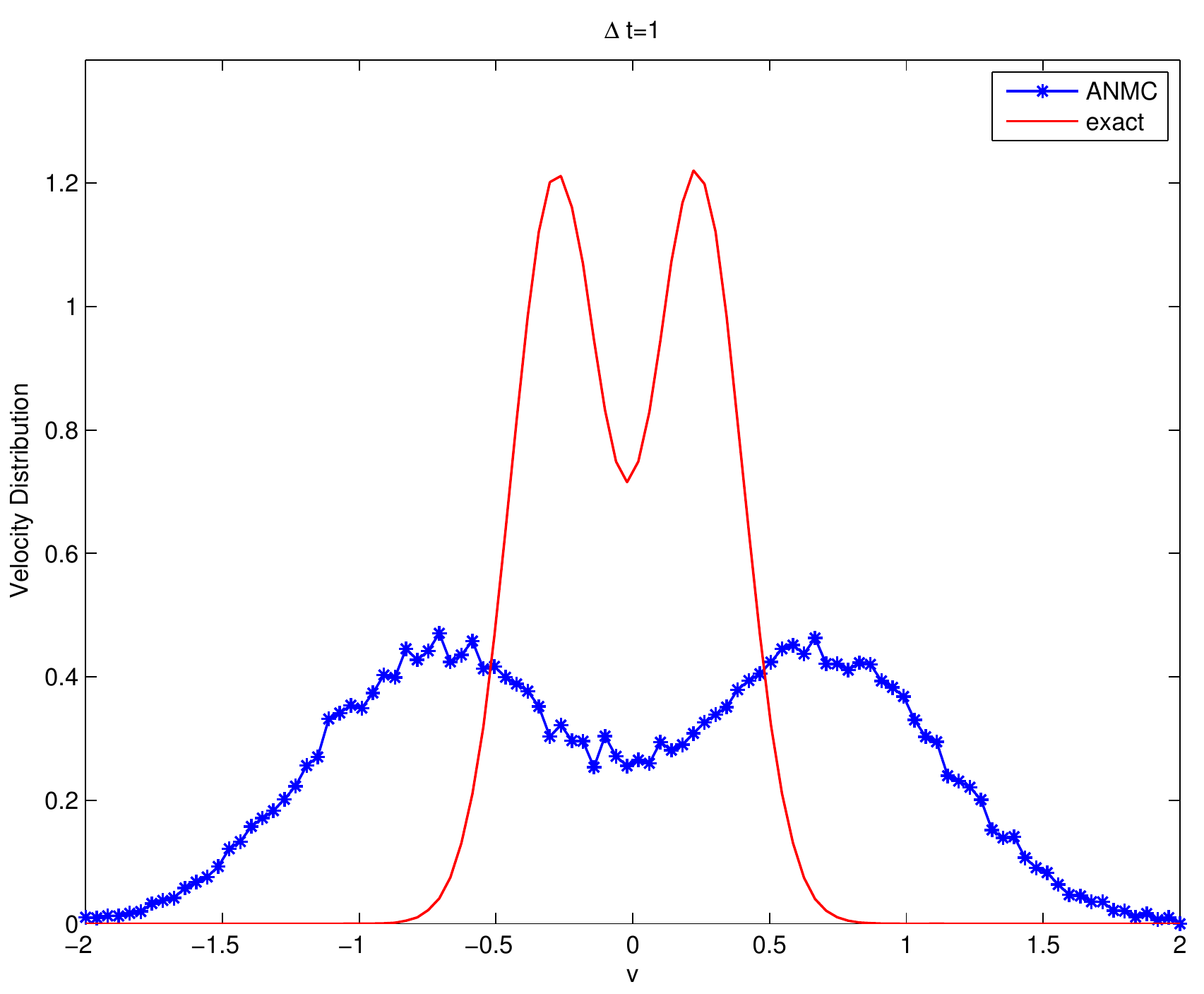}
\includegraphics[width=0.45\textwidth]{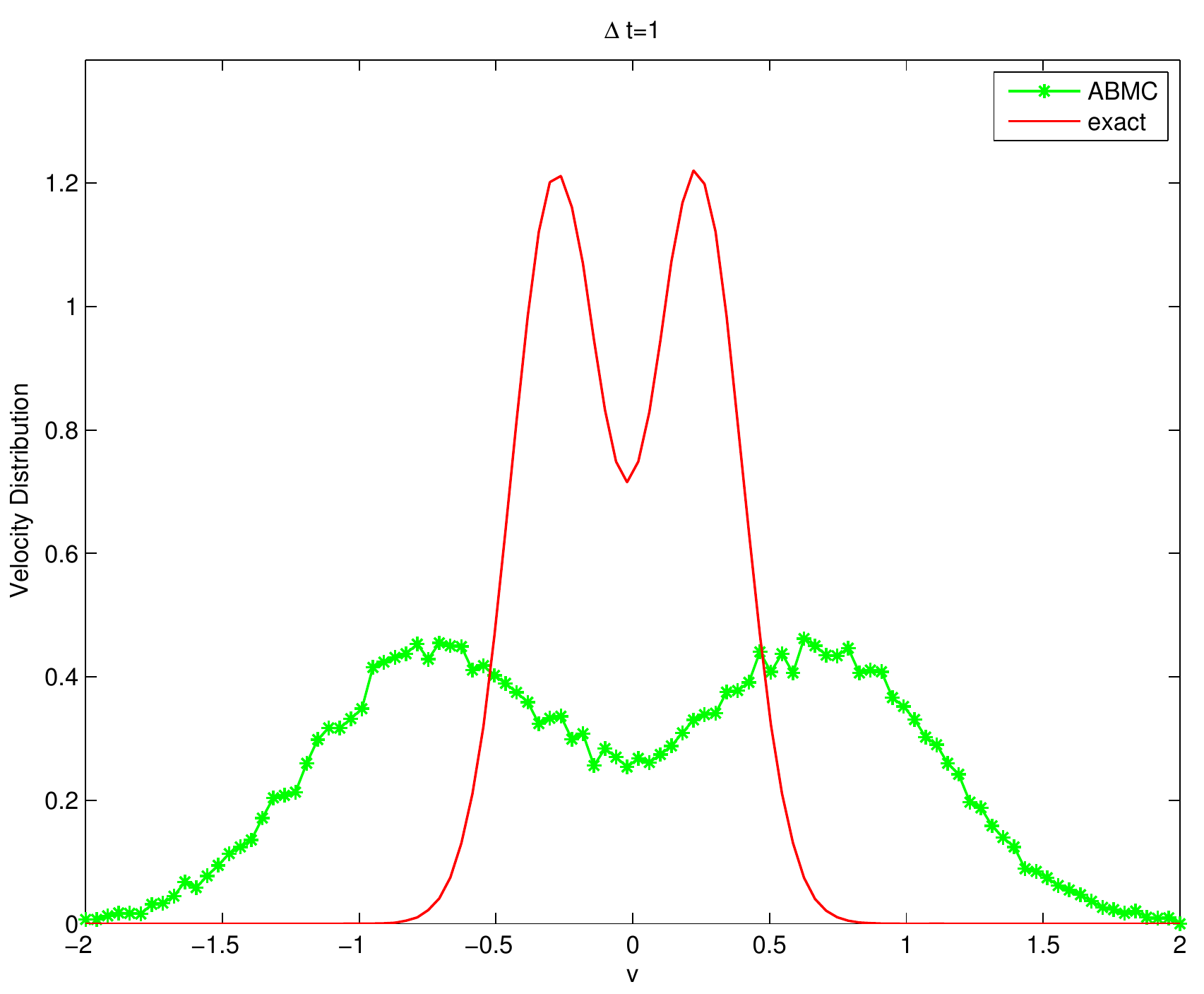}
\\
\includegraphics[width=0.45\textwidth]{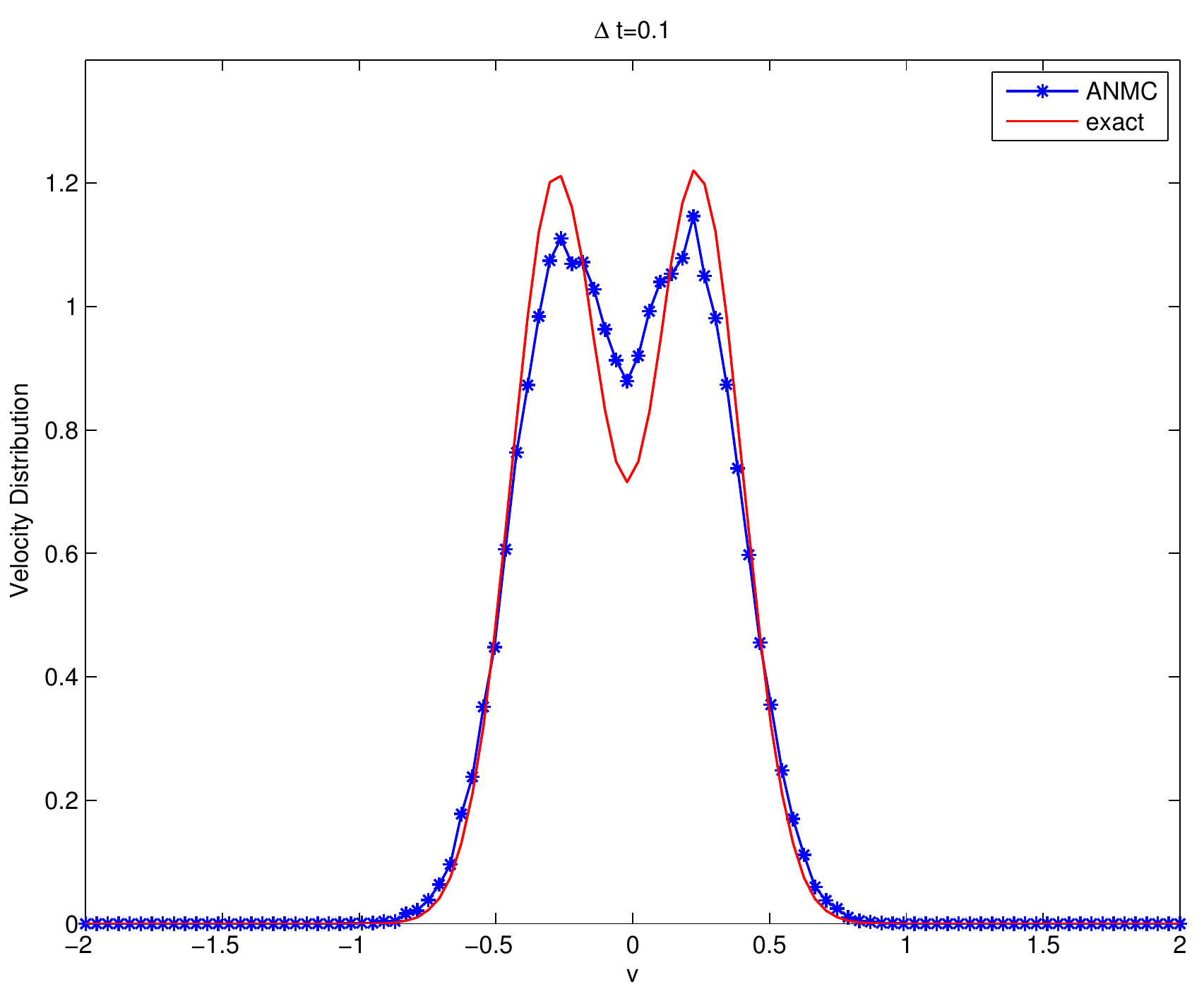}
\includegraphics[width=0.45\textwidth]{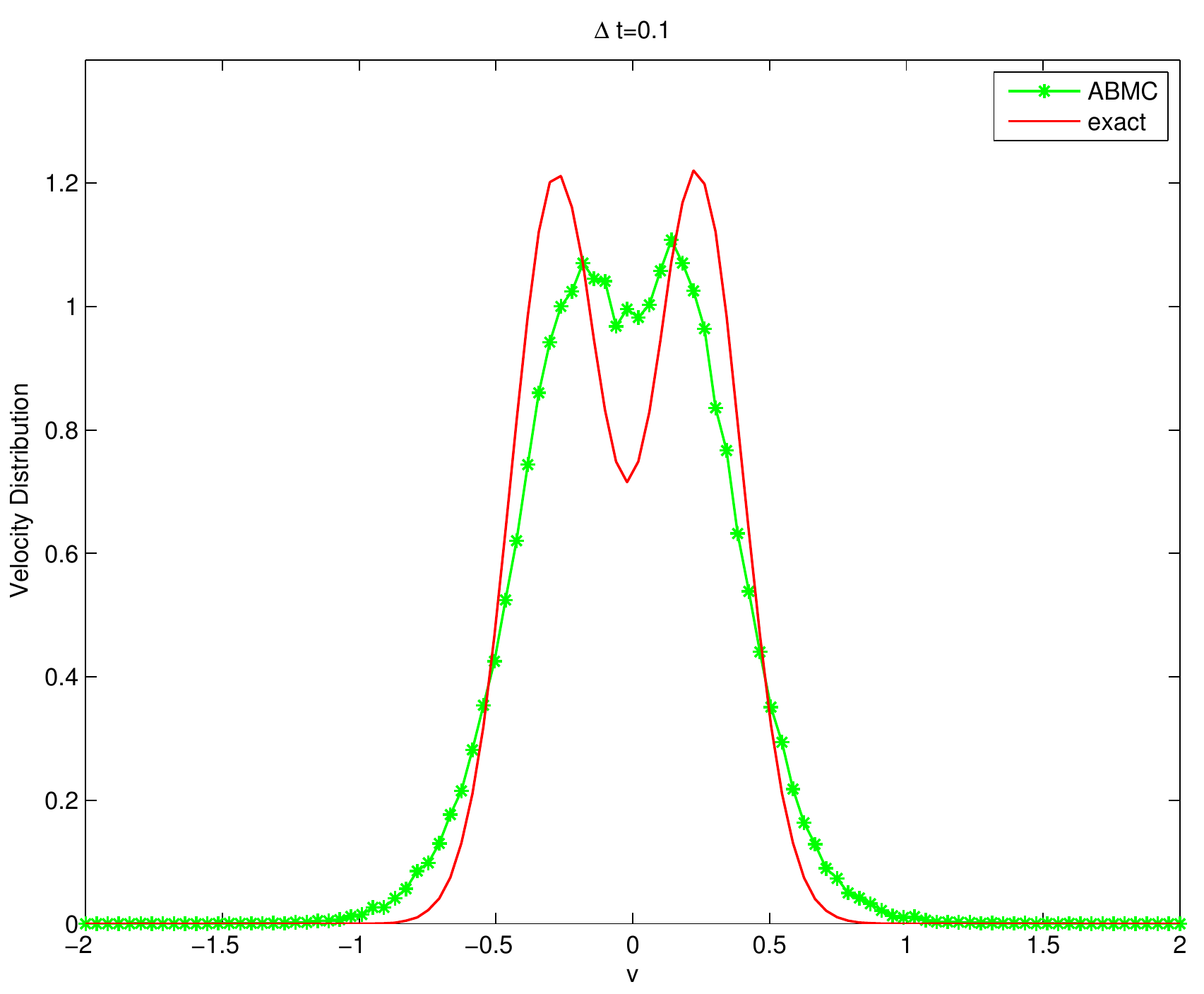}
\\
\includegraphics[width=0.45\textwidth]{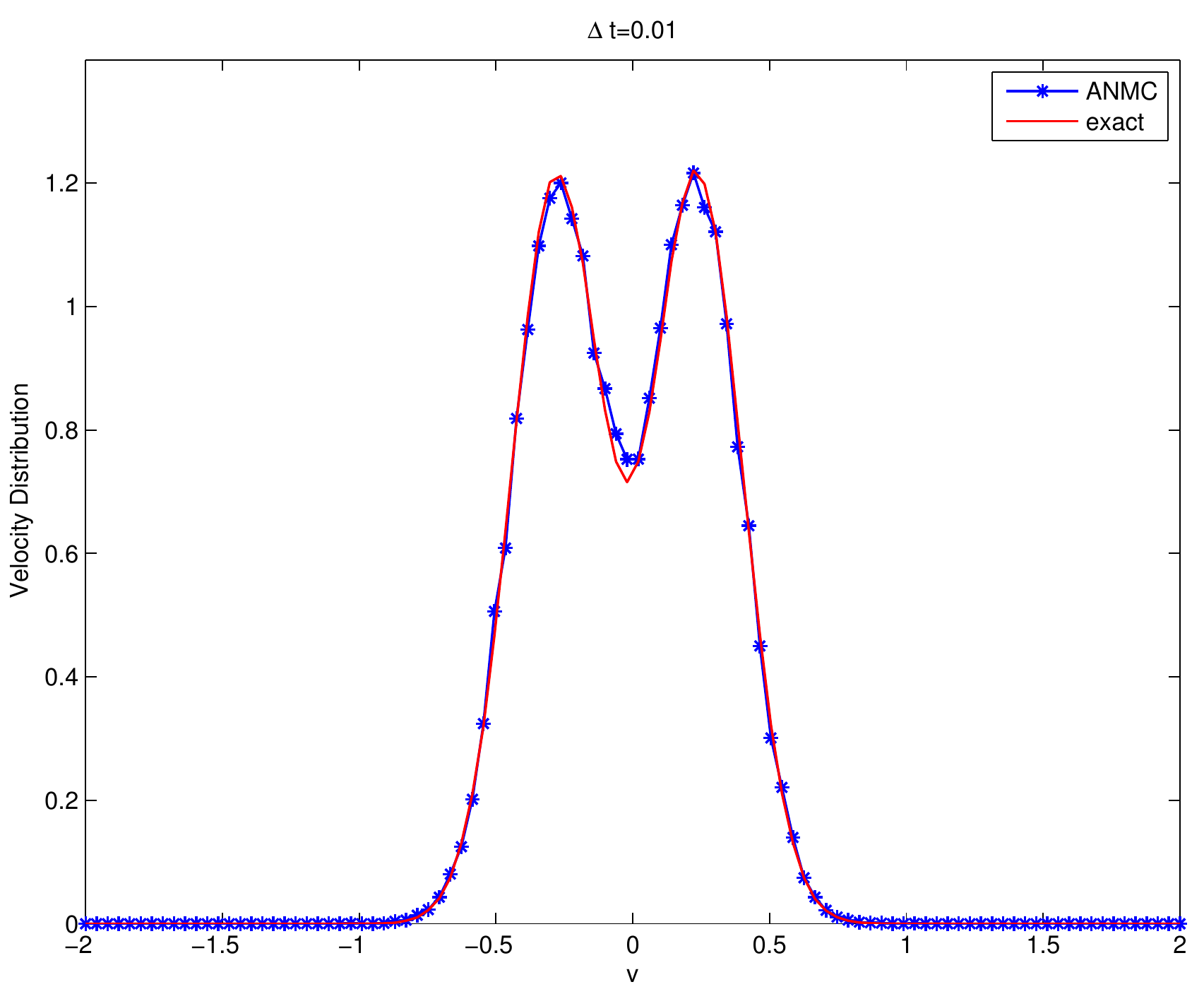}
\includegraphics[width=0.45\textwidth]{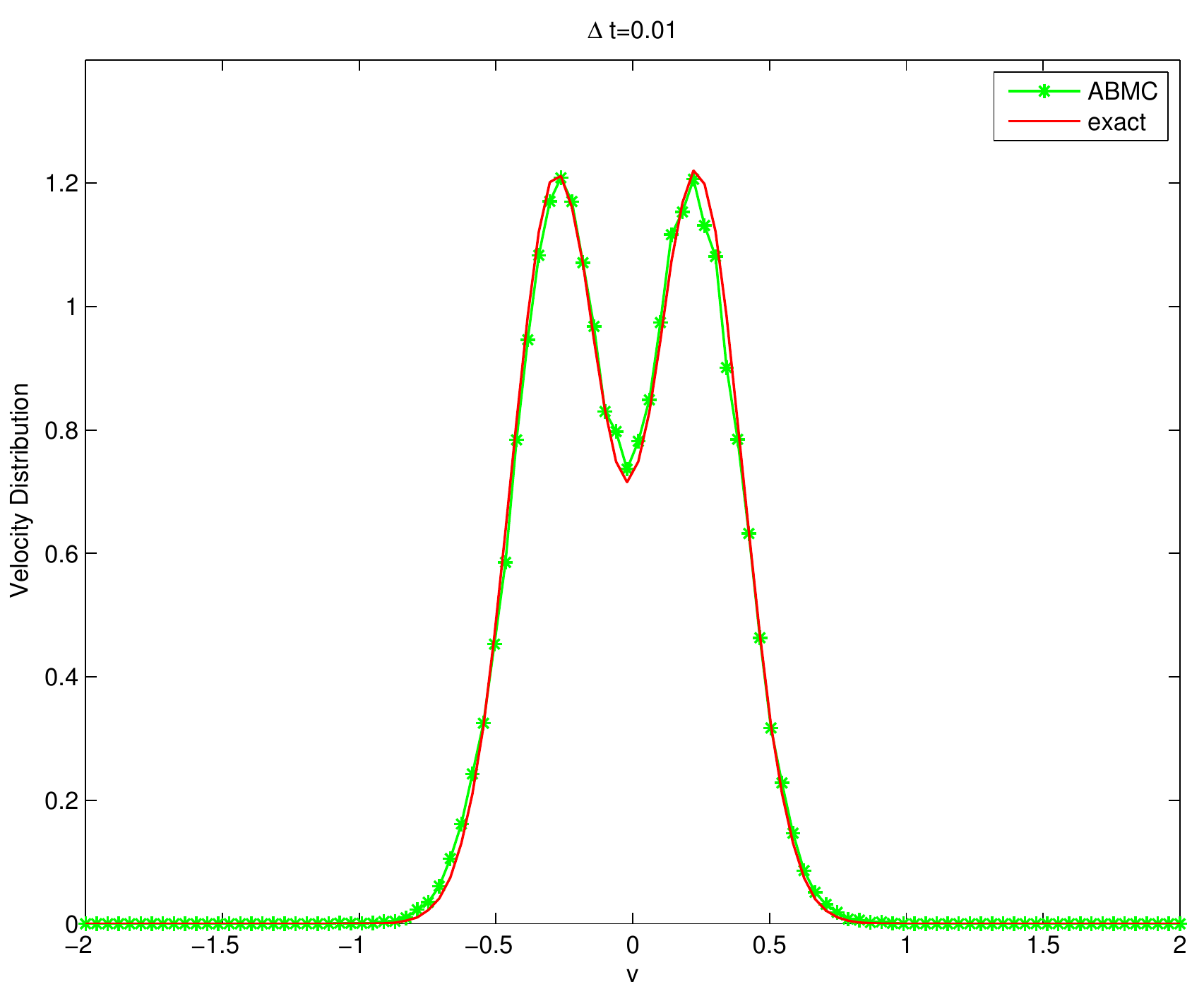}
    \caption{Convergence to the exact solution (continuous line) of the velocity profiles calculated with $ANMC$ (left) and $ABMC$ (right) algorithms. From the top to bottom, $\Delta t=\varepsilon$ with $\varepsilon=1,0.1,0.01$.}
    \label{fig:Accuracy}
\end{figure}

\begin{figure}[ht]
\centering
\includegraphics[width=0.45\textwidth]{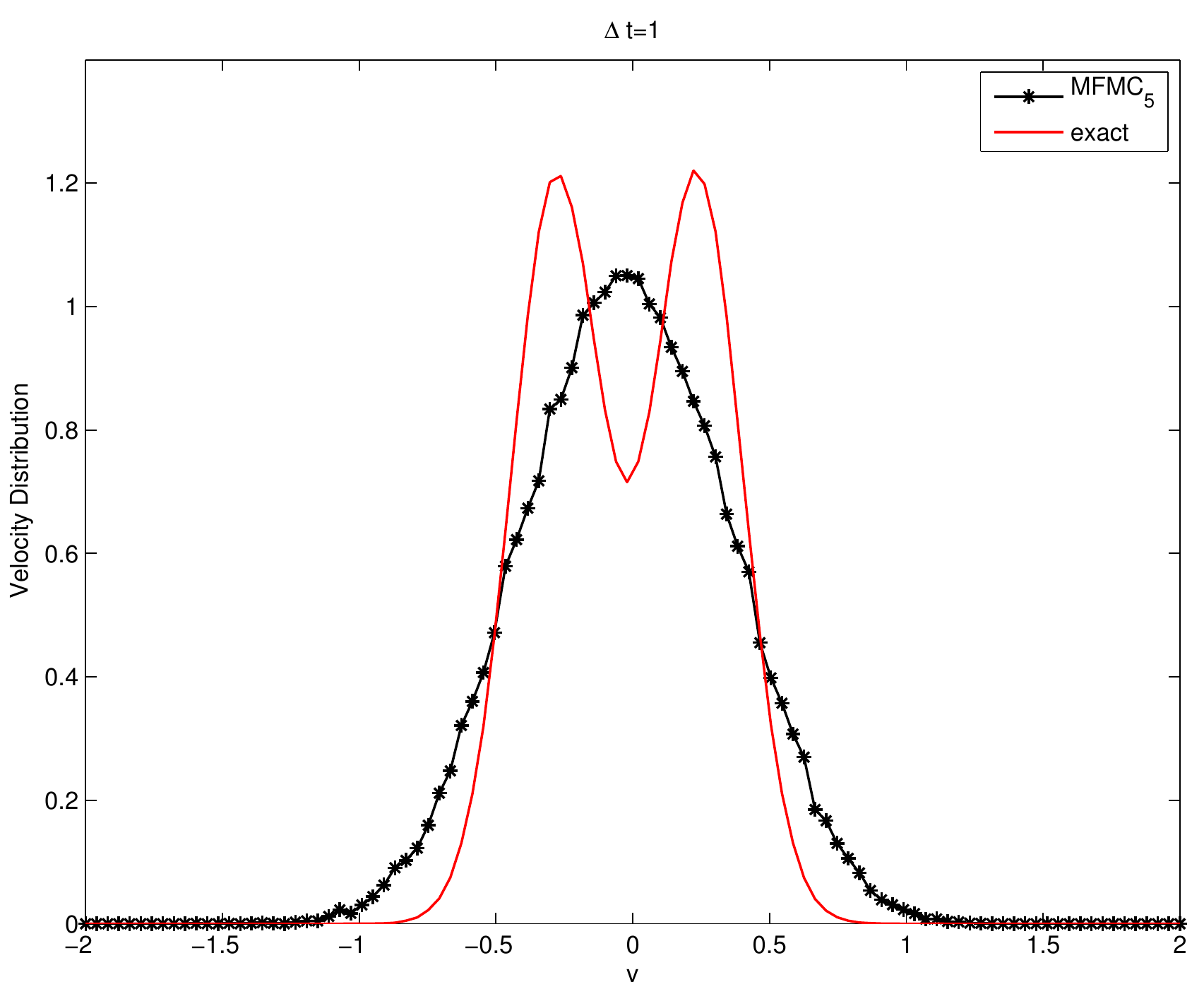}
\includegraphics[width=0.45\textwidth]{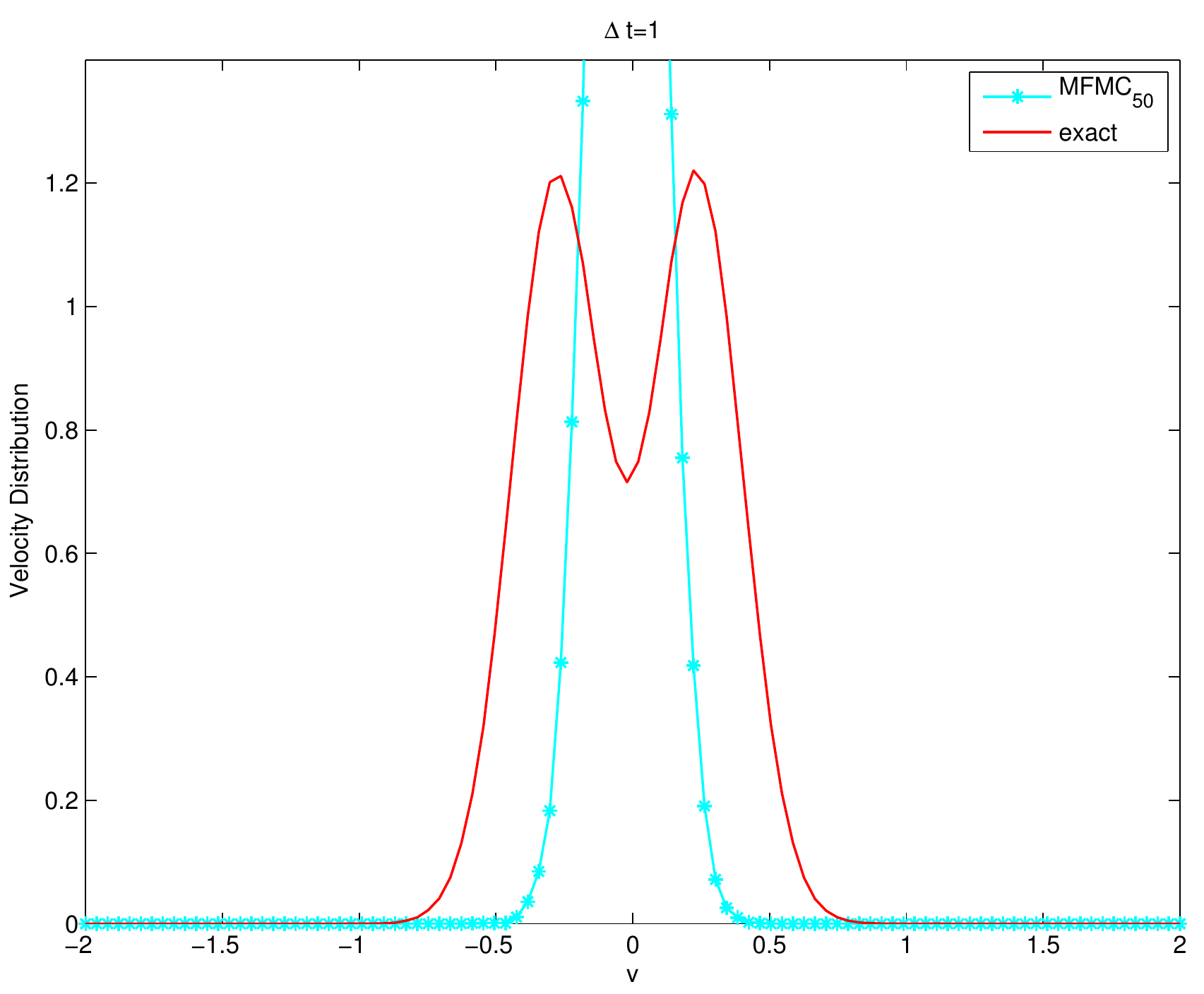}
\\
\includegraphics[width=0.45\textwidth]{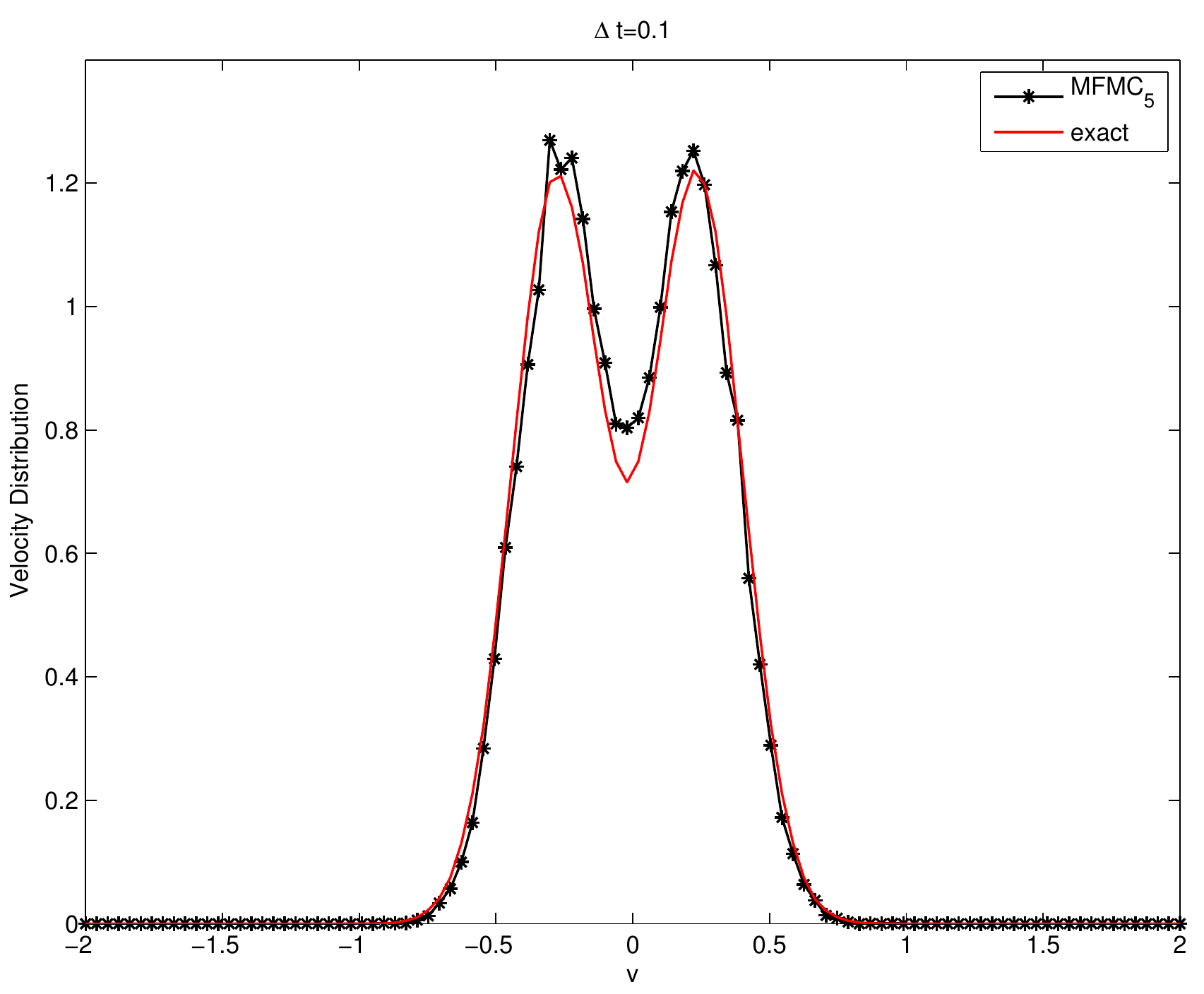}
\includegraphics[width=0.45\textwidth]{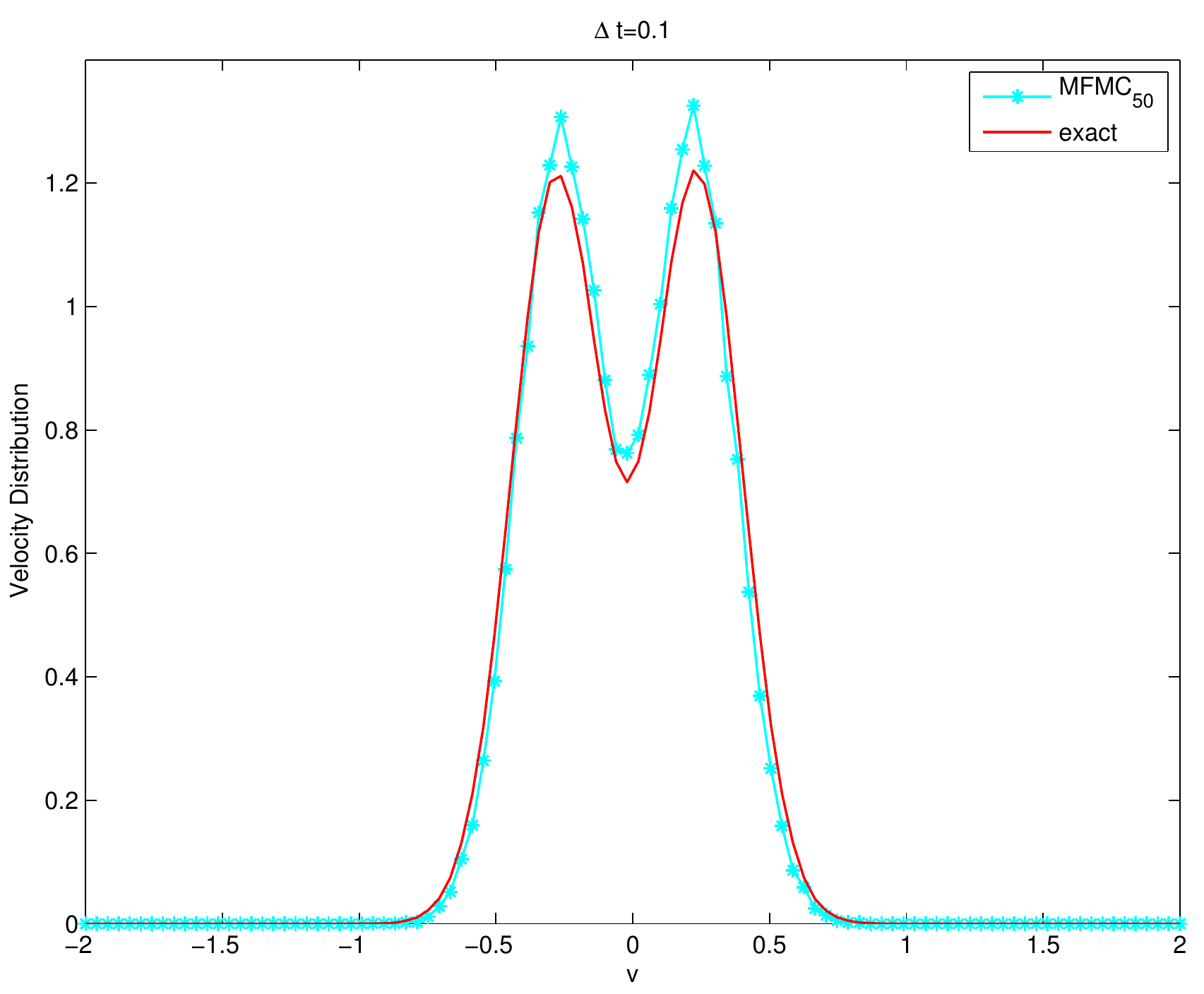}
\\
\includegraphics[width=0.45\textwidth]{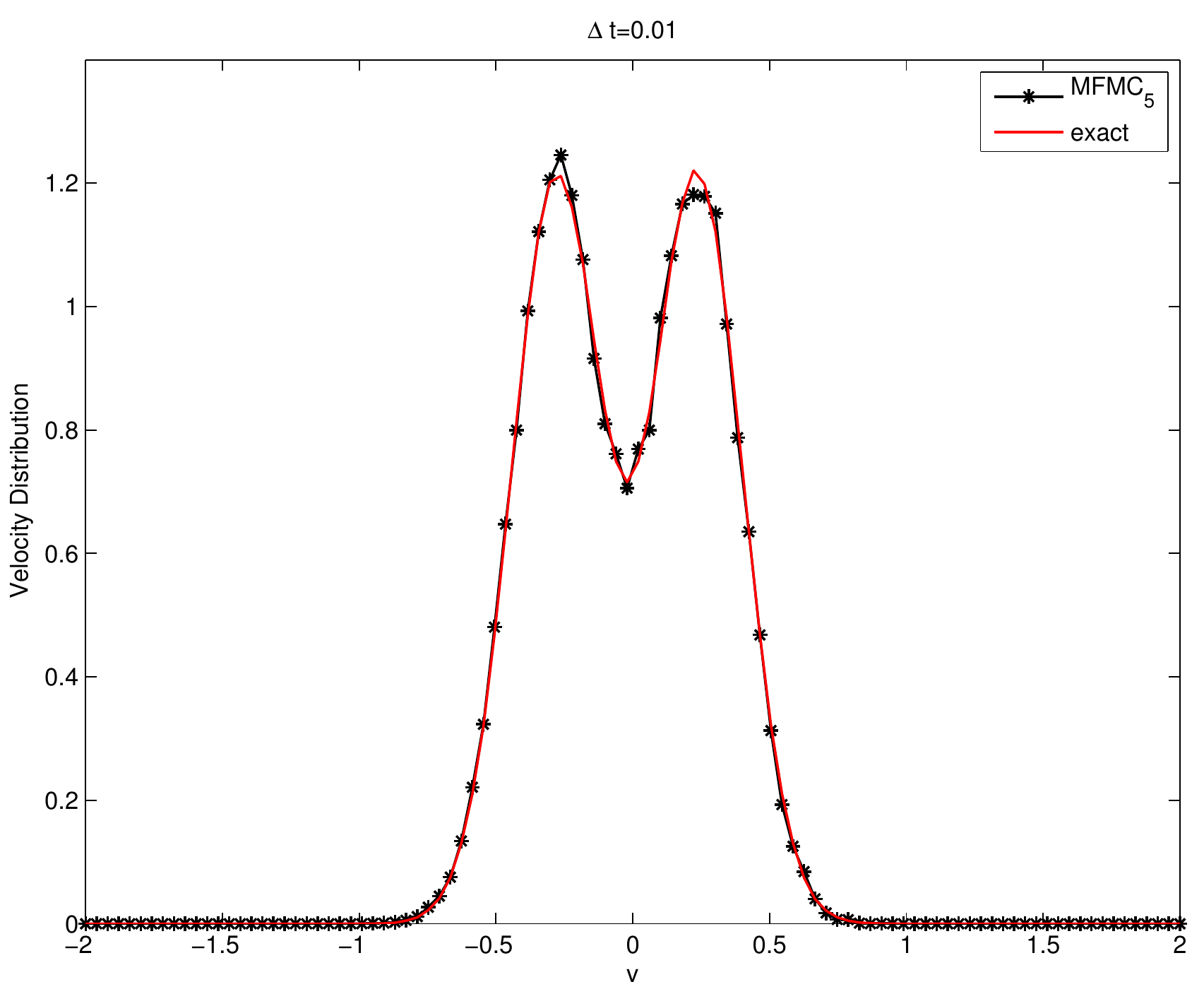}
\includegraphics[width=0.45\textwidth]{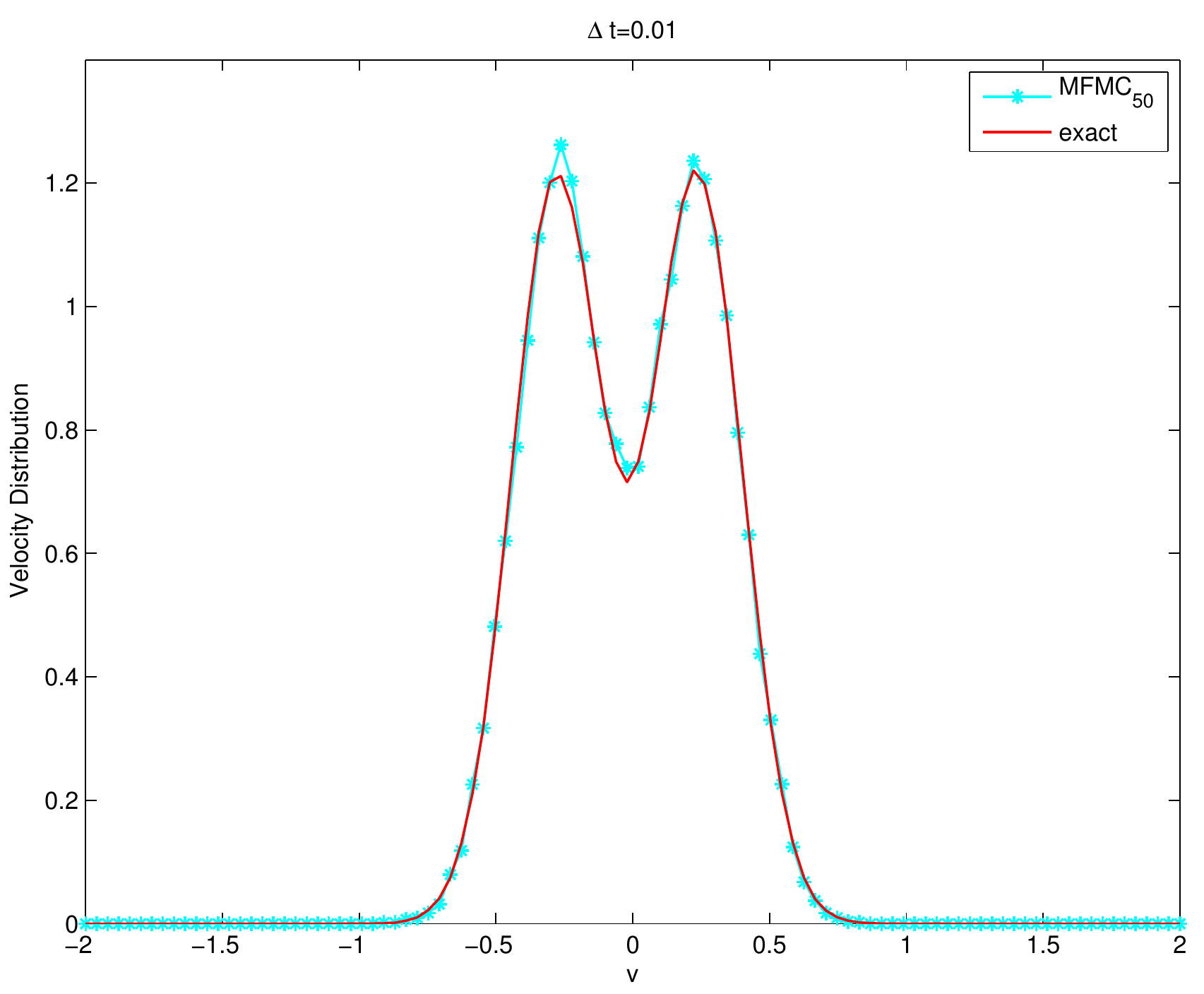}
\\
    \caption{$MFMC_{M}$ algorithms compared with $ANMC$ method, at different time steps. From the top $\Delta t=1$, $\Delta t=0.1$ and $\Delta t =0.01$. On the left column $M=5$ on the left $M=50$.}
    \label{fig:match}
\end{figure}
In this section we first compare accuracy and computational cost of some of the different methods and then illustrate their performance on different two-dimensional and three-dimensional numerical examples. We use the following notations: $ANMC$ (Algorithm \ref{ANMCS}), $ABMC$ (Algorithm \ref{ABMC}), and $MFMC_{M}$ (Algorithm \ref{MFMC} for a given $M$).

\begin{figure}[ht]
    \centering
\includegraphics[scale=0.51]{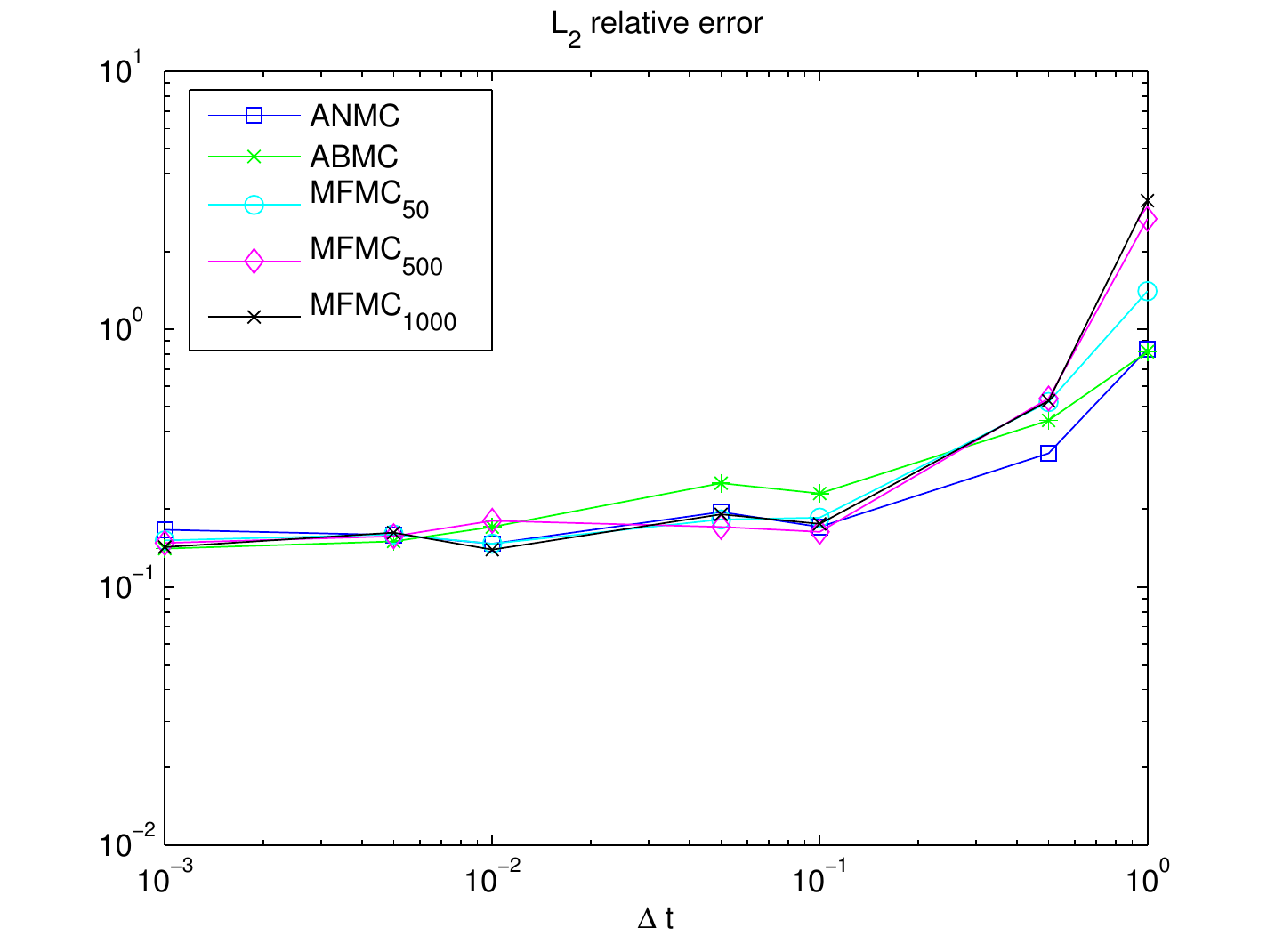}
\includegraphics[scale=0.51]{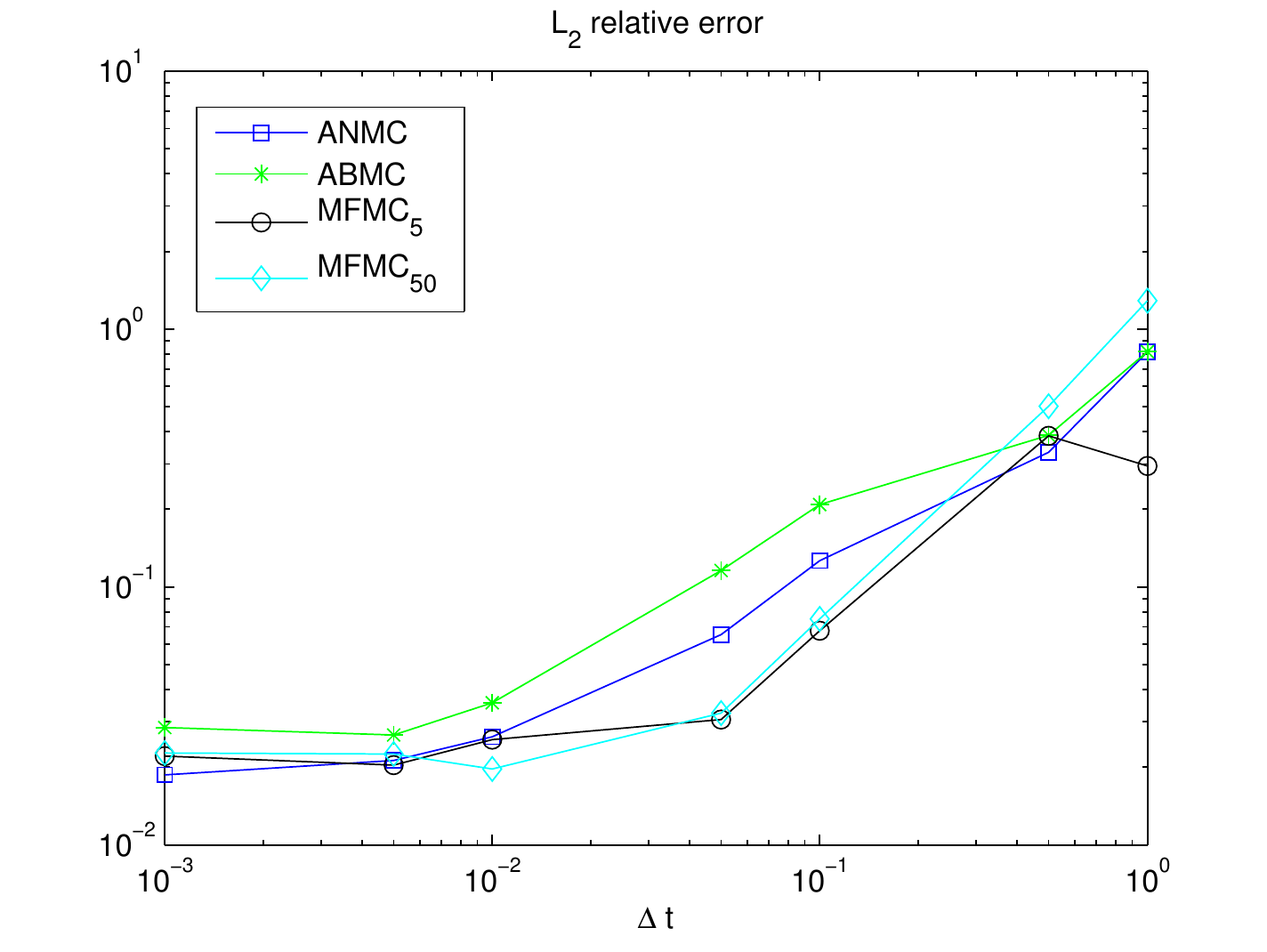}
\\
    \caption{Relative errors in the $L_2$ norm at $T=1$ for the different methods as a function of $\Delta t=\varepsilon$. On the left the error is computed with $N=1000$ particles, on the right the same test is performed with $N=50000$ particles.}
    \label{fig:Erconf}
\end{figure}
\subsection{Accuracy considerations}
Here we compare the accuracy of the different algorithms studied for a simple space homogeneous situation. In fact, since the algorithms differ only in the binary interaction dynamic the homogeneous step is the natural setting for comparing the various approaches in term of accuracy.

We consider the standard Cucker-Smale dynamic. Since we do not have any space dependence we assume $H(|x_i-x_j|) \equiv 1$ for each $i,j$. Thus there is no difference in this test case between formulations (\ref{Q_Boltz}) and (\ref{Q_Boltz2}) and the relative simulation schemes.

We take $N=50000$ individuals and at the initial time the velocity is distributed as the sum of two gaussian $$f_0(v)=\frac{1}{\sqrt{2\pi}\sigma_v}\left(e^{\displaystyle-\frac{(v+v_0)^2}{2\sigma_v^2}}+e^{\displaystyle-\frac{(v-v_0)^2}{2\sigma_v^2}}\right),$$
with $v_0=0.7$, $\sigma_v=\sqrt{0.2}$. 


The results obtained for $ANMC$ and $ABMC$ with $\varepsilon=1, 0.1, 0.01$ at time $T=1$ are reported in Figure \ref{fig:Accuracy}. The reference solution is computed using the microscopic model which in this simple situation can be solved exactly and gives
\[
v_i(t)=v_i(0)e^{-t}+(1-e^{-t})\bar{v},\qquad \bar{v}=\frac1{N}\sum_{j=1}^n v_j(0).
\]
As expected convergence towards the exact solution is observed for both methods. In particular for $\varepsilon=0.01$ the results are in good agreement with the direct solution of the microscopic model.

Next in Figure \ref{fig:match} we compare the behavior of the $MFMC_M$ method with $ANMC$ for the same values of the time step. A considerable difference is observed for large values of $\Delta t$ and both methods are poorly accurate. On the other hand for smaller values of $\Delta t$ they both converge towards the reference solution.

Finally in Figure \ref{fig:Erconf} we report the $L_2$-norm of the error for $ANMC$, $ABMC$ and $MFMC_M$ for various $M$ as a function of $\Delta t=\varepsilon$. We compare the convergence of the schemes with different number of particles $N=1000$ and $N=50000$.
Note that in both cases the convergence rate of the schemes is rather close and for  $\varepsilon=\Delta t < t^*$, the statistical error dominates the time error so that we observe a saturation effect, where $t^*\approx 0.1$ for $N=1000$ and $t^*\approx 0.01$ for $N=50000$.



\subsection{Computational considerations and 1D simulations}

Next we want to compare the computational cost of the different binary interaction methods for solving the kinetic equation (\ref{kinetic}), when compared to the direct numerical solution of the original system (\ref{model}).

We consider the same one-dimensional test problem as in~\cite{MR2744704} for the Cucker-Smale dynamic. The initial distribution is given by $$f_0(x,v)=\frac{1}{2\pi\sigma_x\sigma_v}e^{\displaystyle\frac{-x^2}{2\sigma_x^2}}\left(e^{\displaystyle\frac{-(v+v_0)^2}{2\sigma_v^2}}+e^{\displaystyle\frac{-(v-v_0)^2}{2\sigma_v^2}}\right),$$
for $v_0=2.5$, $\sigma_v=\sqrt{0.1}$ and $\sigma_x=\sqrt{2}$.

\begin{figure}[H]
\centering
\includegraphics[scale=0.4]{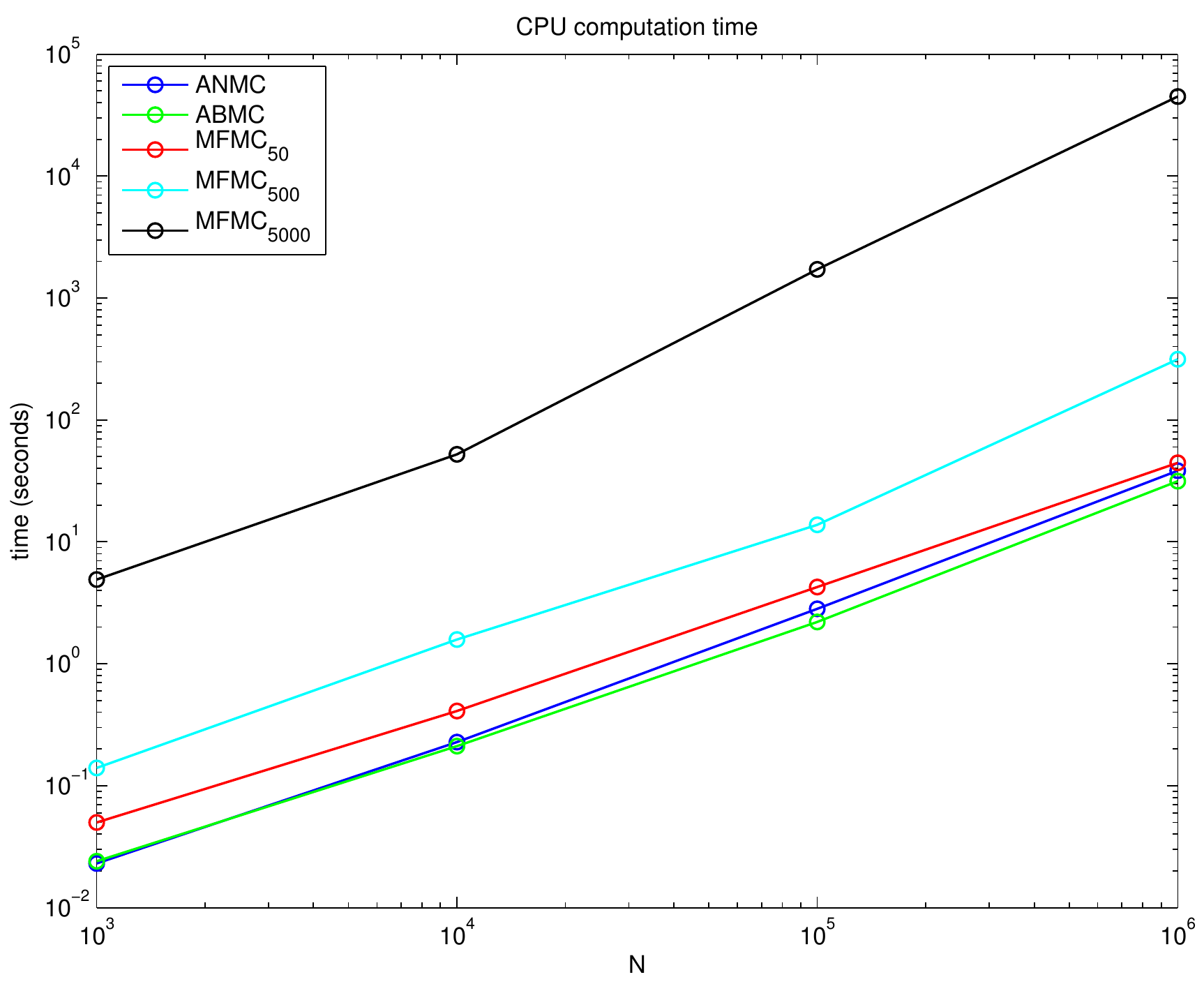}
\caption{Comparison of the computational times for the different methods. For each method the time step is equal to $\Delta t=0.01$.}
\label{fig:CPUtime}
\end{figure}

The computational time for the different methods at $T=1$ using $\varepsilon=0.01$ and different number of individuals is reported in Table~\ref{tab:ComputationalCost}. Simulations have been performed on a Intel Core $I7$ dual-core machine using Matlab. The $O(N)$ cost of $ANMC$ and $ABMC$ is evident. The same results are also reported in Figure~\ref{fig:CPUtime} which shows the linear growth of the various Monte Carlo algorithms.


\begin{table}[ht]
\centering
    \begin{tabular}{c|c|c|c|c|}
\hline
$N$ &$10^3$ & $10^4$ & $10^5$ & $10^6$
\\
\hline $ANMC$ & $0.02$ s & $0.23$ s & $2.82$ s& $3.83\times10^1$ s
\\
\hline $ABMC$ & $0.02$ s & $0.21$ s & $2.20$ s & $3.14\times10^1$ s
\\
\hline $MFMC_{50}$ & $0.05$ s & $0.41$ s & $4.26$ s & $4.44\times10^1$ s
\\
\hline $MFMC_{500}$ & $0.14$ s & $1.58$ s & $1.33\times10^1$ s & $3.14\times10^2$ s
\\
\hline
 $MFMC_{5000}$ & $5.00$ s & $5.20\times10^1 $ s & $1.71\times10^3$ s & $4.49\times10^4$ s
\\
\hline
\\
\end{tabular}
\caption{Computational times for the different methods with various values of $N$. The final time is $T=1$ and the scaling factor $\varepsilon=\Delta t$ is fixed at $10^{-2}$.}
    \label{tab:ComputationalCost}
\end{table}

Finally we report the phase-space plots of the previous 1D problem obtained using the perception cone (\ref{HHH}). Clearly the parameter $\theta$ has no meaning in the one-dimensional case, so that the perception limitation concerns only the capability to detect other individuals on the left and on the right over the line.

\begin{figure}[H]
\centering
\includegraphics[scale=0.4]{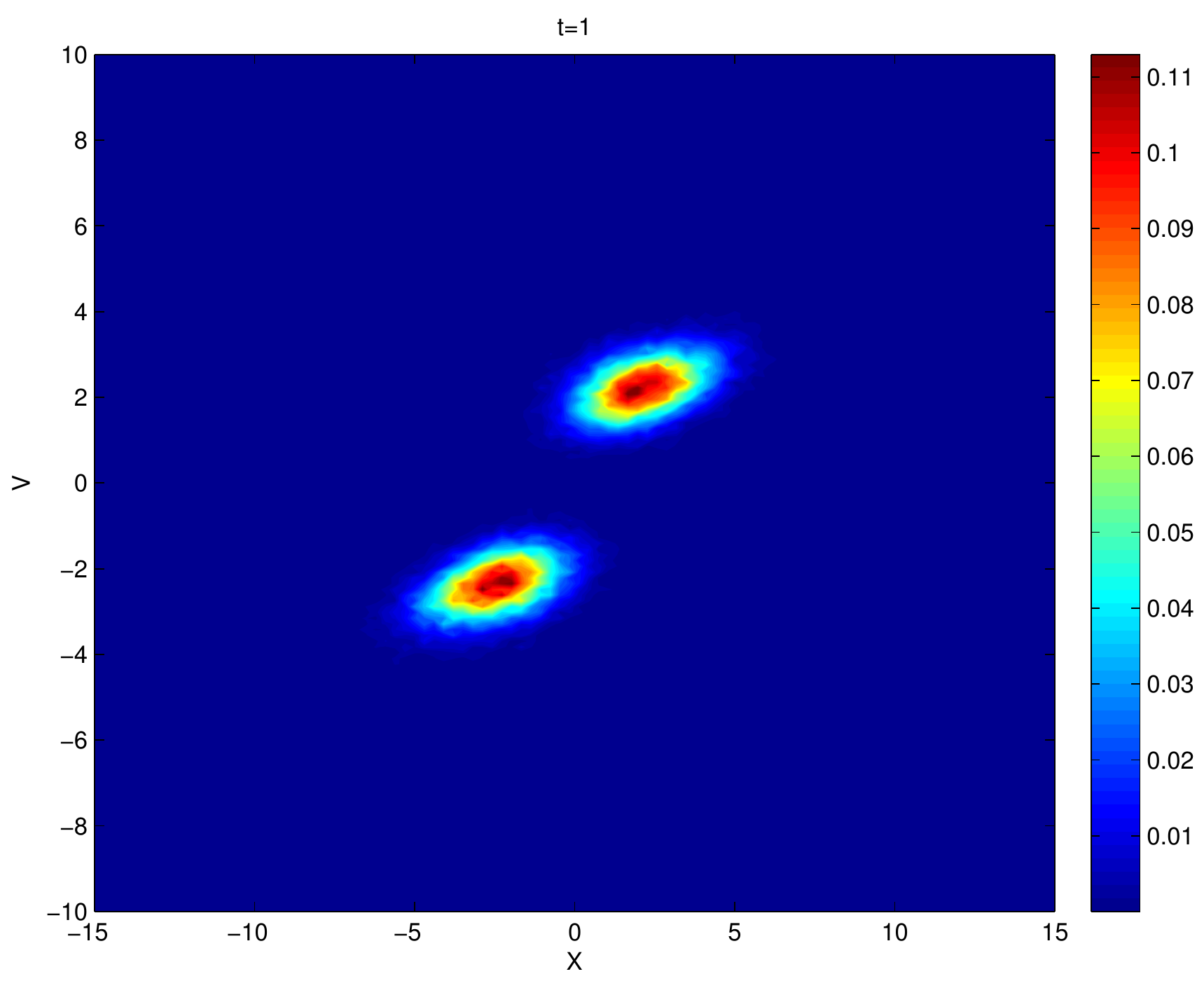}
\includegraphics[scale=0.4]{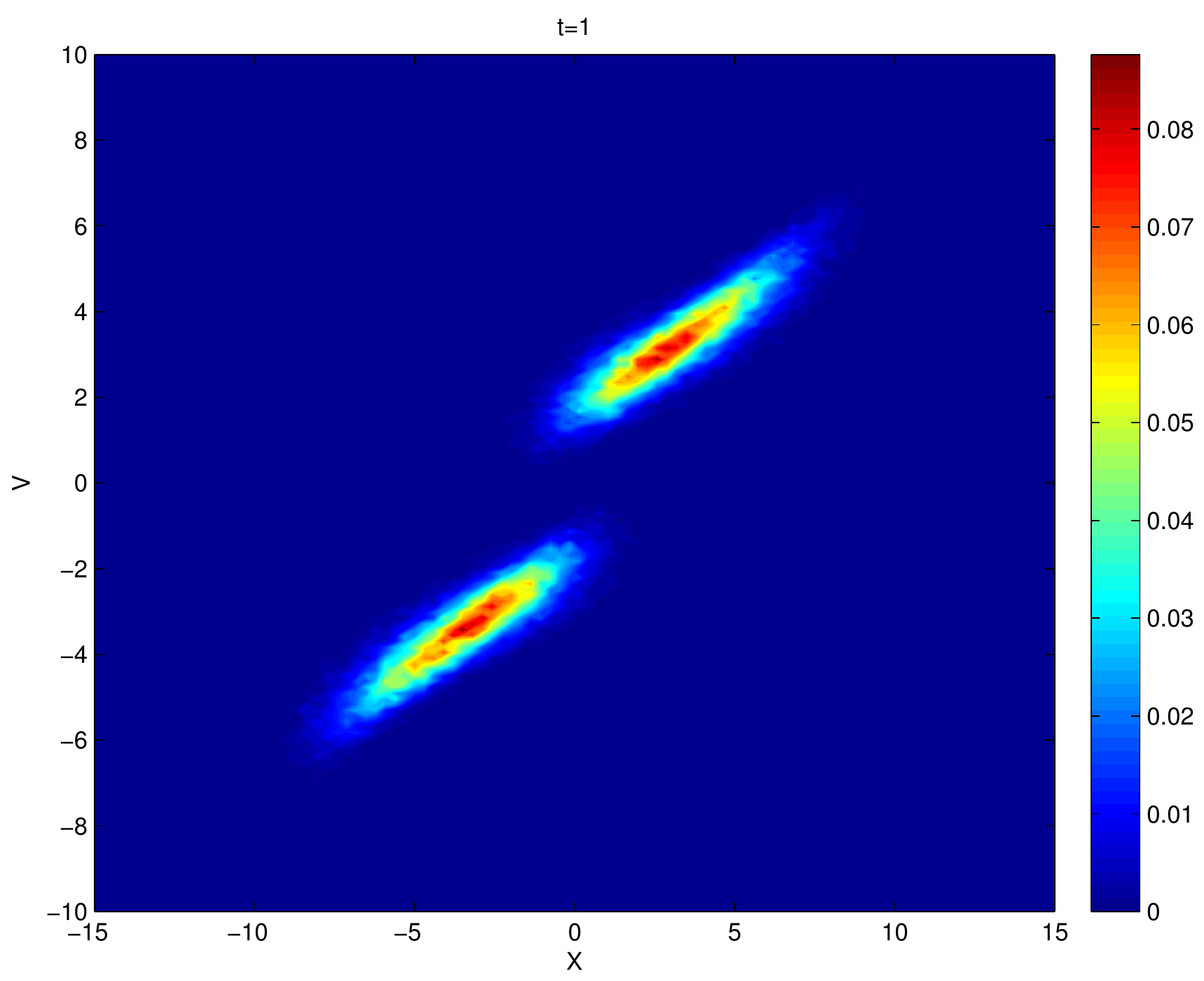}
\\
\includegraphics[scale=0.4]{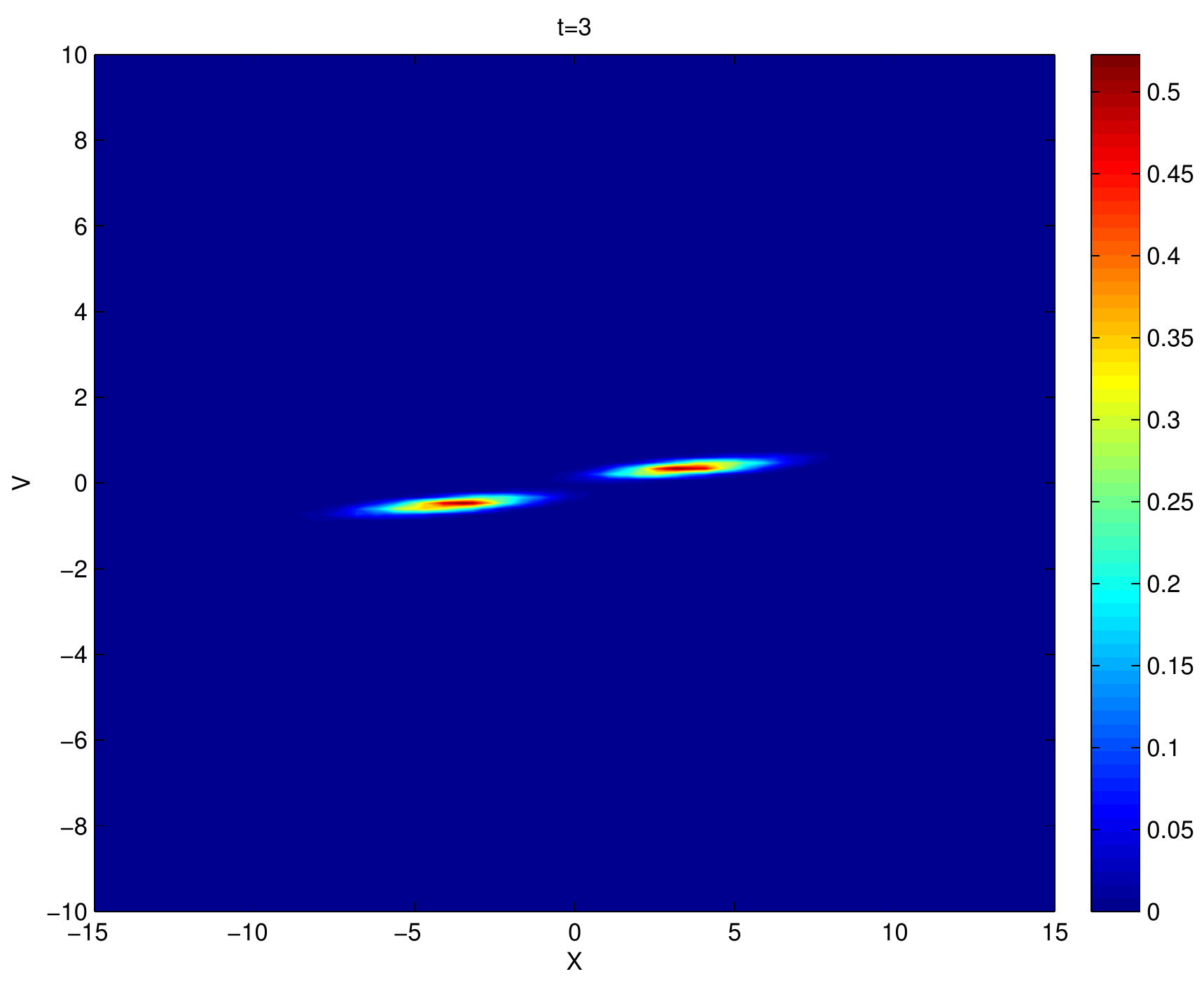}
\includegraphics[scale=0.4]{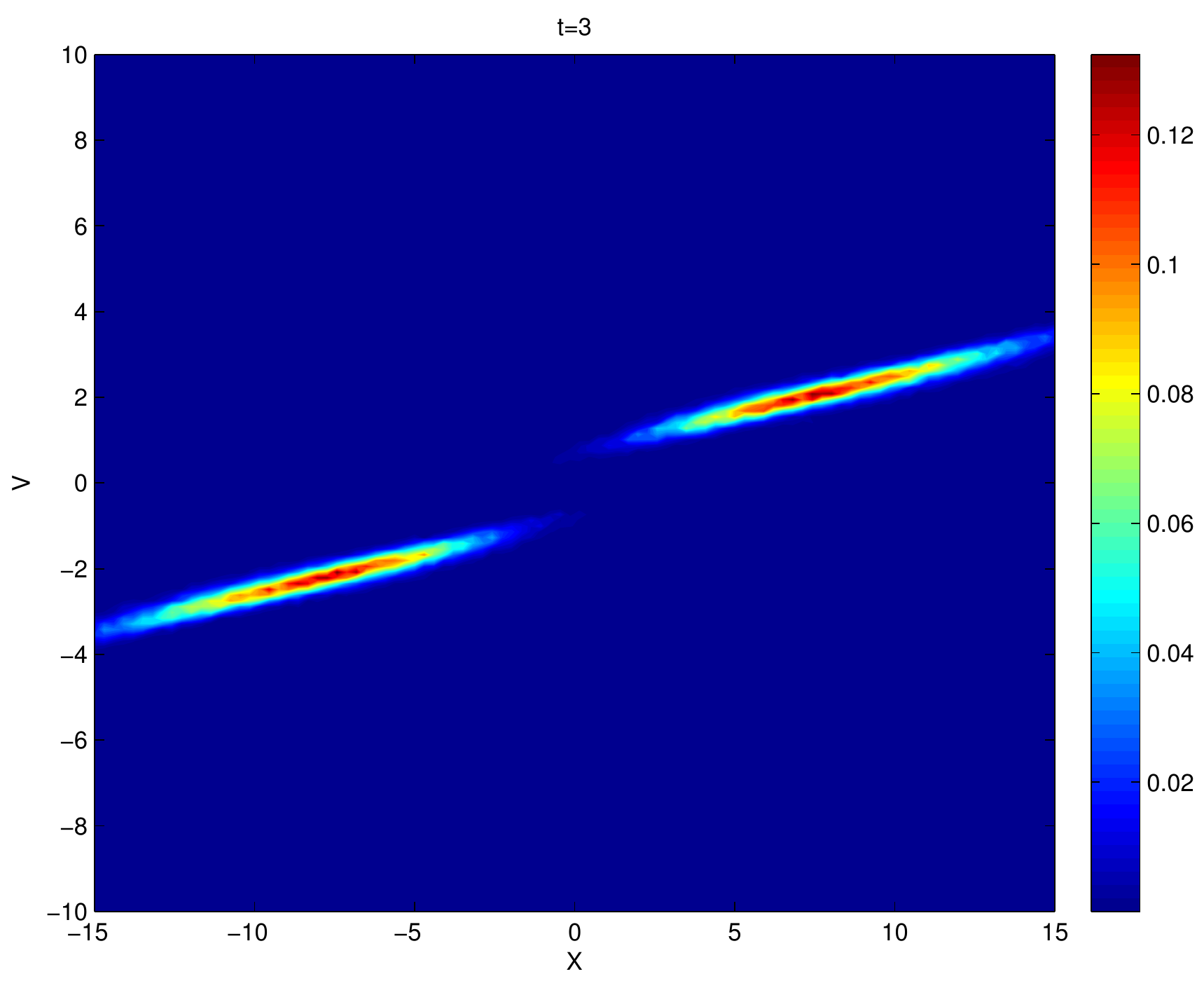}
\\
\includegraphics[scale=0.4]{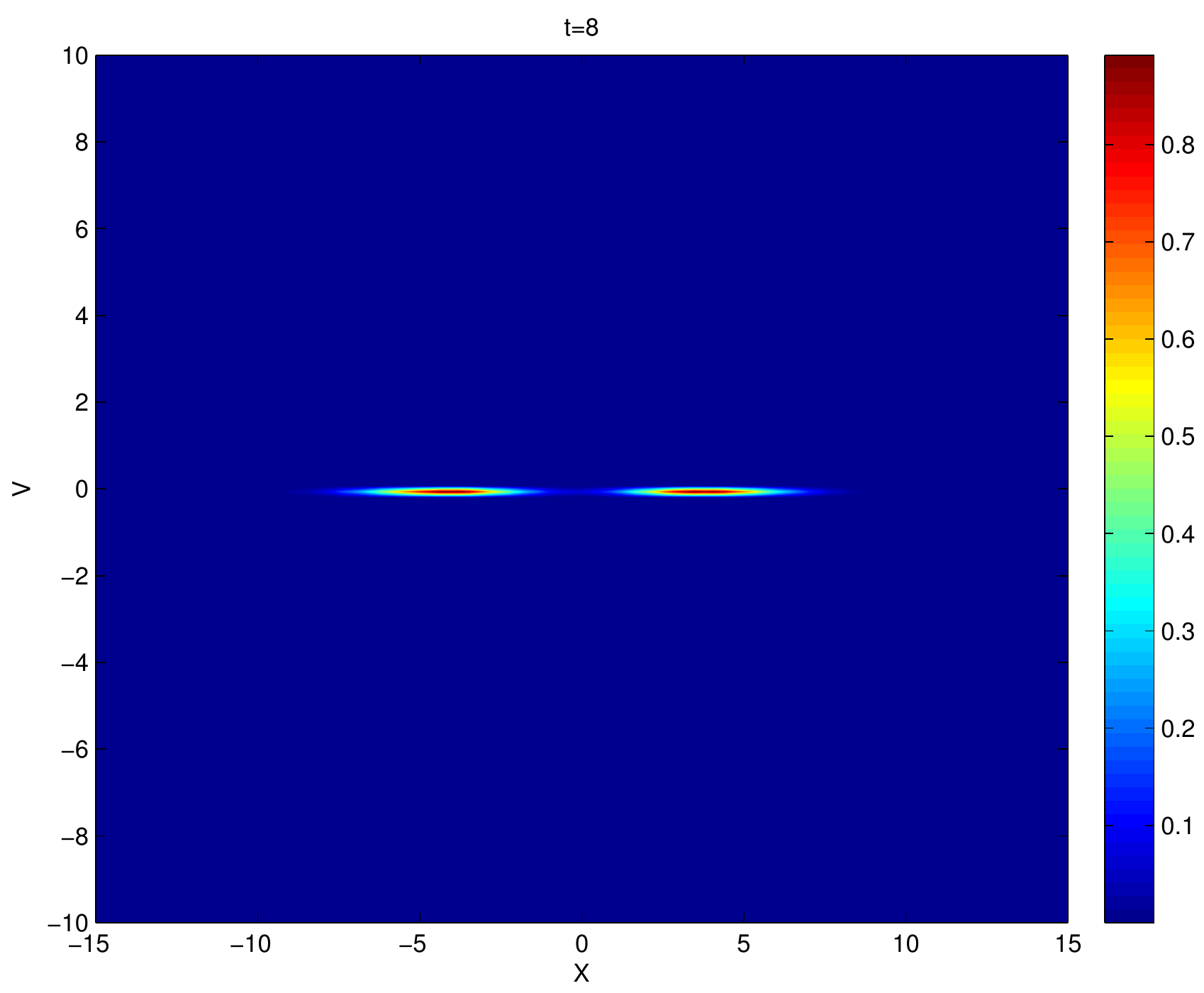}
\includegraphics[scale=0.4]{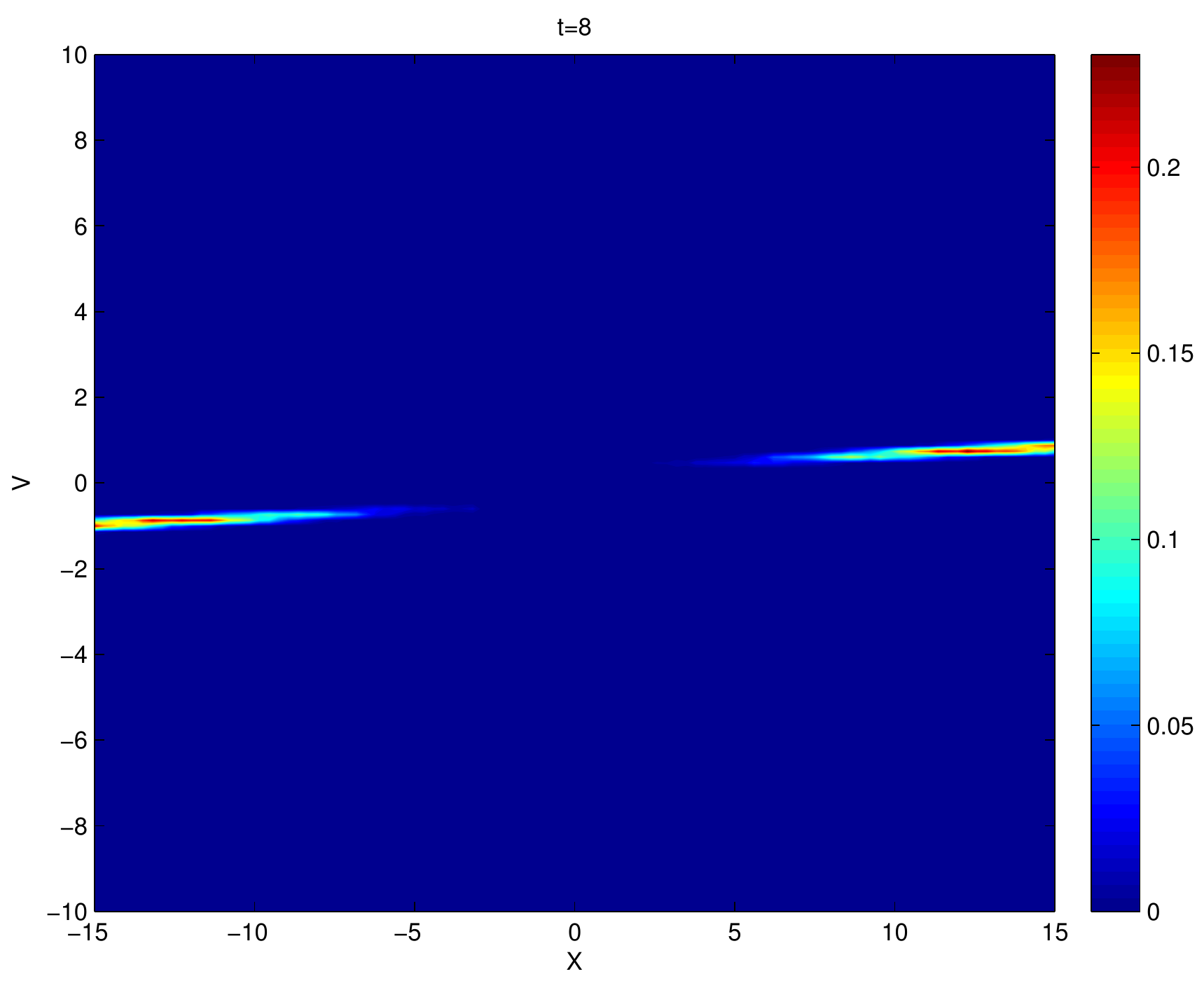}
\caption{1D Cucker-Smale flocking in the phase space. On the left without perception limitations, on the right with a perception bound characterized by $p_1=1$ and $p_2=0.5$}
    \label{fig:cs1d}
\end{figure}

We compare the evolution in the phase space of two different cases: the classical Cucker-Smale model (non visual limitation $p_1=p_2=1$) and the weighted visual cone with $p_1=1$ and $p_2=0.5$. 
The results are reported in Figure~\ref{fig:cs1d}.

The simulations have been computed using $ABMC$ with $\Delta t=\varepsilon=0.01$. The number $N$ of individuals is $N=50000$, with $\gamma=0.05$, that is unconditional flock
condition. The phase space representation is obtained using a space-velocity grid of $100\times 150$ cells over the box $[-15,15]\times[-10,10]$. The results are in good agreement with the one presented in~\cite{MR2744704}. Note how perception limitations reduce the flocking tendency of the group of individuals by creating two different flocks moving towards left and right respectively.

\subsection{2D Simulations}
\paragraph{Cucker-Smale dynamics} A generalization of the previous test in two-dimension is obtained by considering a group of $N$ individuals with position $(x,y)\in\mathbb{R}^2$, initially distributed as 
\[
f_0(x,y,v_x,v_y)=g_0(x,y) h_0(v_x,v_y),\] where $$g_0(x,y)=\frac{1}{2\pi\sigma^2}\exp\{-(x^2+y^2)/2\sigma^2\},\qquad h(r)=\frac{1}{2\pi\nu^2}\left(e^{\displaystyle\frac{-(r+v_0)^2}{2\nu^2}}+e^{\displaystyle\frac{-(r-v_0)^2}{2\nu^2}}\right),$$ 
with $r=|(v_x,v_y)|$, $v_0=10$, $\sigma=\sqrt{2}$ and $\nu=\sqrt{0.1}$.
In the following simulations we use $N=100000$ particles and the limited perception cone defined by (\ref{HHH}). 

We compare the evolution of the space density for different choices of the perception parameters and at different times. In the test case considered the parameters for the perception cone are $\theta=4/3\pi$, $p_1=1$ and $p_2=0.04$.
To reconstruct the probability density function in the space we use a $100\times 100$ grid on $[-20,20]\times[-20,20]$. In each figure we also add the velocity flux to illustrate the flock direction. We report the results computed with $ABMC$ method and $\Delta t=\varepsilon=0.01$, similar results are obtained with the other stochastic binary algorithms.

At $T=30$ the final flocking structure is reached. It is remarkable that in absence of perception limitations we obtain a perfect circular ring moving at constant speed.   
\begin{figure}[H]
\centering
		\includegraphics[scale=0.4]{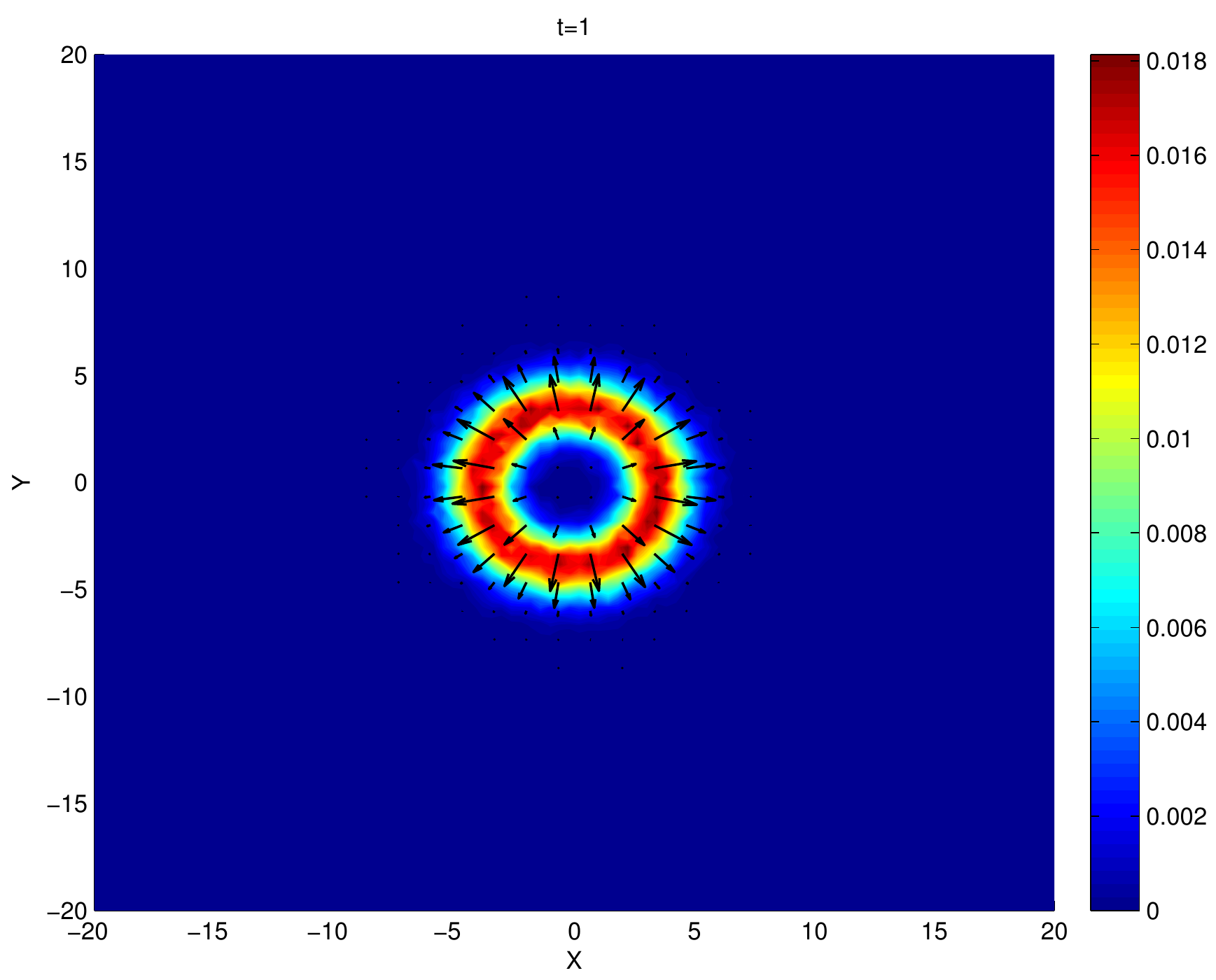}
		\includegraphics[scale=0.4]{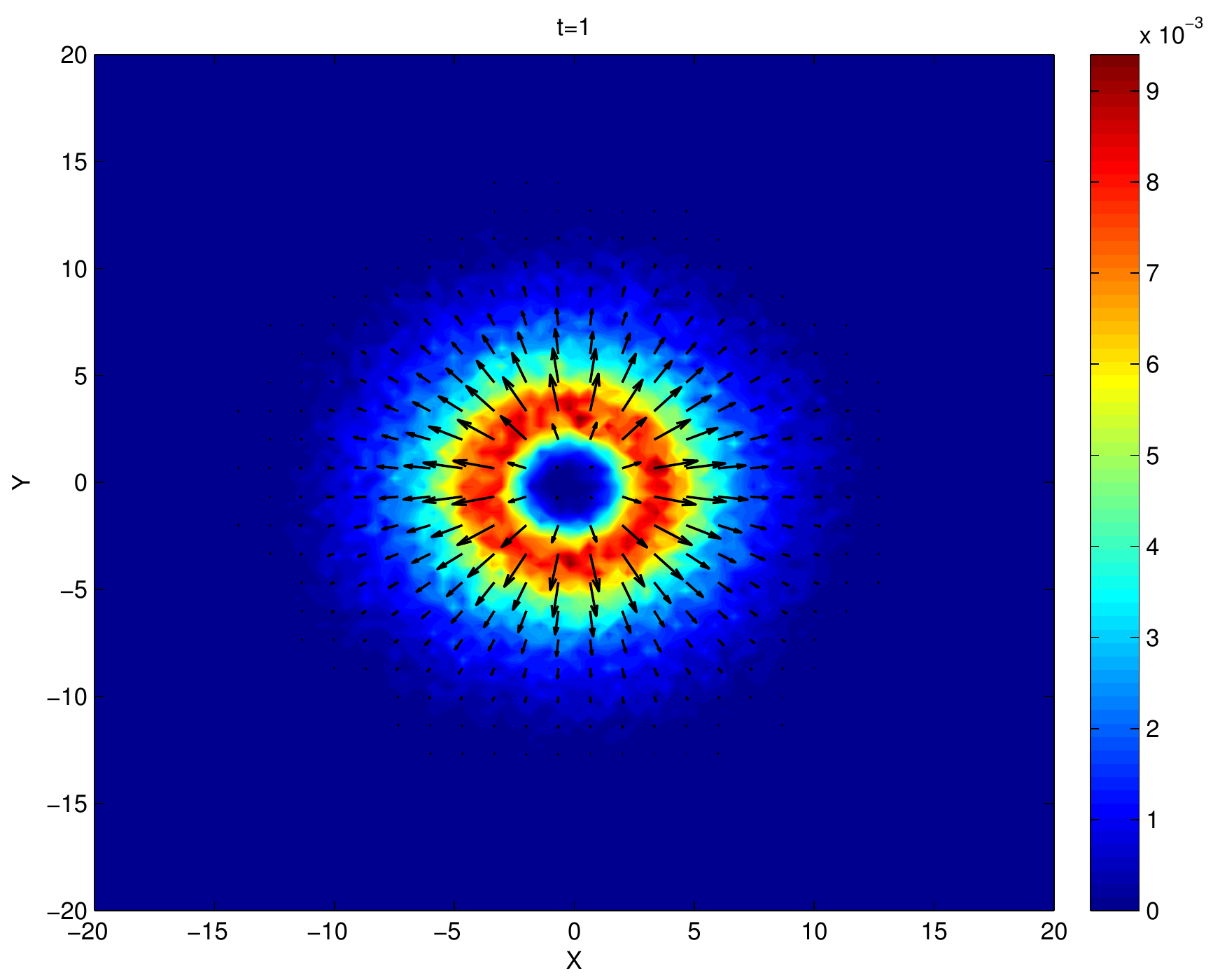}
		\\
		\includegraphics[scale=0.4]{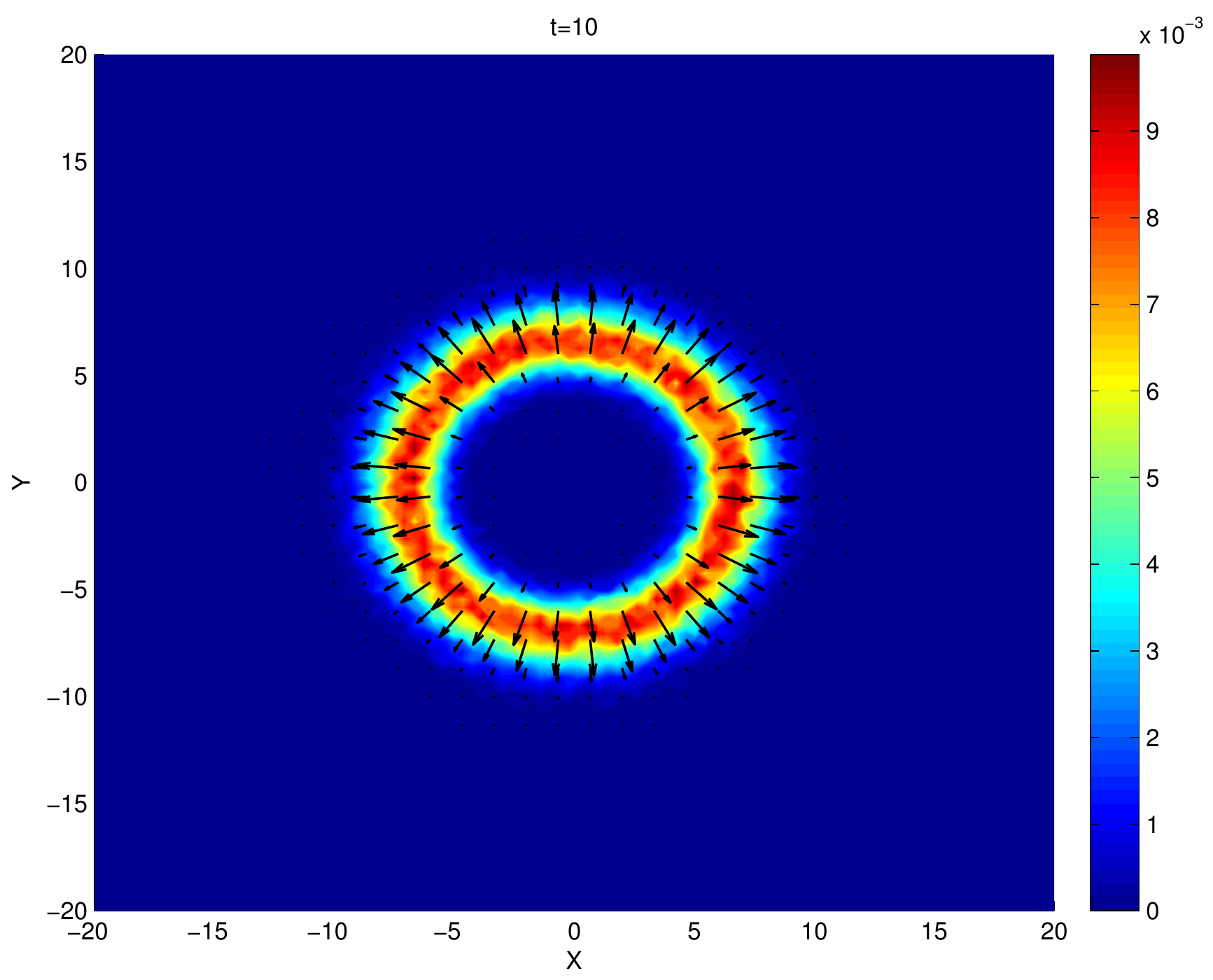}
		\includegraphics[scale=0.4]{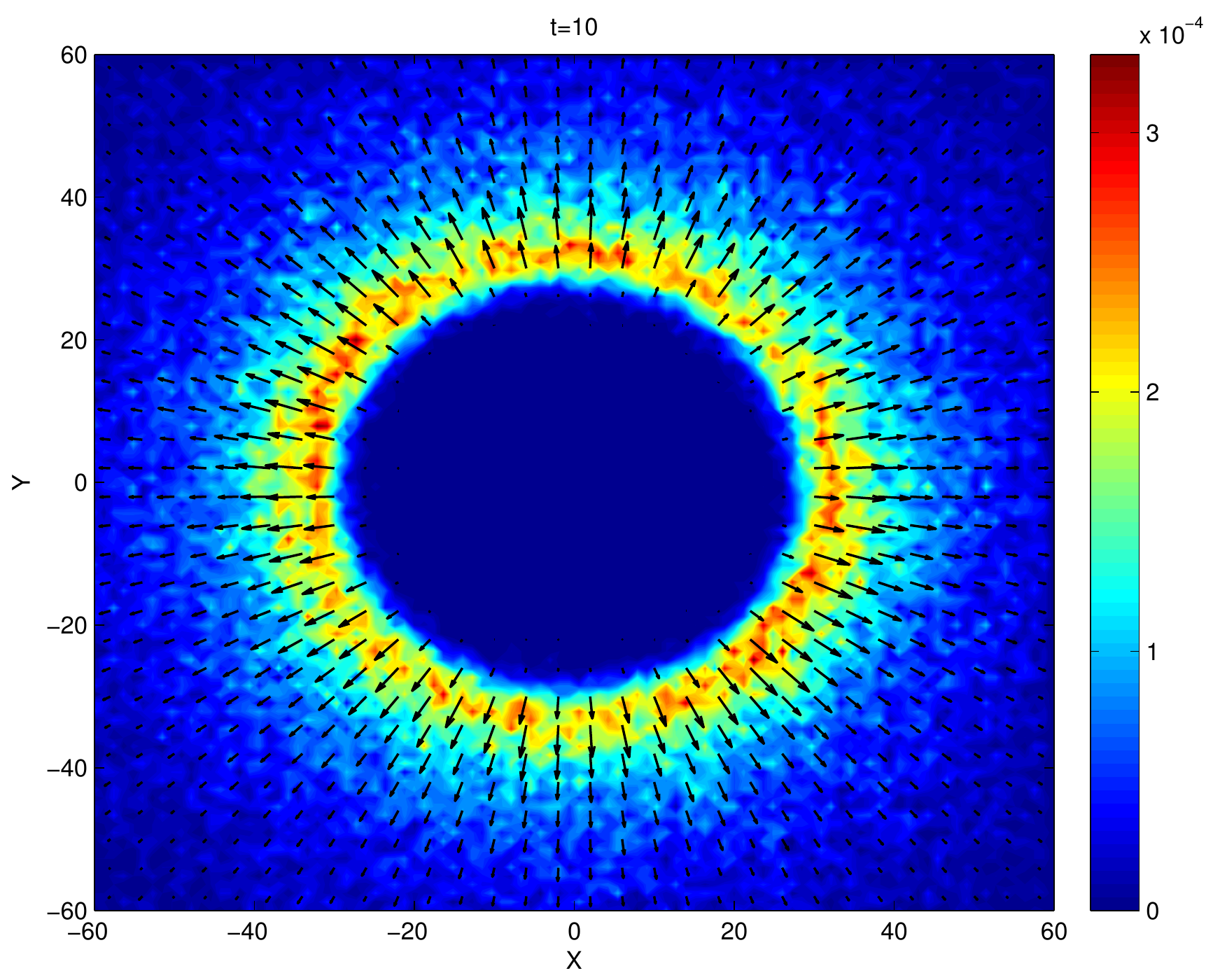}
		\\
		\includegraphics[scale=0.4]{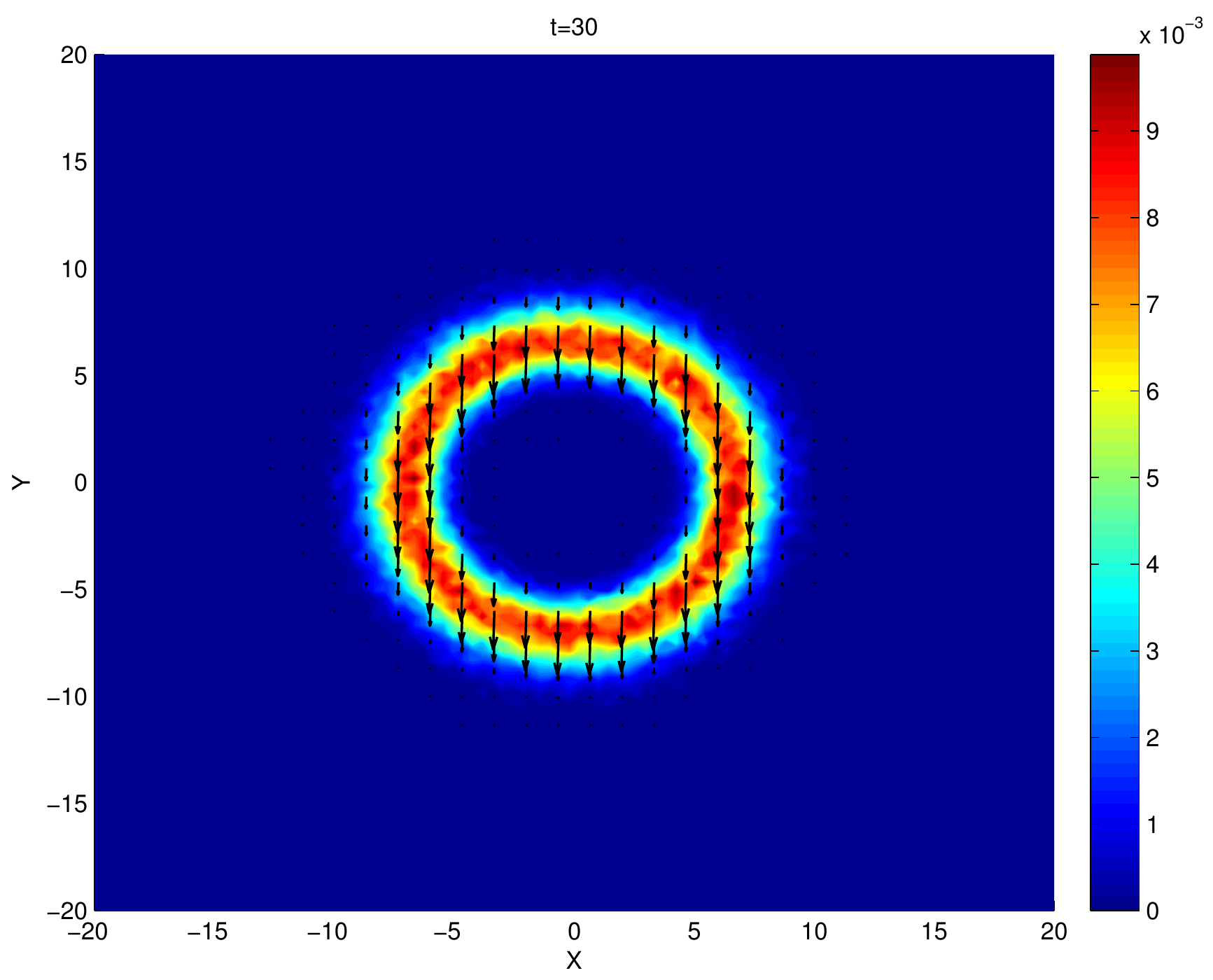}
		\includegraphics[scale=0.4]{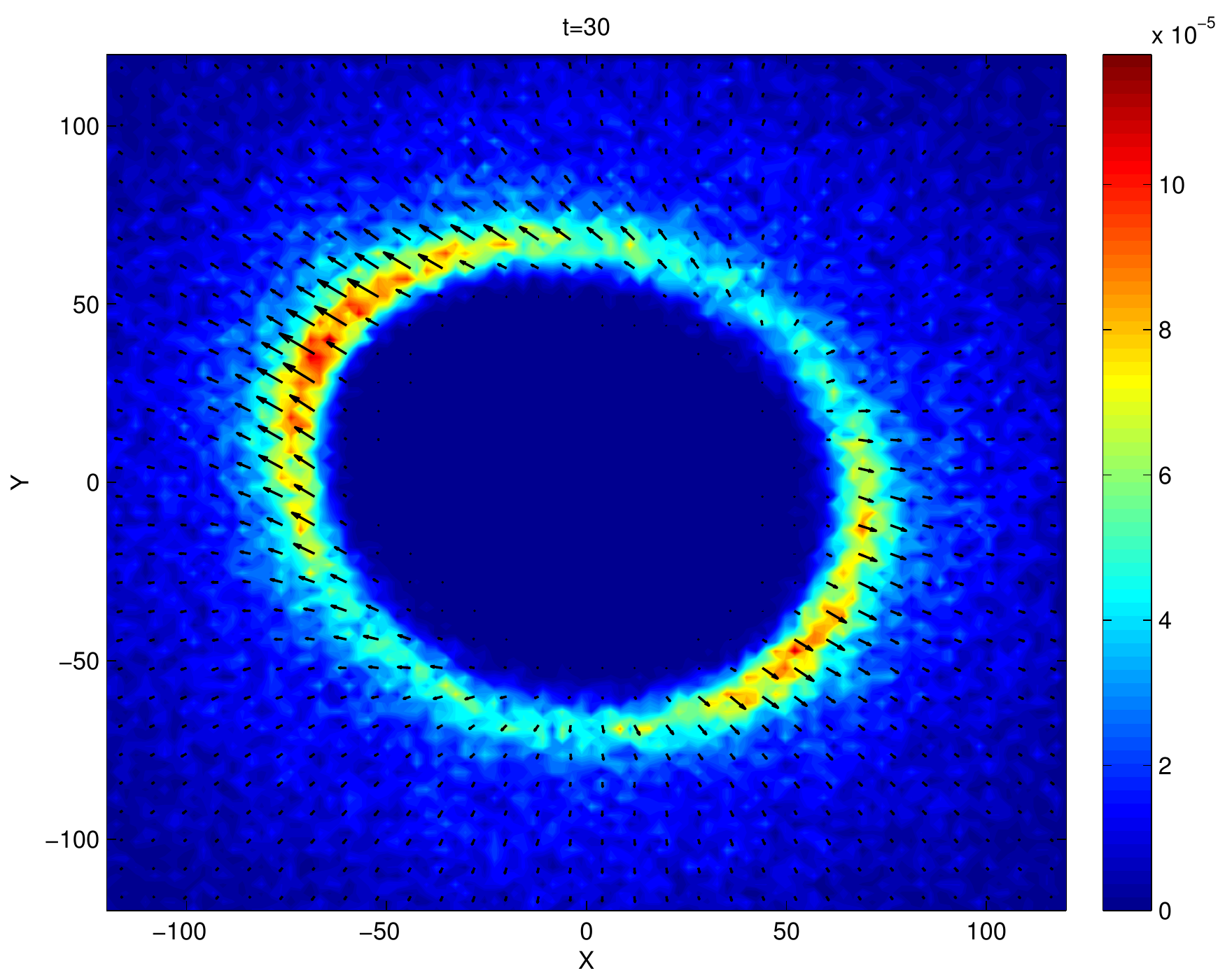}		
\label{fig:MidDispCirc}
\caption{2D Cucker-Smale flocking. On the left without perception limitations, on the right with perception cone with $\theta=4/3\pi$, $p_1=1$ and $p_2=0.04$.}
\end{figure}
In contrast when we introduce limitations the flocking behavior is less evident and the groups splits in two flocks moving in opposite directions.
Finally we can also modify the previous example to create more complex patterns, but with the same basic structure.  

\begin{figure}[H]
\centering
\includegraphics[scale=0.4]{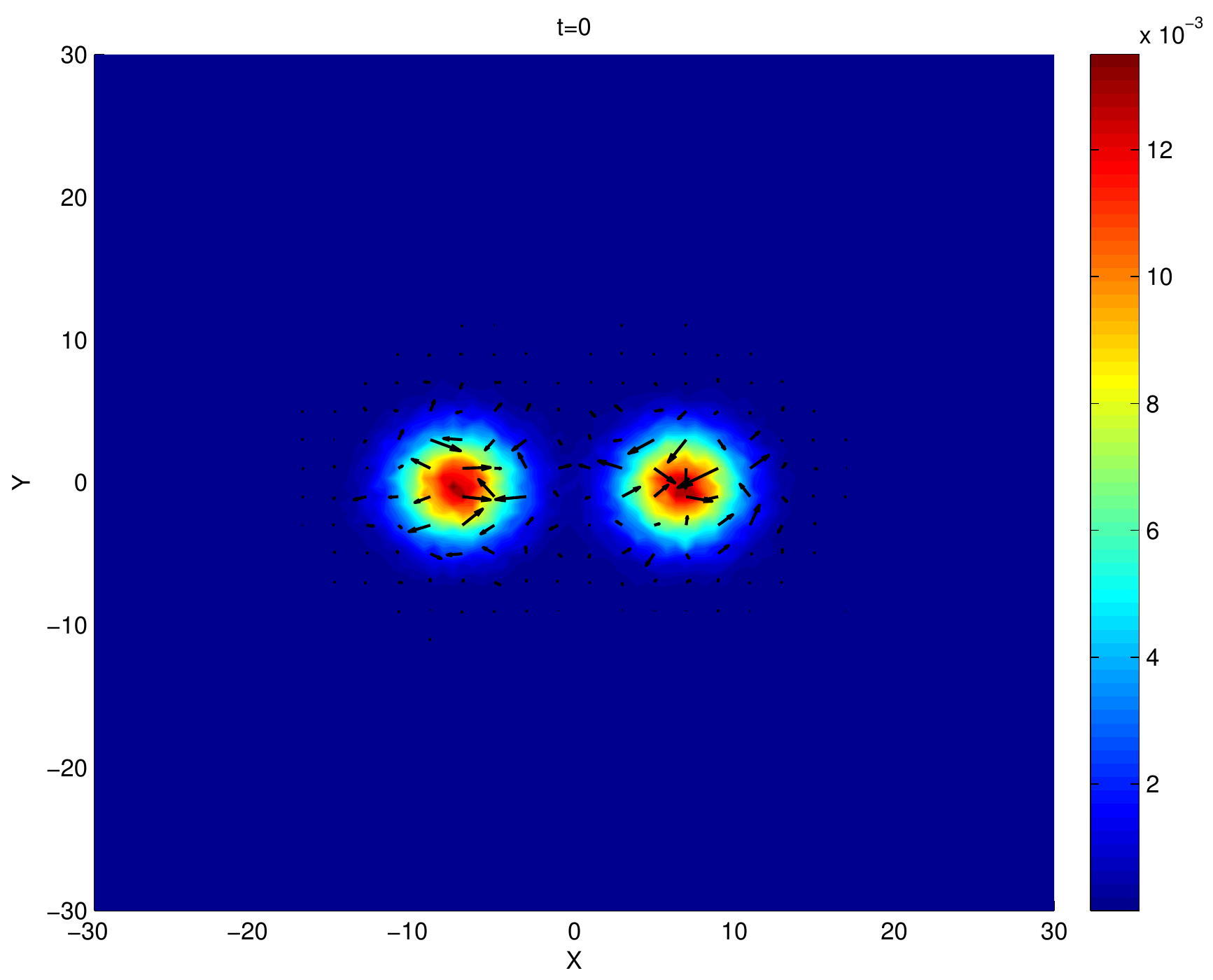}
\includegraphics[scale=0.4]{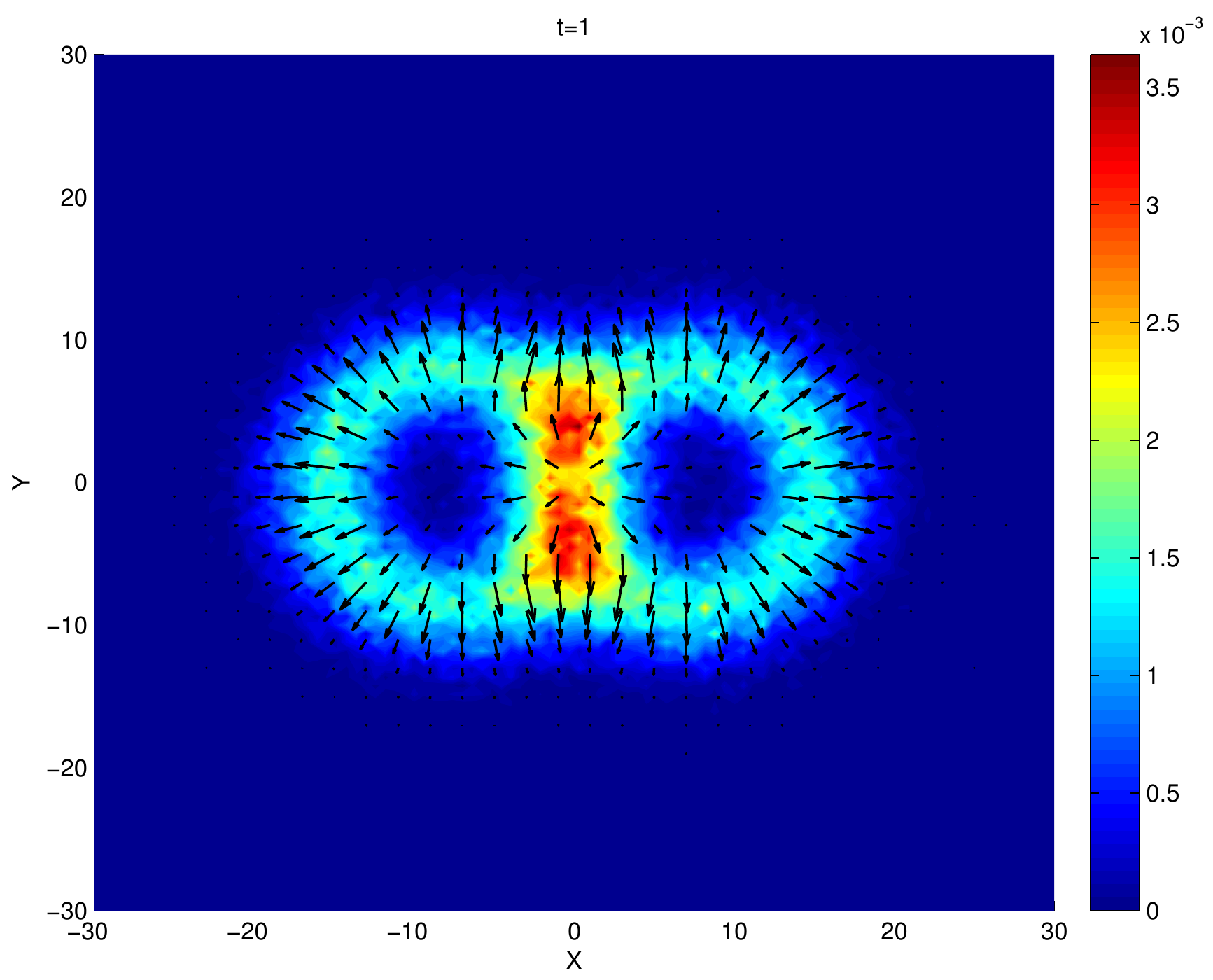}
\\
\includegraphics[scale=0.4]{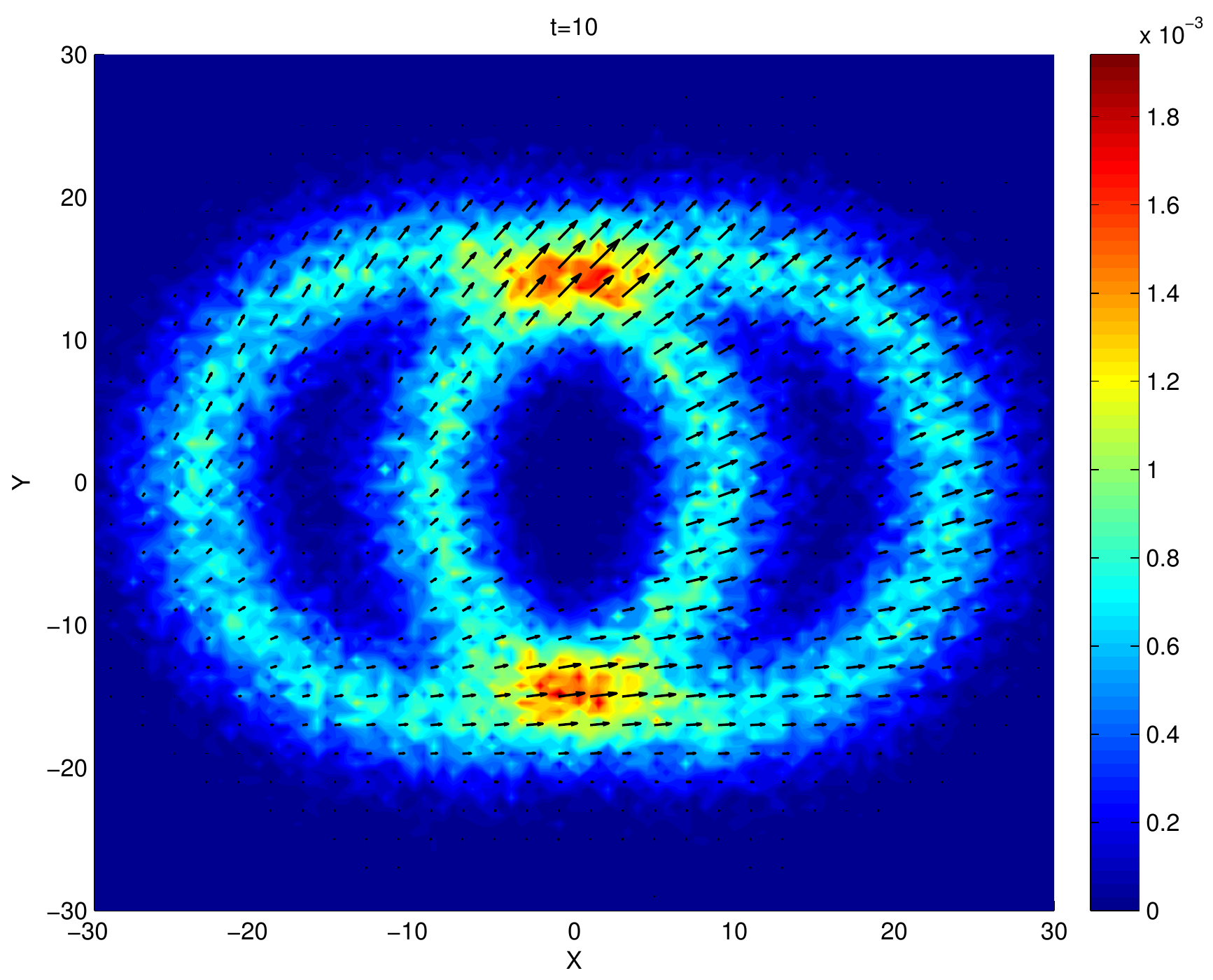}
\includegraphics[scale=0.4]{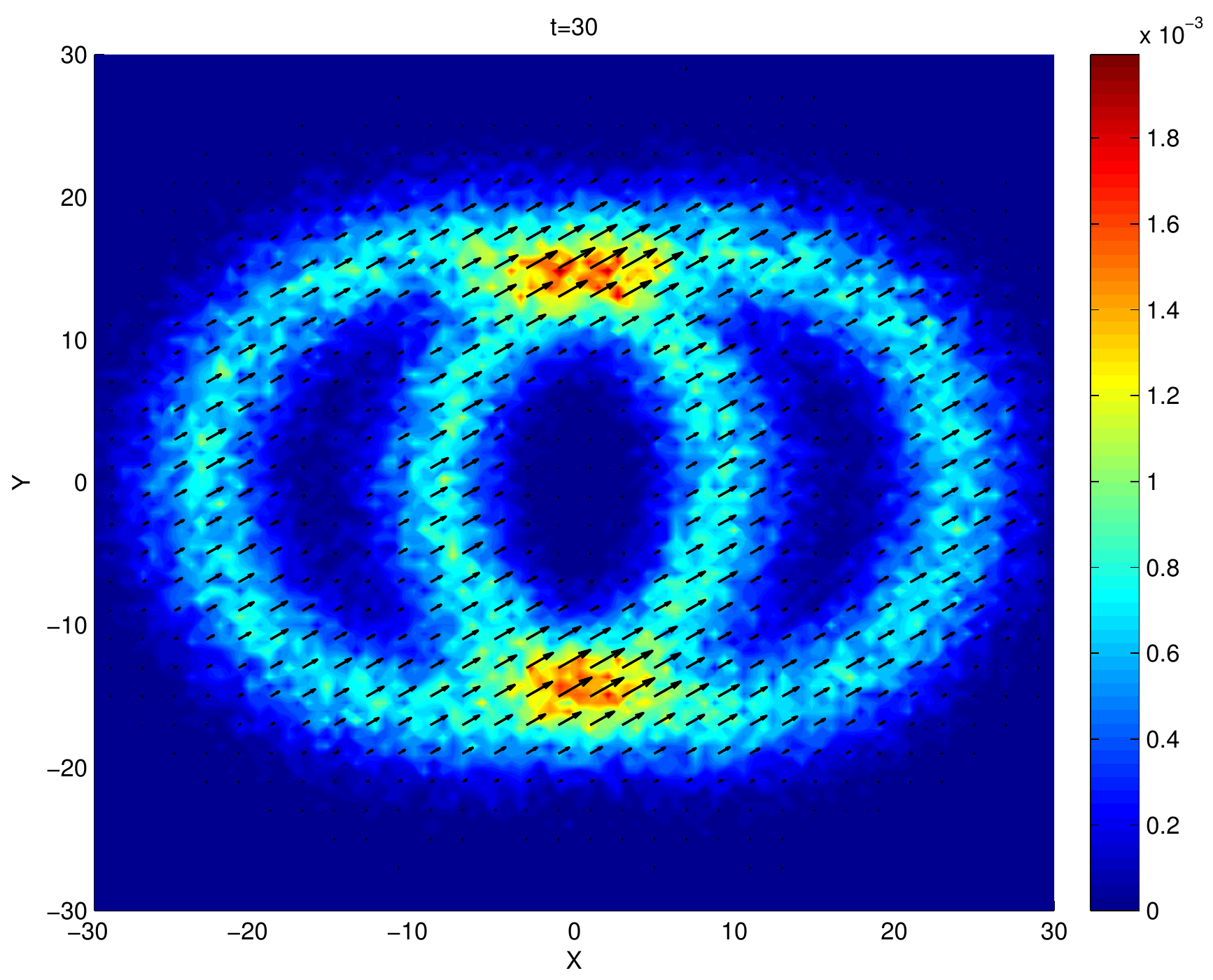}
    \caption{2D Cucker-Smale dynamics. Spatial density of two flock that merge together.}
    \label{fig:Birad}
\end{figure}


The initial distribution now is given by $$g_0(x,y,v_x,v_y)=f_0(x+m,y,v_x,v_y)+f_0(x-m,y,v_x,v_y),$$ where $f_0$ is defined as before, and $m=7$. We report the results obtained in absence of perception cone. The final flocking state is reached at $t\approx 30$ (see figure \ref{fig:Birad}). 

\paragraph{D'Orsogna, Bertozzi model et al. dynamic}
Next we want to simulate the D'Orsogna Bertozzi model et al. model with the aim to reproduce the typical mill dynamics as in \cite{MR2507454, carrillo2010self, MR2765734} but using the Boltzmann kinetic approximation.

Mills and double mills are typical emergence phenomena in school of fishes and flock of birds which travel in a compact circular motion, for example, in order to protect themselves from predator attacks.
At first we work in the twodimensional space taking into account $N=100000$ individuals. According to the interaction described in (\ref{model_Bert}), we consider the long-range attraction and short-range repulsion.

\begin{figure}[H]
\centering
\includegraphics[scale=0.4]{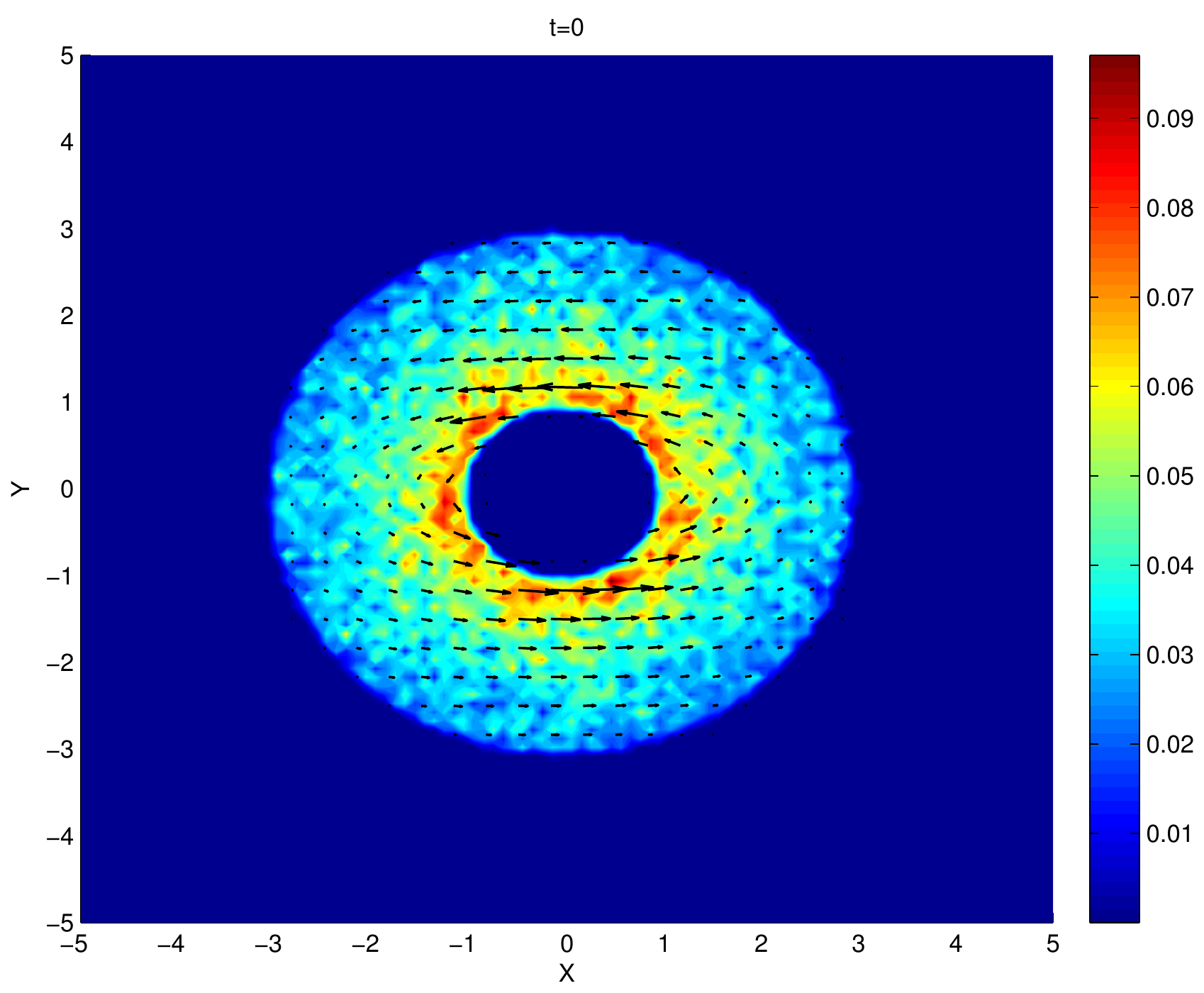}
\includegraphics[scale=0.4]{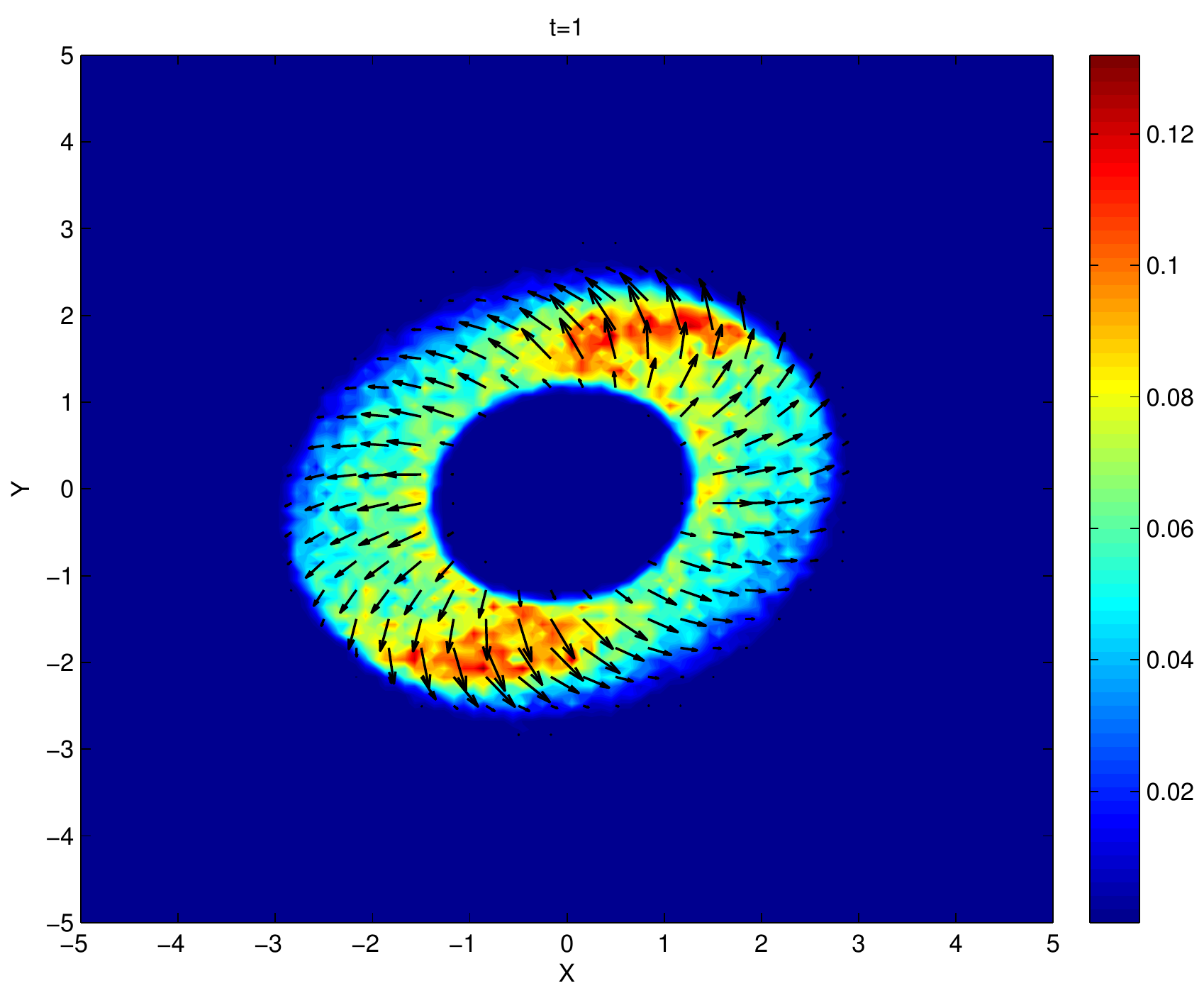}
\\
\includegraphics[scale=0.4]{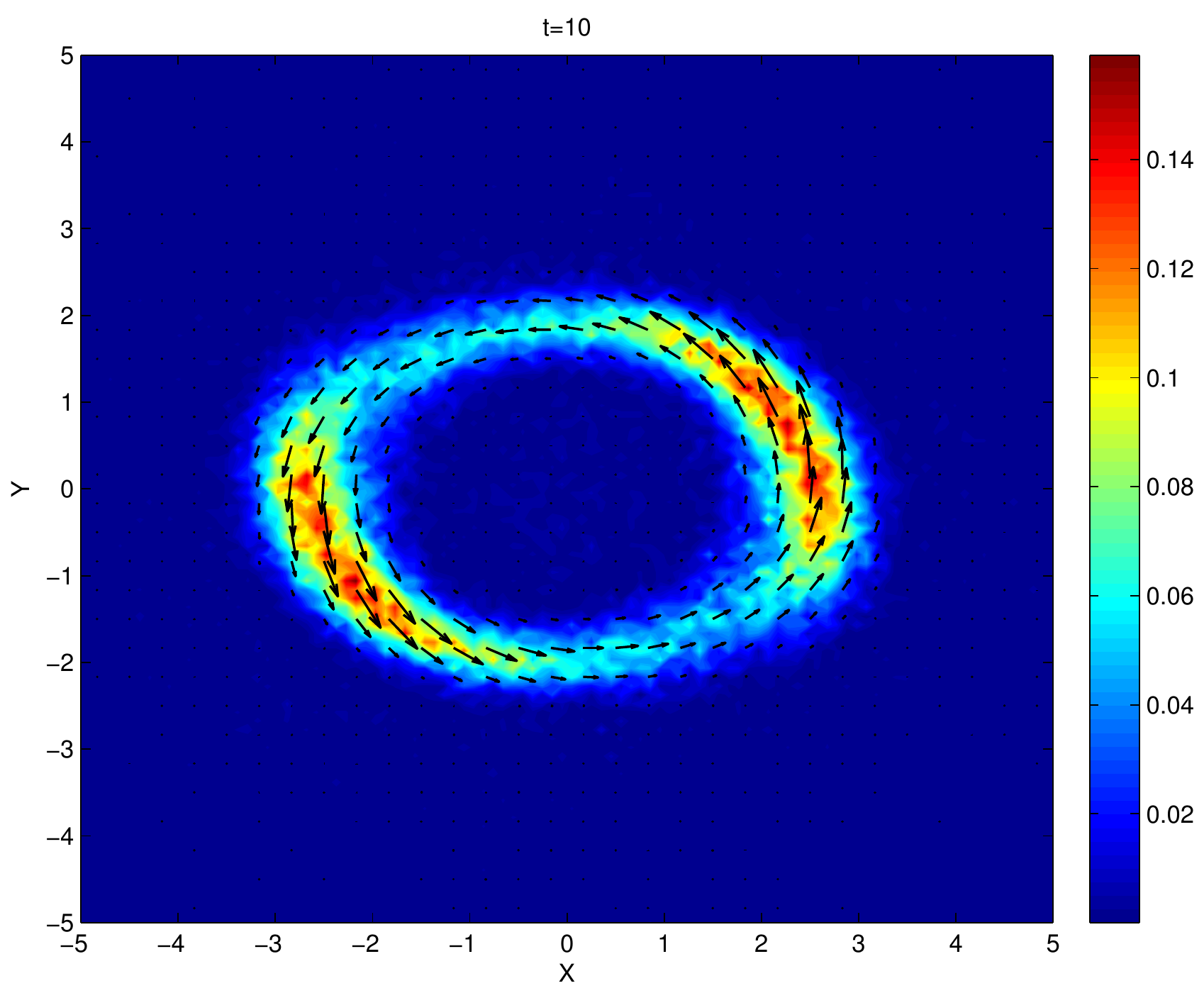}
\includegraphics[scale=0.4]{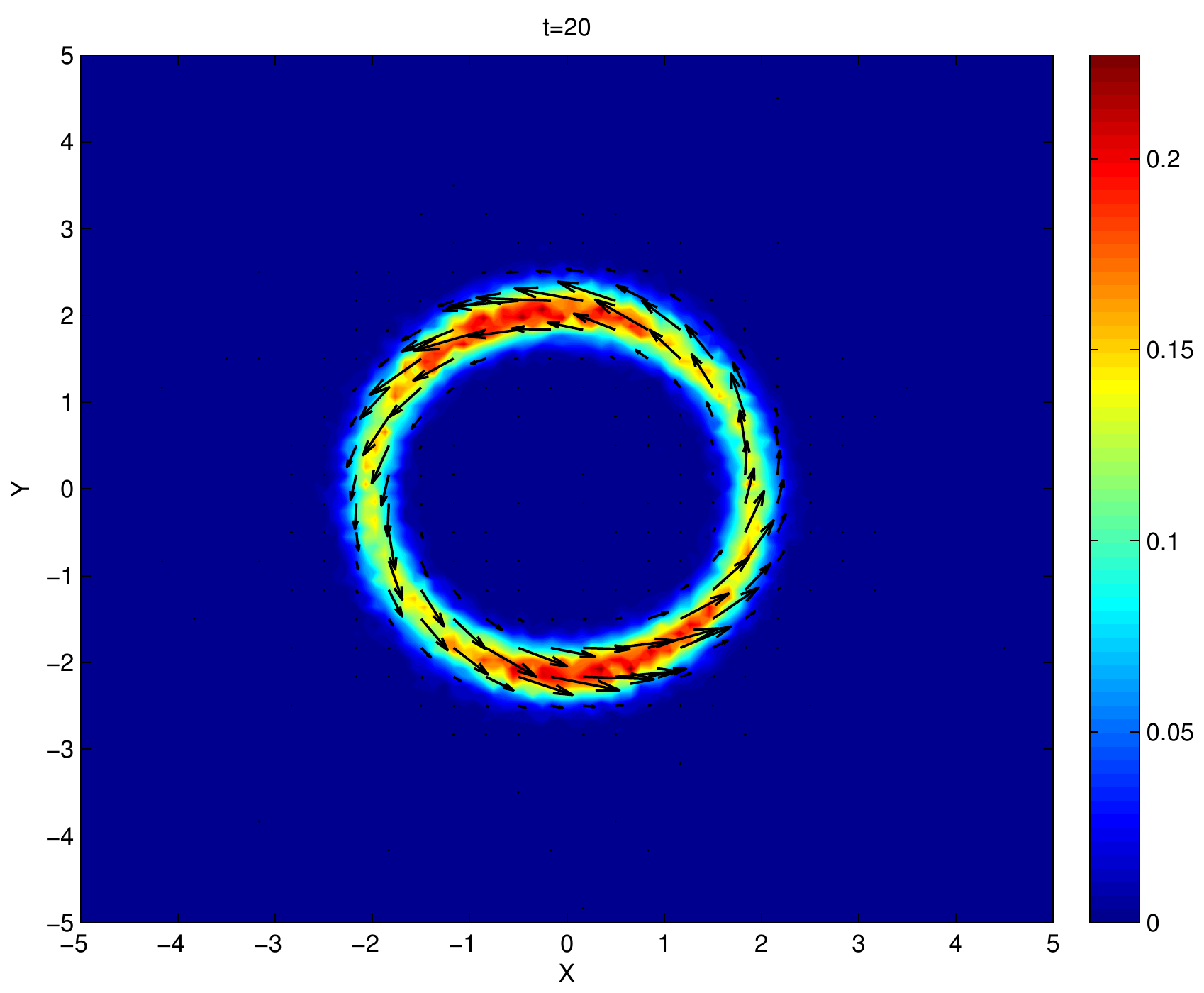}
    \caption{2D Mill in D'Orsogna-Bertozzi et al. model at various times. Parameters in the attraction-repulsion potential are such that $C=C_R/C_A=30$, $l=l_R/l_A=0.3$, $\alpha=0.7$, $\beta=0.05$. Final configuration is reached after $t=20$}
    \label{fig:Mill2}
\end{figure}

In figure (\ref{fig:Mill2}) the initial data is uniformly distributed on a twodimensional torus, with a circular motion.
The evolution shows how the attraction and repulsion forces stretch the mill reaching after $t=20$ a condition of equilibrium in a stable circular motion as a single mill.

\begin{figure}[H]
\centering
\includegraphics[scale=0.4]{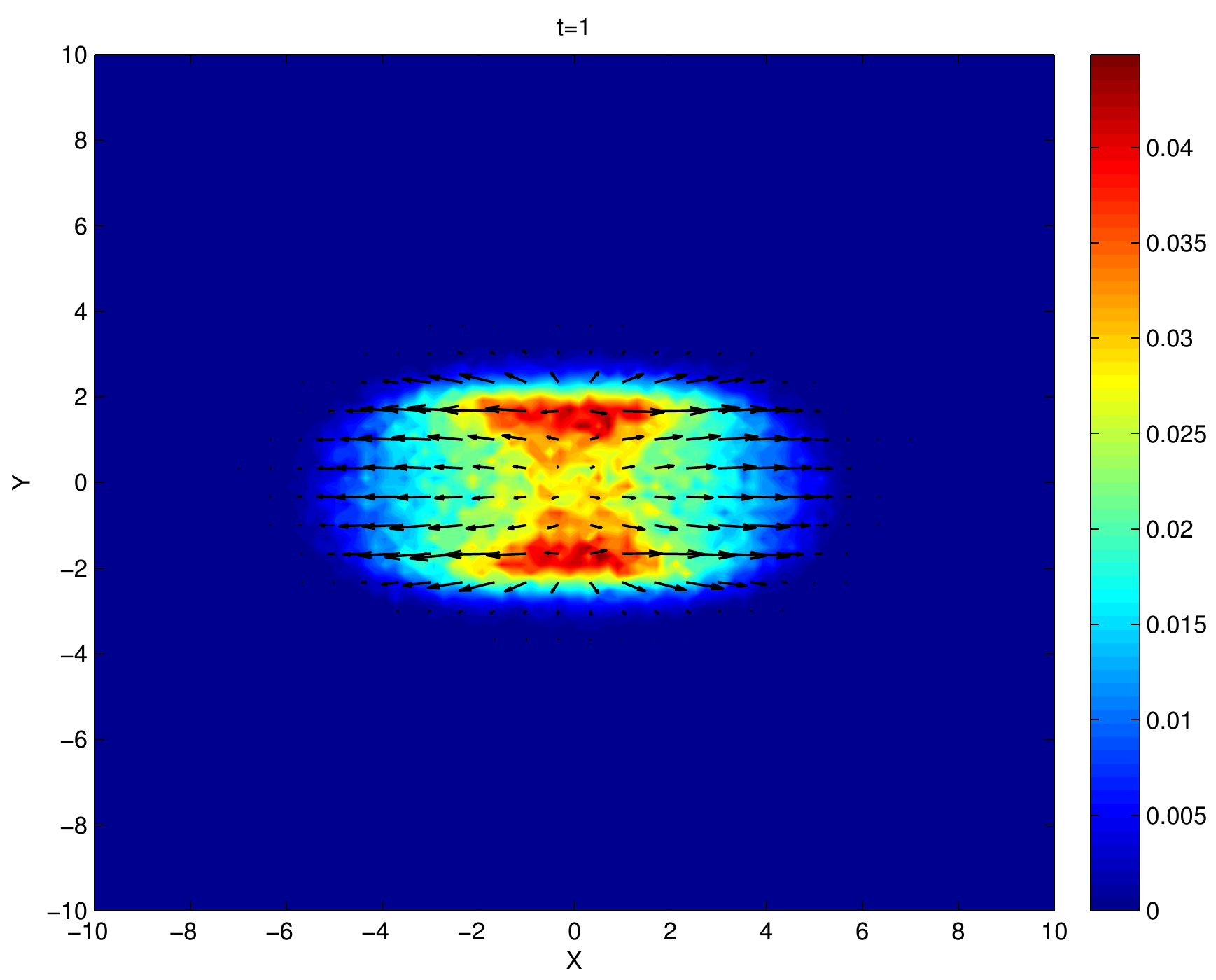}
\includegraphics[scale=0.4]{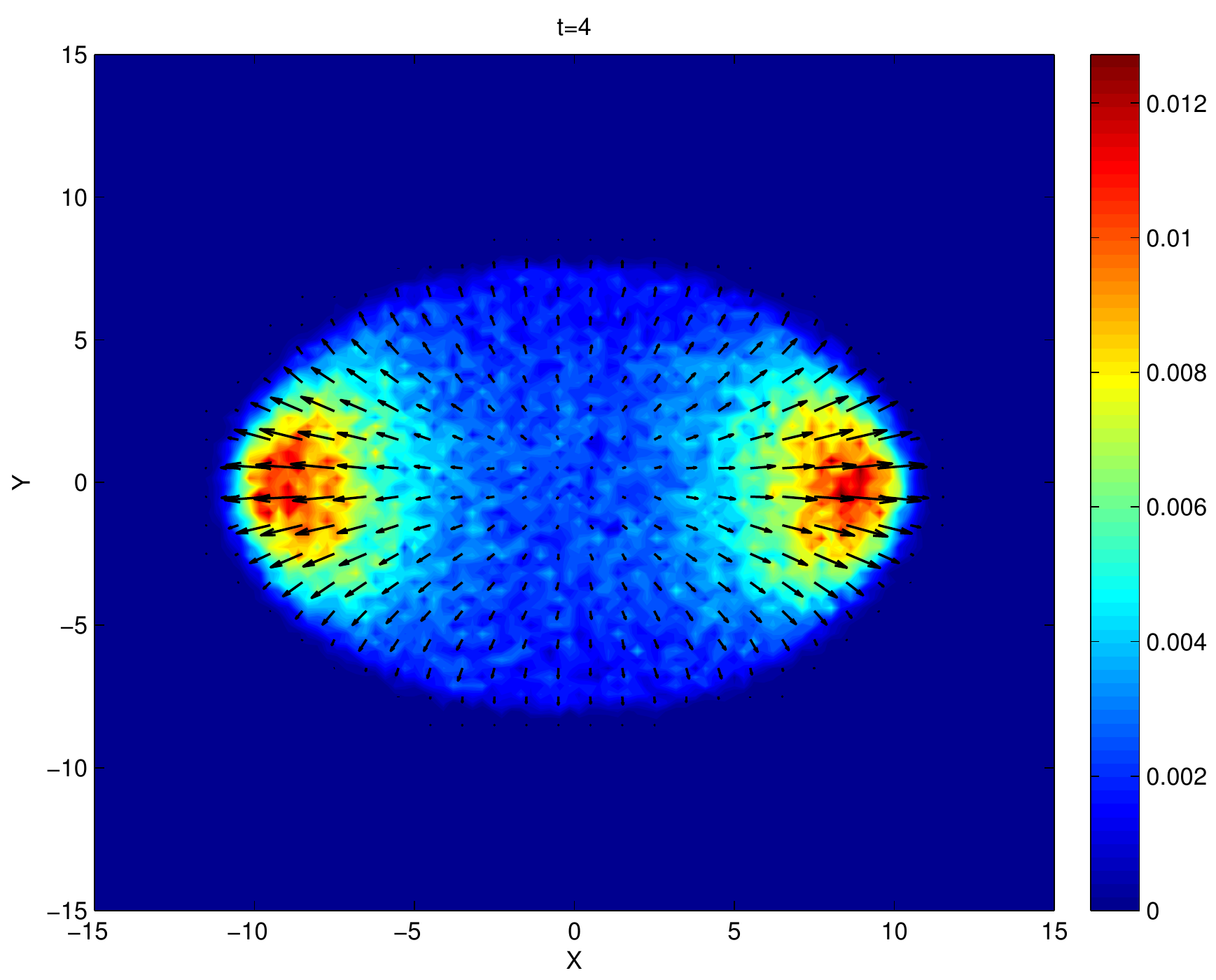}
\\
\includegraphics[scale=0.4]{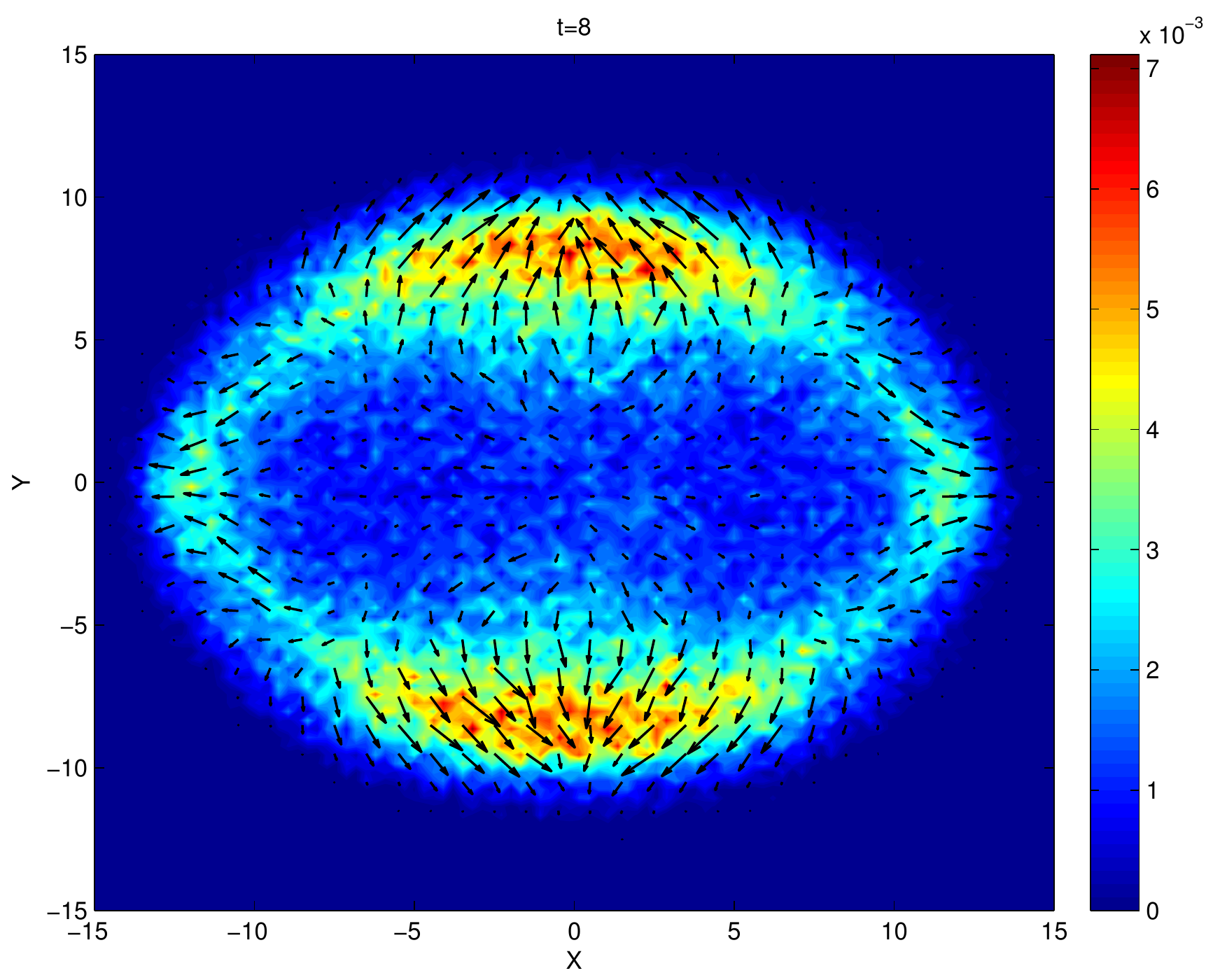}
\includegraphics[scale=0.4]{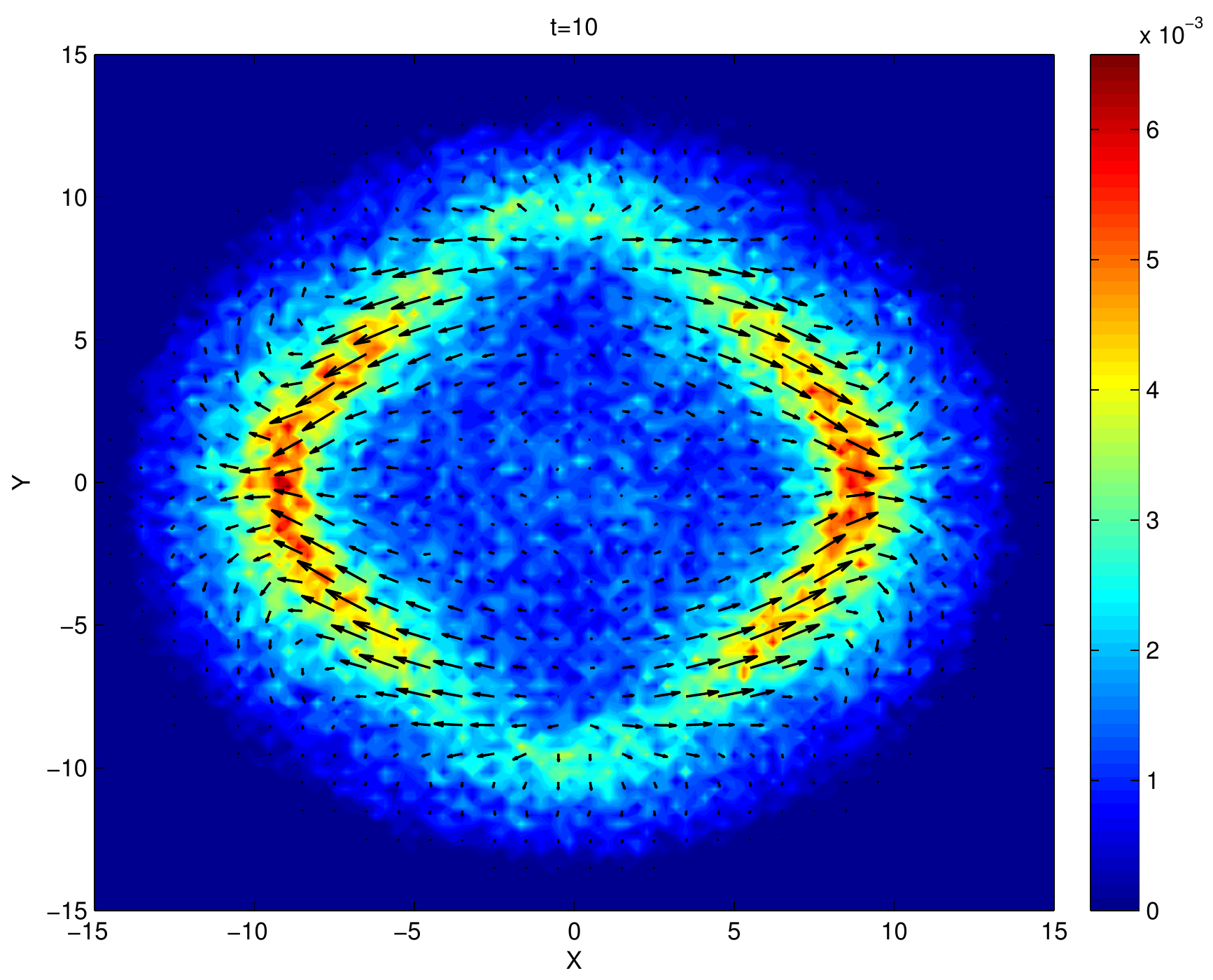}
\\
\includegraphics[scale=0.4]{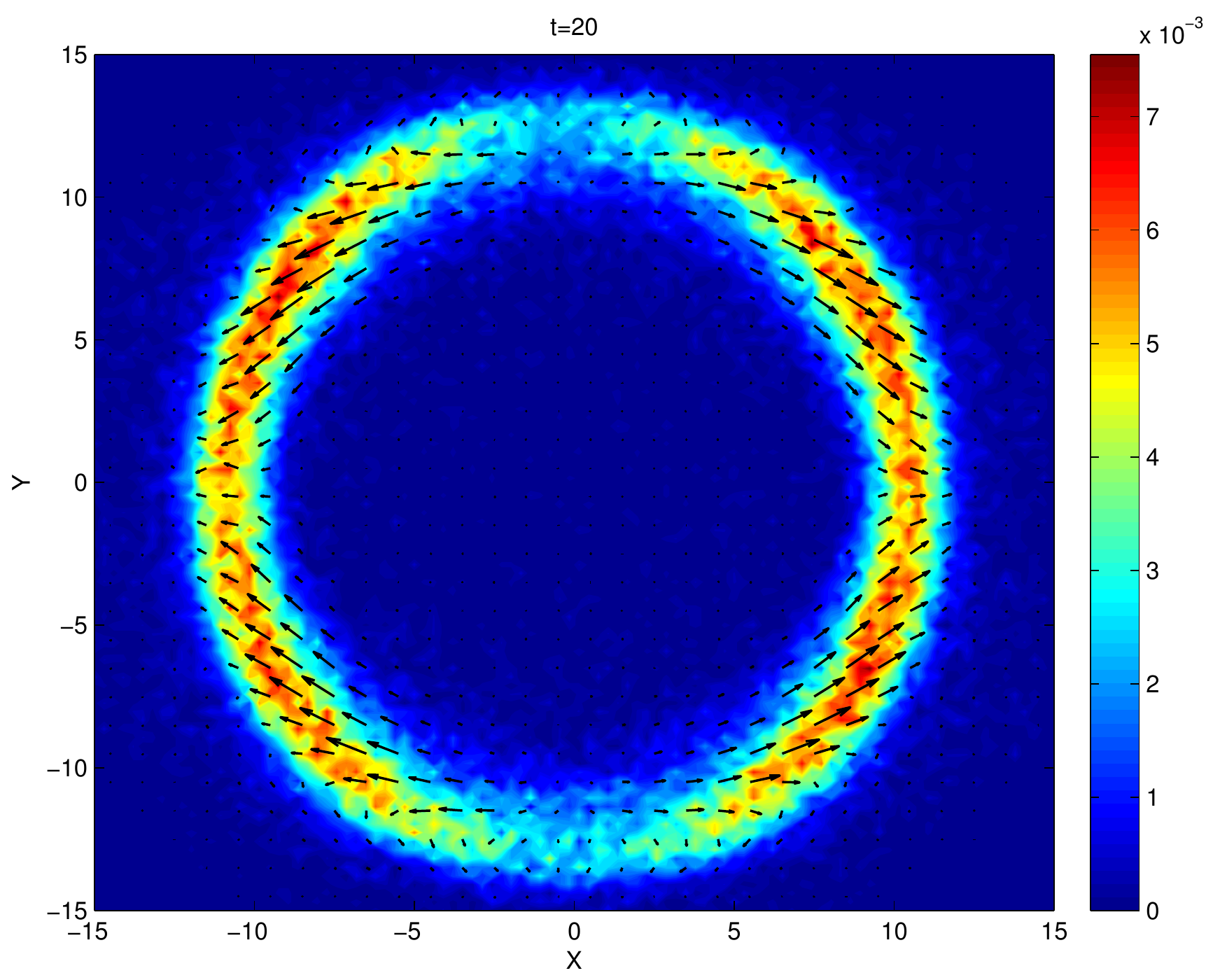}
\includegraphics[scale=0.4]{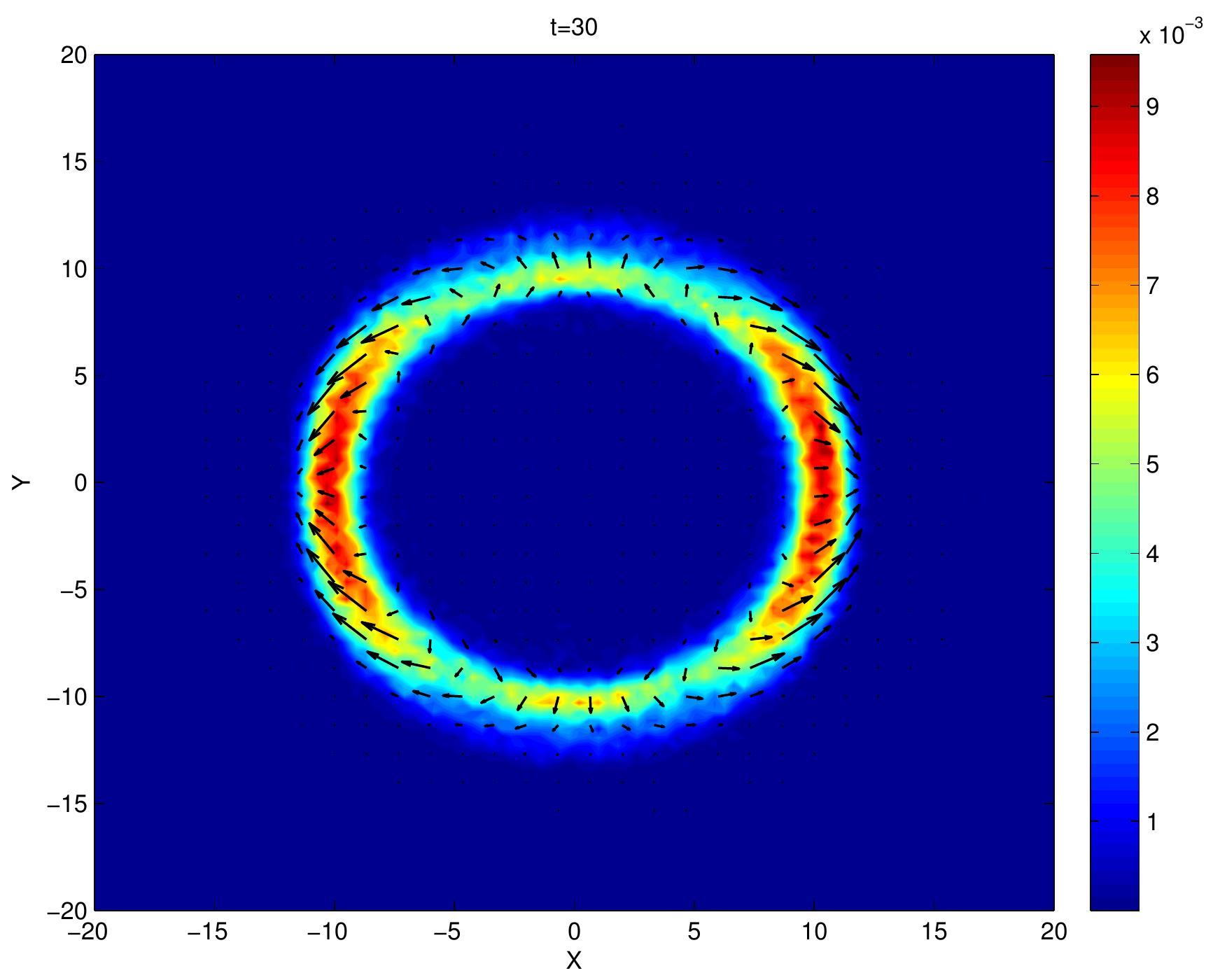}
\\
    \caption{2D Double Mill in D'Orsogna-Bertozzi et al. model at various times. Parameters in the attraction-repulsion potential are such that $C=C_R/C_A=1.6$, $l=l_R/l_A=0.025$, $\alpha=0.7$, $\beta=0.05$. Final configuration is reached after $t=30$}
    \label{fig:DblMill}
\end{figure}

In figure (\ref{fig:DblMill}), we instead consider the following initial data
$$f_0(x,y,v_x,v_y)=\frac{1}{4\pi^2\sigma^2}e^{\displaystyle-\frac{x^2+y^2}{2\sigma^2}}\left(e^{\displaystyle-\frac{(v_x+v_0)^2}{2}}+e^{\displaystyle-\frac{(v_x-v_0)^2}{2}}\right),$$ where $\sigma=\sqrt{2}$ and $v_0=0.5$. Thus density in space is a normal distribution centered in zero and velocity distribution has two main directions left and right. The evolution  computed with $ABMC$ and $\Delta t=\varepsilon$ shows that equilibrium is reached after $t=30$ in a stable double mill formation.

\subsection{3D simulations}
Finally we present some three dimensional simulations for the models taking into account the different effects of the thee zone dynamic. All the simulations have been performed with $ABMC$ and $\Delta t=\varepsilon=0.01$.

\paragraph{Bertozzi-D'Orsogna et al. model} First we consider the tridimensional extension of the previous simulation for the Bertozzi-D'Orsogna model et al. Initial data is uniformly distributed in space on a 3D-torus, and initial velocity is described by a circular motion in the $(x,y)$ components in $z$ direction initial velocity has no influence.

\begin{figure}[!h]
\centering
\includegraphics[scale=0.45]{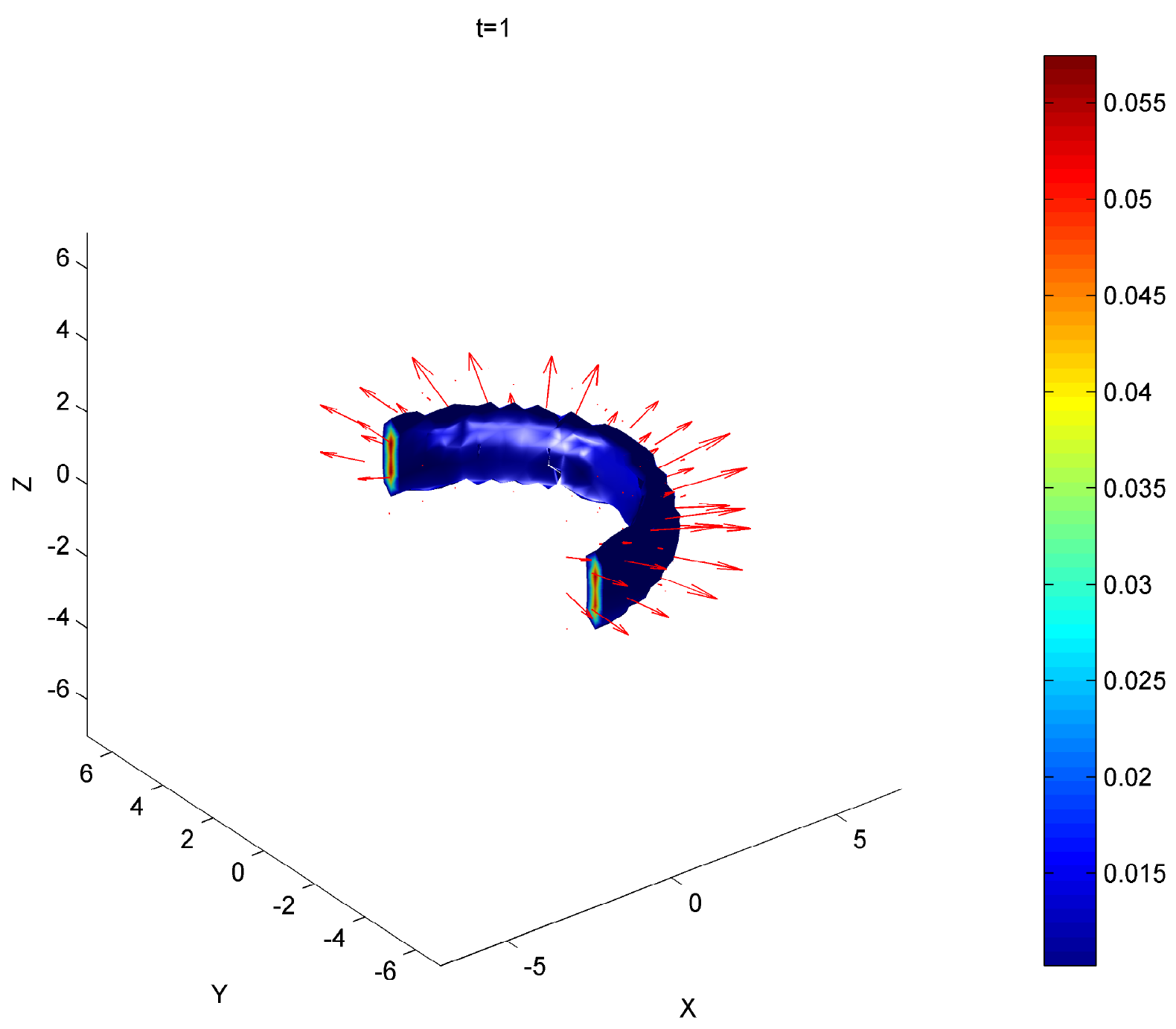}
\includegraphics[scale=0.45]{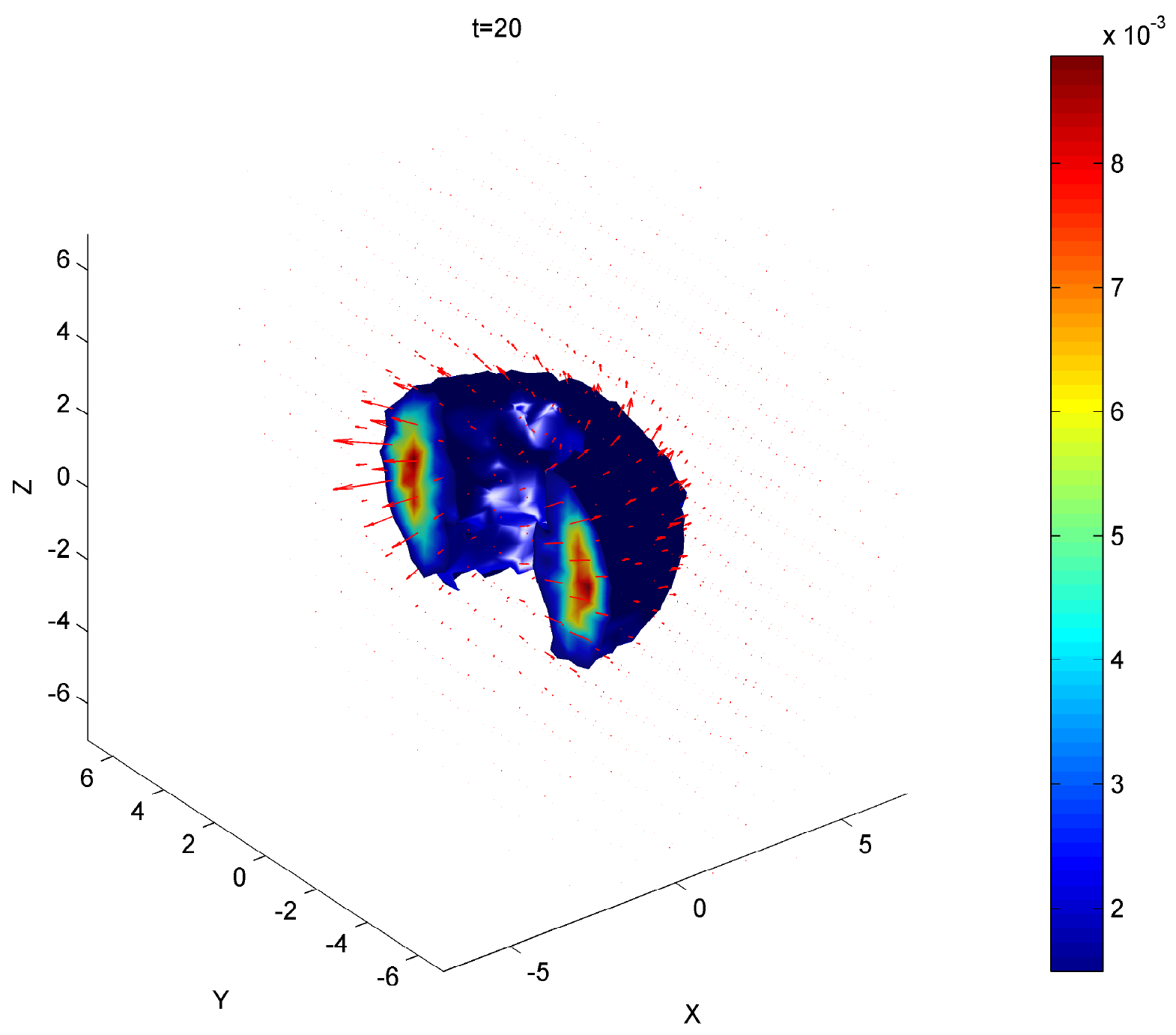}
\\
\includegraphics[scale=0.45]{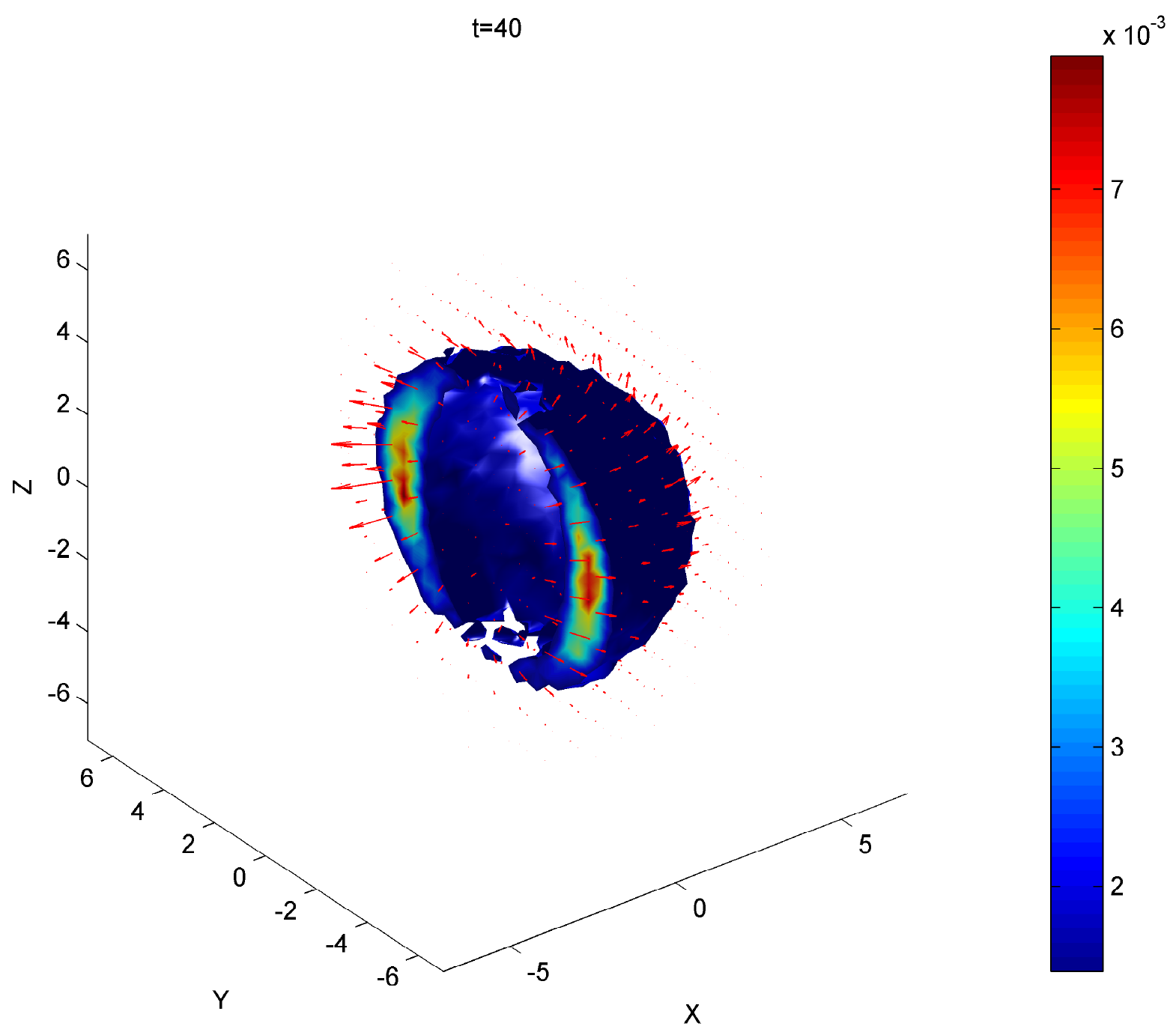}
\includegraphics[scale=0.45]{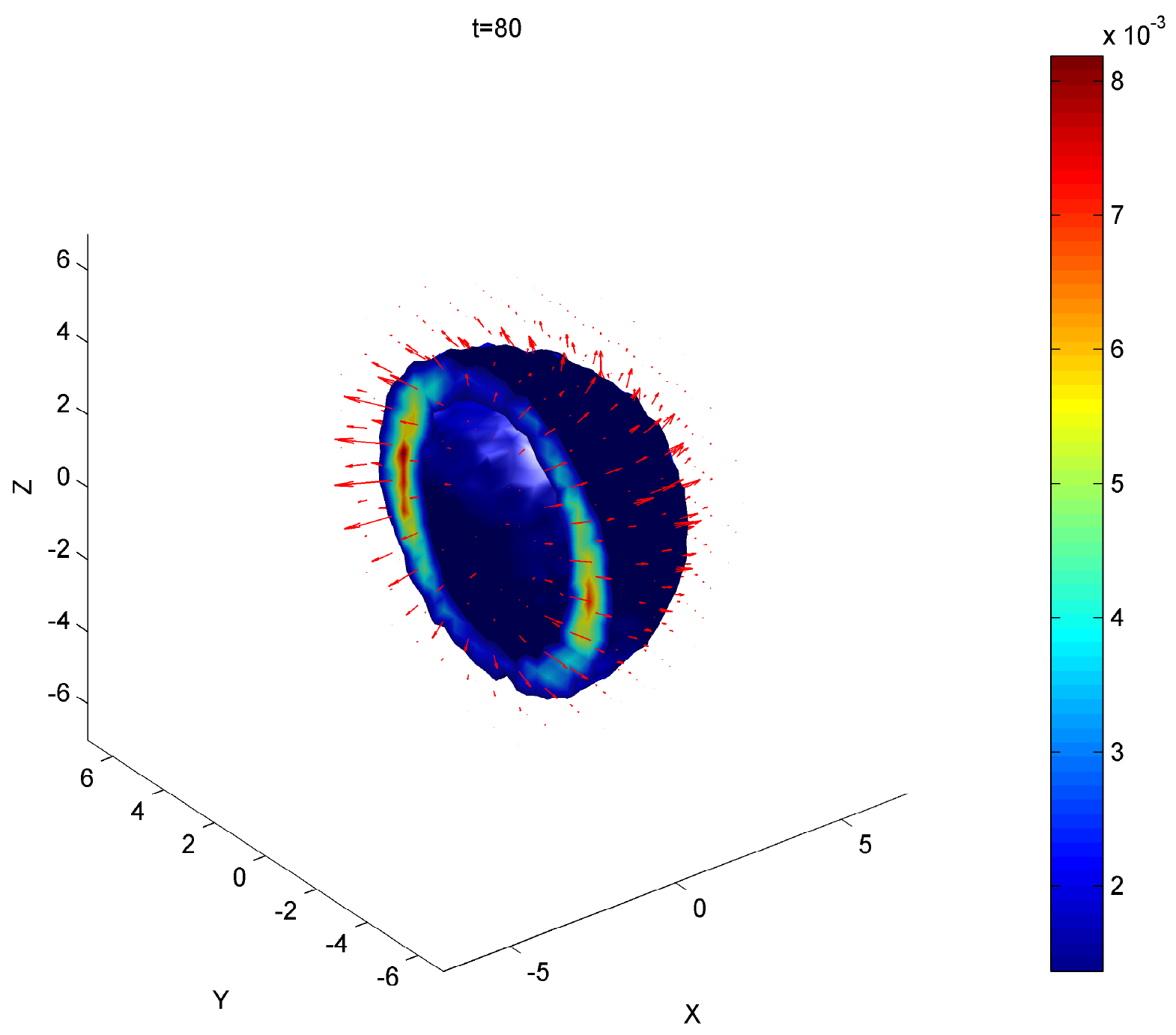}
    \caption{Evolution of the 3D mill in D'Orsogna-Bertozzi et al. model at different times.}
    \label{fig:Mill3d}
\end{figure}

We present the evolution of the swarm mass density and the vectorial field. The equilibrium reached after $t=80$ is a ball-shaped flock with mass concentrated on the border and empty zones in the middle, that is the typical configuration observed for a mill of a fish school.

Simulation is made taking in account N=200000 particles, and reconstructing the probability density function in the space we use a 3D grid with $\Delta x \times \Delta y\times\Delta z=100\times 100\times 100$.

\begin{figure}[!ht]
	\centering
		\includegraphics[scale=0.5]{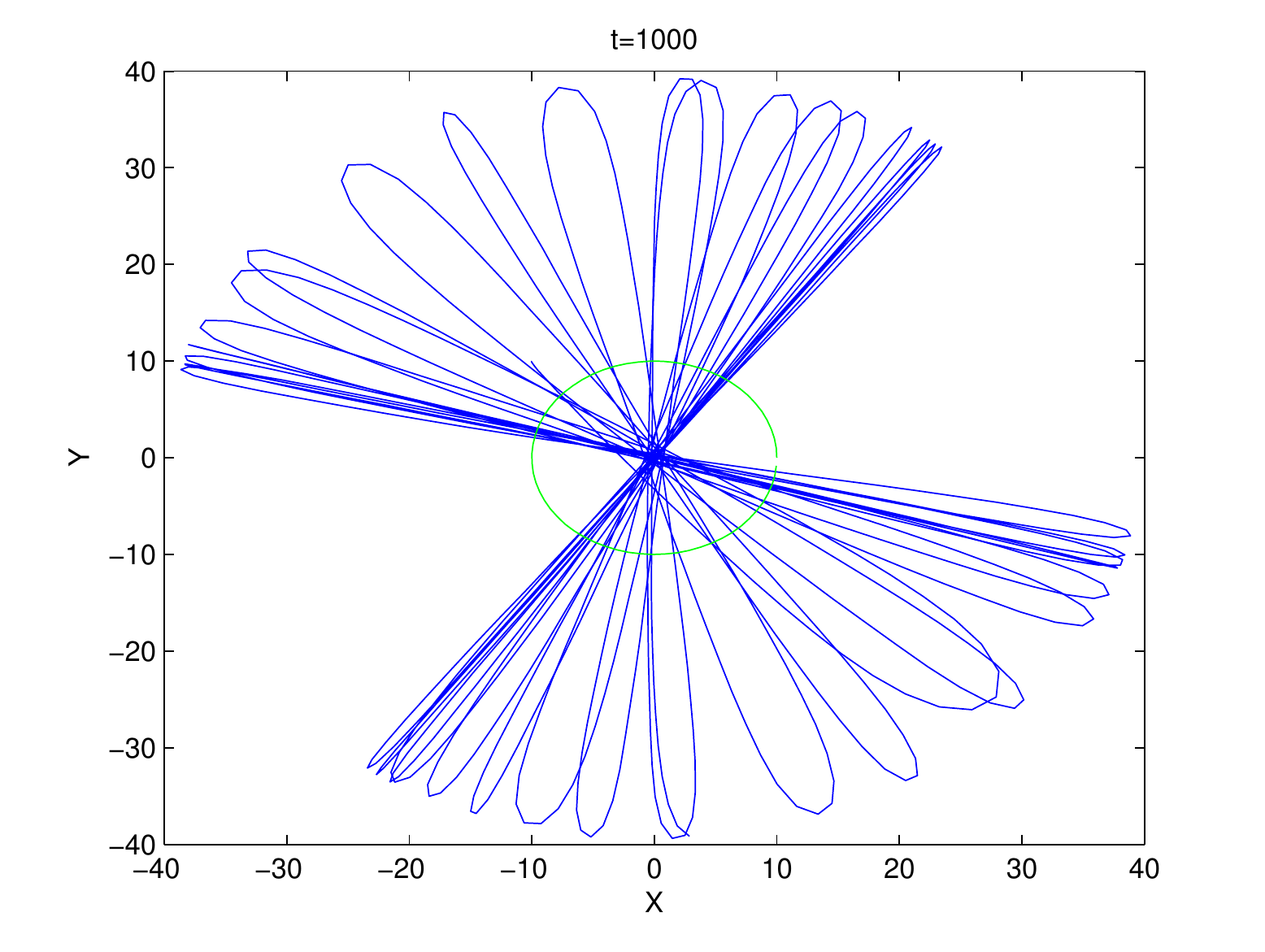}
	\label{fig:Roost_bar}
	\caption{Trajectory of the center of mass in the roosting dynamic}
\end{figure}

\paragraph{Roosting Force}
Accordingly to the work \cite{carrillo2010self} we introduce in the D'Orsogna-Bertozzi model a roosting force term.
The term expresses essentially the tendency of a flock or a school of fishes to stay around a certain zone. Such zones usually are of food interest or where birds settle their nests.
Different approaches can be used to model this biological behavior, see for example \cite{hildenbrandt2010self, ballerini2008interaction}.

Mathematically speaking such term can be described by the introduction of a force term of the type
\begin{equation}
F_{roost}=-\left[v_i^{\perp}\cdot\nabla\phi(x_i)\right]v_i^{\perp}.
\end{equation}
Such force gives the individuals a tendency to move towards the origin, for a suitable function $\phi$. 
Here $\phi$, called \emph{roosting potential} is a function $\phi:\mathbb{R}^d\longrightarrow\mathbb{R}$. In the simulation we take $$\phi(x)=\frac{d}{4}\left(\frac{|x|}{R_{roost}}\right)^4,$$ where $R_{roost}$ gives the roosting area radius, and $b$ is a constant weight. Other choice of this roost term are of course possible, we refer the interested reader to \cite{carrillo2010self}.

Starting from the following initial data $$f_0(x,y,z,v_x,v_y,v_z)=\frac{1}{4\pi^2\sqrt{2\pi}\sigma^3\nu^2}\exp\left\{\frac{1}{2\sigma^2}\left[(x-x_0)^2+(y-y_0)^2+(z-z_0)^2\right]+\frac{1}{2\nu^2}\left[v_x^2+v_y^2\right]\right\},$$ with $(x_0,y_0,z_0)=(-10,10,5)$, after a certain time the simulation shows a flock in stable equilibrium as an orbital motion around the roosting zone.

\begin{figure}[ht]
\centering
\includegraphics[scale=0.45]{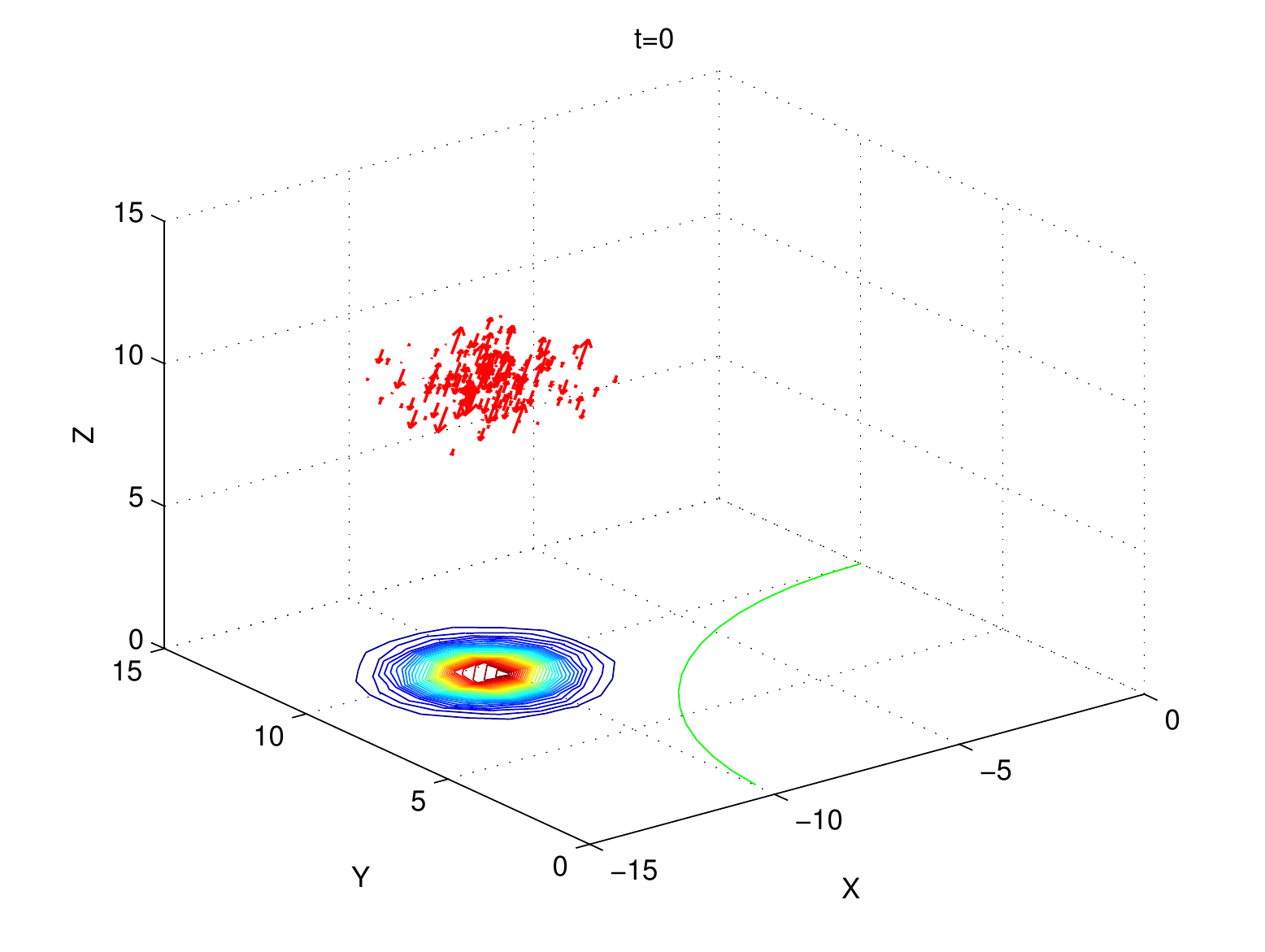}
\includegraphics[scale=0.45]{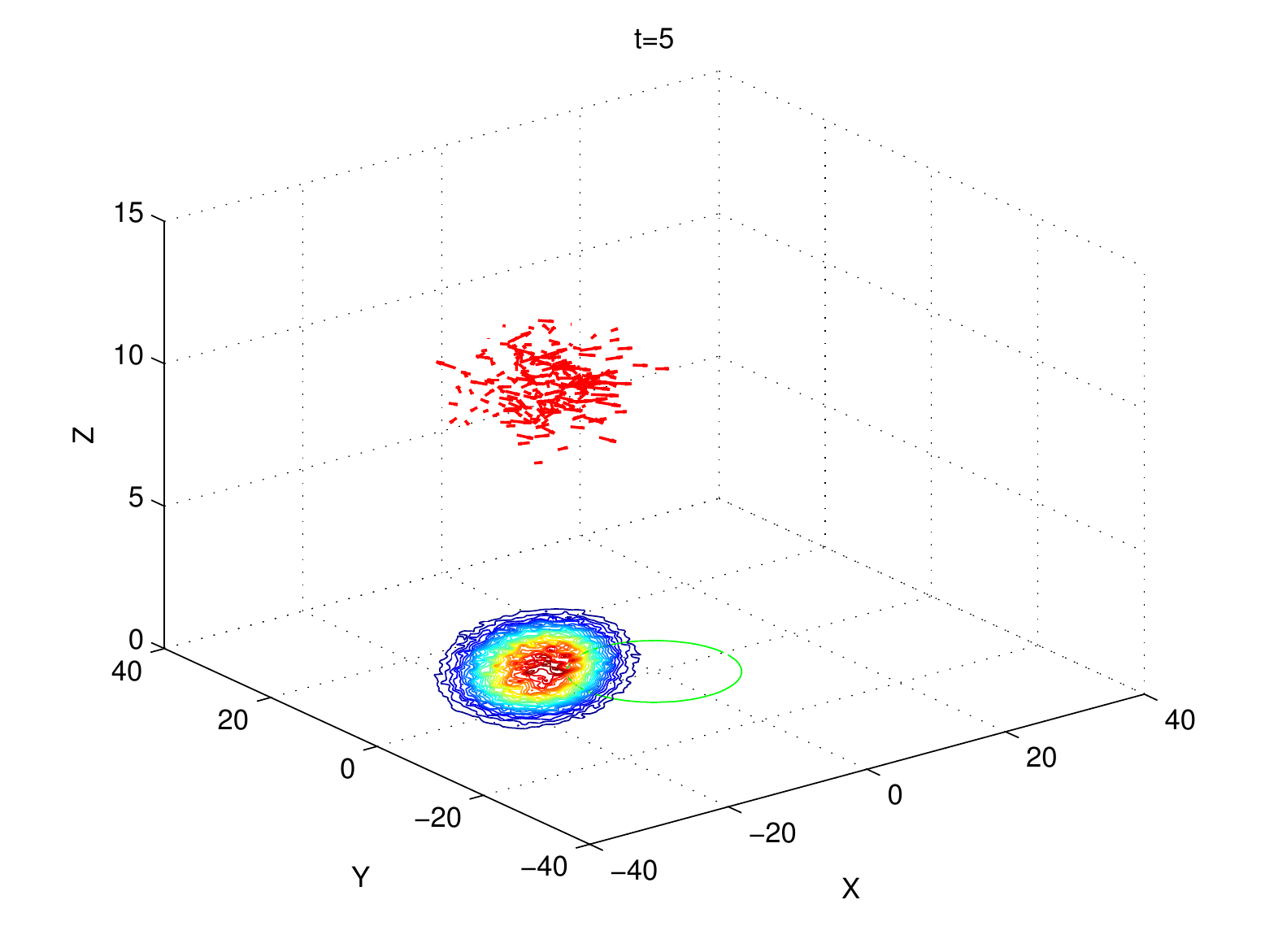}
\\
\includegraphics[scale=0.45]{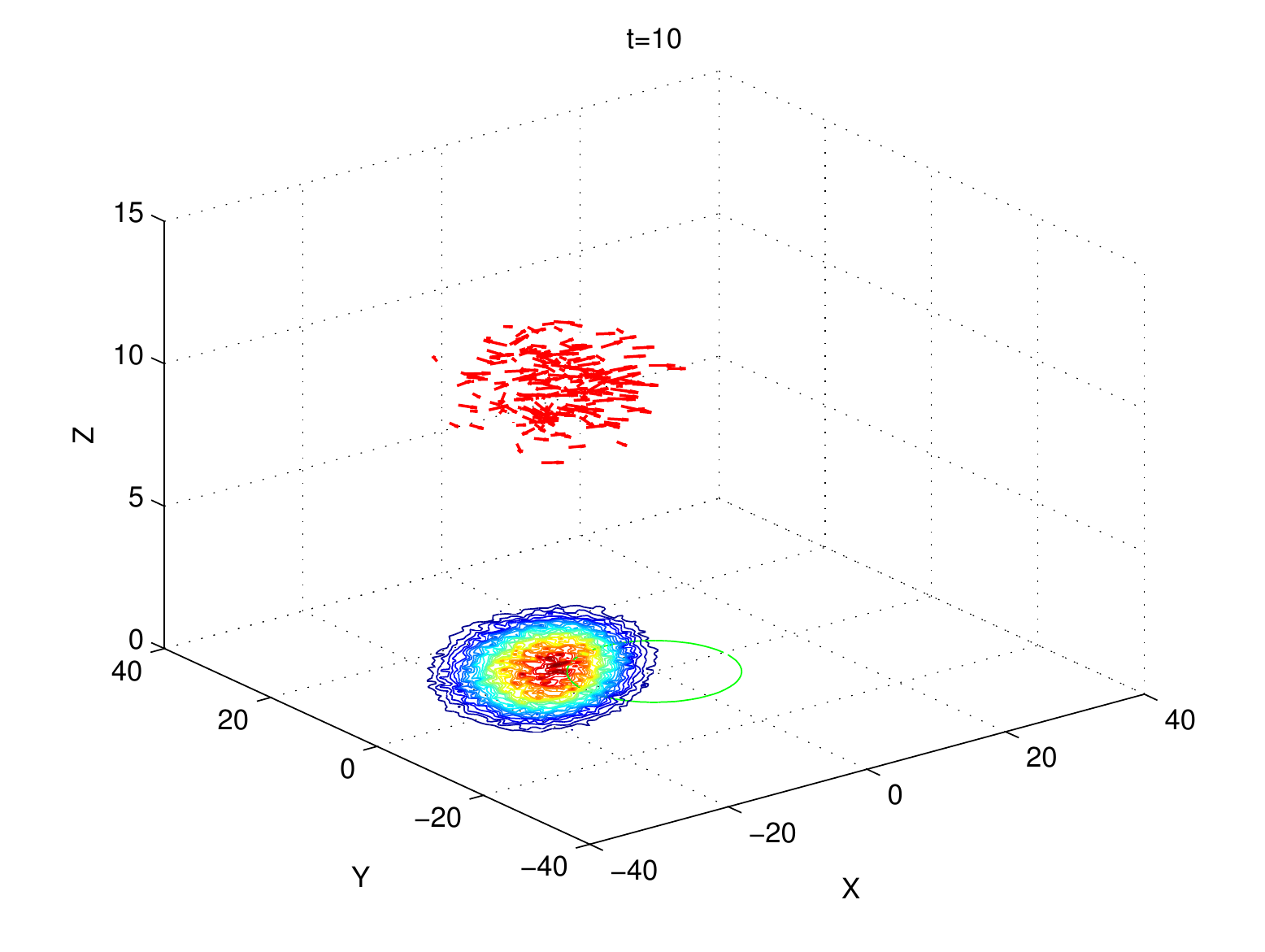}
\includegraphics[scale=0.45]{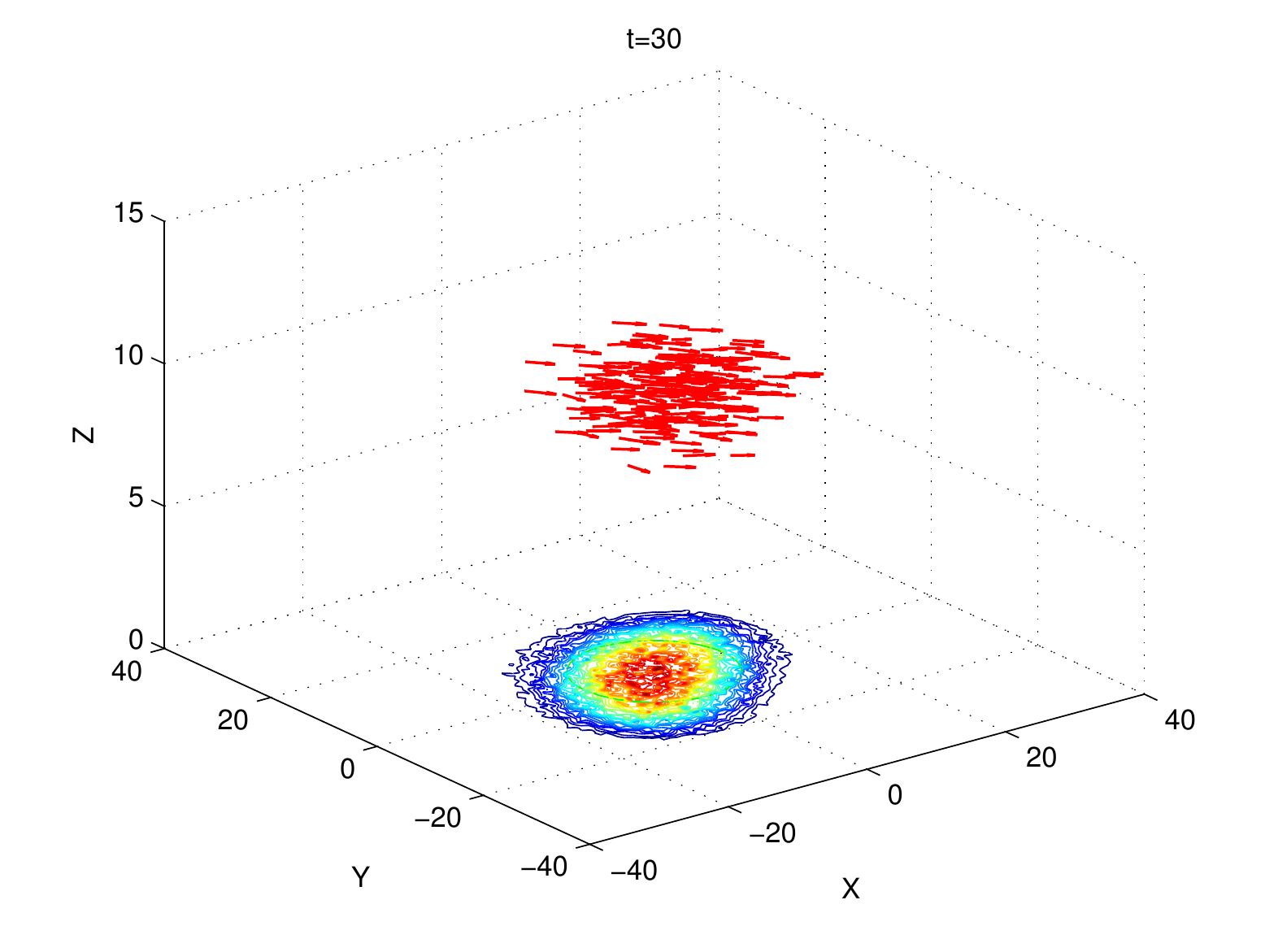}
\\
\includegraphics[scale=0.45]{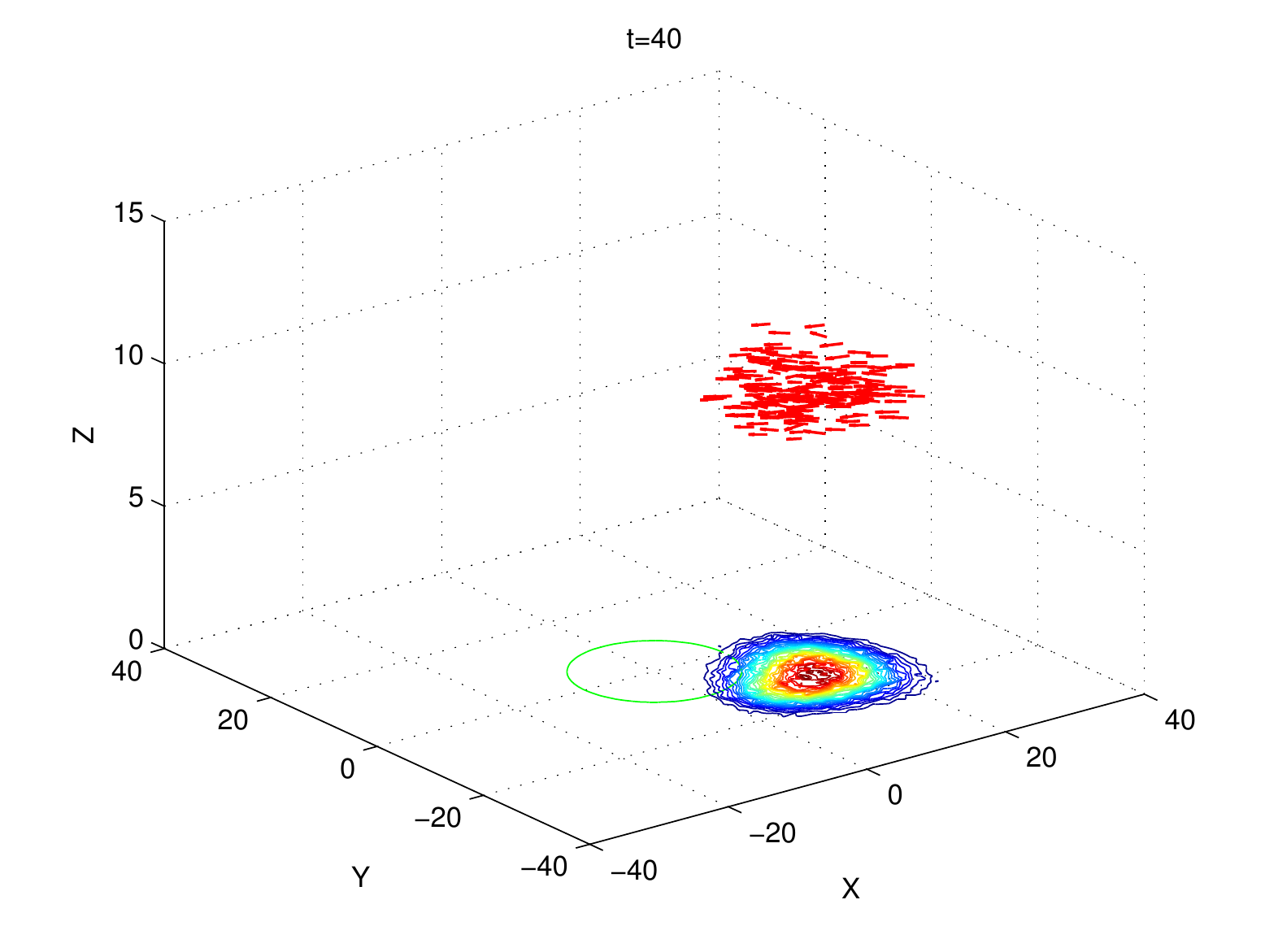}
\includegraphics[scale=0.45]{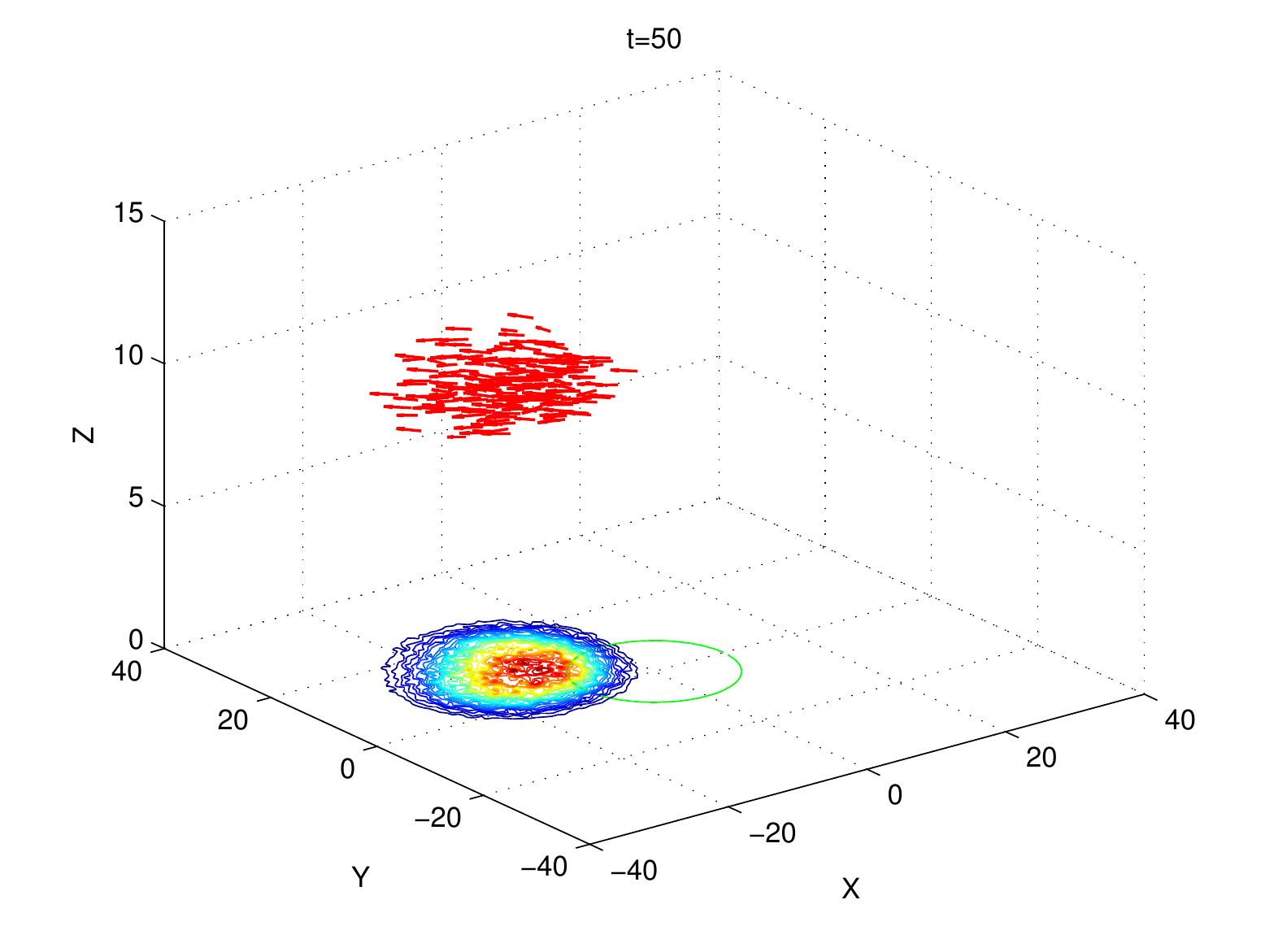}
    \caption{Evolution of the flock in 3D space, subject to a roosting force. The red arrow denotes the flock direction sampled from the initial population of $N=200000$ individuals, the green circle represents the roosting area. We also add the density distribution of the whole flock projected over the plane $(x,y)$.}
    \label{fig:Roost}
\end{figure}

The simulation takes in account the following parameters $C=C_R/C_A=30$, $l=l_R/l_A=3/5$, $\alpha=0.7$, $\beta=0.05$ and the term of roosting force with parameters $R_{roost}=10$ and $d=1/10$. 
For a long time simulation the center of mass describe the trajectory depicted in figure \ref{fig:Roost_bar}. Some configurations of the flock at different times obtained using $N=200000$ individuals are reported in figure \ref{fig:Roost}.

\section{Conclusion}
Mathematical modeling of collective behavior involves the interaction of several individuals (of the order of millions) which may be computationally highly demanding. Here we focus on models for flocking and swarming where the particle interactions implies an $O(N^2)$ cost for $N$ interacting objects. Using a probabilistic description based on a Boltzmann equation we show how it is possible to evaluate the interaction dynamic in only $O(N)$ computations. In particular we derive different approximation methods depending on a small parameter $\varepsilon$. 

The building block of the method is given by classical binary collision simulations techniques for rarefied gas dynamic. Beside the presence of a further scaling parameter the resulting algorithms are fully meshless and can be applied to several different microscopic flocking/swarming models. Applications of the present ideas to other interacting particle systems and comparison with fast multipole methods are under study and will be presented elsewhere.

\section*{Acknowledgements}
The authors would like to thank ICERM at Brown University for the kind hospitality during the 2011 Fall Semester Program on "Kinetic Theory and Computation" where part of this work has been carried on.

\end{document}